\documentclass[longauth]{aa} 
\DeclareUnicodeCharacter{2202}{$delta$}
\usepackage{tablefootnote}
\usepackage{footnote}
\usepackage{subcaption}
\usepackage{graphicx}
\usepackage{tabularx}
\usepackage{txfonts}
\usepackage[colorlinks=true,citecolor=blue,urlcolor=blue]{hyperref}
\usepackage{breakurl}
\begin{document}

\title{The Gaia-ESO Survey: homogenisation of stellar parameters and elemental abundances }


\author{A. Hourihane\inst{\ref{ioa}}
\fnmsep\thanks{aph@ast.cam.ac.uk}
\and P. Fran{\c c}ois\inst{\ref{oparis}}
\and C. C. Worley\inst{\ref{ioa},\ref{ucan}}
\and L. Magrini\inst{\ref{oarcetri}}
\and A. Gonneau\inst{\ref{ioa}}
\and A. R. Casey\inst{\ref{umonash}}
\and G. Gilmore\inst{\ref{ioa}}
\and S. Randich\inst{\ref{oarcetri}}
\and G. G. Sacco\inst{\ref{oarcetri}}
\and A. Recio-Blanco\inst{\ref{uca}}
\and A. J. Korn\inst{\ref{uuppsala-oa}}
\and C. Allende Prieto\inst{\ref{iac},\ref{ulaguna}}
\and R. Smiljanic\inst{\ref{ncac}}
\and R. Blomme\inst{\ref{obelgium}}
\and A. Lanzafame \inst{\ref{ocatania}}
\and A. Bragaglia\inst{\ref{obologna}}
\and N. A. Walton\inst{\ref{ioa}}
\and S. Van Eck\inst{\ref{ulbruxelles}}
\and T. Bensby\inst{\ref{ulund}}
\and A. Frasca\inst{\ref{ocatania}}
\and E. Franciosini\inst{\ref{oarcetri}}
\and F. Damiani\inst{\ref{opalermo}}
\and K. Lind\inst{\ref{ustockholm}}
\and M. Bergemann\inst{\ref{mpia},\ref{nbia}}
\and P. Bonifacio\inst{\ref{oparis-m}}
\and V. Hill\inst{\ref{uca}}
\and A. Lobel\inst{\ref{obelgium}}
\and D. Montes\inst{\ref{ucmadrid}}
\and D. K. Feuillet\inst{\ref{olund}}
\and G. Tautvai{\v s}ien{\. e}\inst{\ref{uvilnius}}
\and G. Guiglion\inst{\ref{mpia}}
\and H. M. Tabernero\inst{\ref{cda-vc}}
\and J. I. Gonz{\'a}lez Hern{\'a}ndez\inst{\ref{iac}}
\and M. Gebran\inst{\ref{notredame}}
\and M. Van der Swaelmen\inst{\ref{oarcetri}}
\and {\u S}. Mikolaitis\inst{\ref{uvilnius-ao}}
\and S. Daflon\inst{\ref{on-mcti}}
\and T. Merle\inst{\ref{ulbruxelles}}
\and T. Morel\inst{\ref{uliege}}
\and J. R. Lewis$^\dagger$\inst{\ref{ioa}}
\and E. A. Gonz{\'a}lez Solares\inst{\ref{ioa}}
\and D. N. A. Murphy\inst{\ref{ioa}}
\and R. D. Jeffries\inst{\ref{ukeele}}
\and R. J. Jackson\inst{\ref{ukeele}}
\and S. Feltzing\inst{\ref{olund}}
\and T. Prusti\inst{\ref{esa}}
\and G. Carraro\inst{\ref{upadova}}
\and K. Biazzo\inst{\ref{oroma}}
\and L. Prisinzano\inst{\ref{opalermo}}
\and P. Jofr{\'e}\inst{\ref{udp}}
\and S. Zaggia\inst{\ref{opadova}}
\and A. Drazdauskas\inst{\ref{uvilnius-ao}}
\and E. Stonkut{\'e}\inst{\ref{uvilnius-ao}}
\and E. Marfil\inst{\ref{cda-vc}}
\and F. Jim{\'e}nez-Esteban\inst{\ref{cda-vc}}
\and L. Mahy\inst{\ref{obelgium}}
\and M. L. Guti{\'e}rrez Albarr{\'a}n\inst{\ref{ucmadrid}}
\and S. R. Berlanas\inst{\ref{ualicante},\ref{ukeele}}
\and W. Santos\inst{\ref{on-mcti}}
\and L. Morbidelli\inst{\ref{oarcetri}}
\and L. Spina\inst{\ref{opadova}}
\and R. Minkevi\v{c}i\={u}t\.{e}\inst{\ref{uvilnius-ao}}
}

\institute{Institute of Astronomy, University of Cambridge, Madingley Road, Cambridge CB3 0HA, United Kingdom \\ \email{aph@ast.cam.ac.uk} \label{ioa}
\and GEPI, Observatoire de Paris, PSL Research University, CNRS, Paris Diderot Sorbonne Paris, 61 avenue de l'Observatoire, 75014, Paris, France \label{oparis}∂
\and School of Physical and Chemical Sciences --- Te Kura Mat\={u}, University of Canterbury, Private Bag 4800, Christchurch 8140, New Zealand\label{ucan}
\and INAF - Osservatorio Astrofisico di Arcetri, Largo E. Fermi, 5, 50125, Firenze, Italy\label{oarcetri}
\and School of Physics \& Astronomy, Monash University, Wellington Road, Clayton 3800, Victoria, Australia\label{umonash}
\and Universit{\'e} C{\^o}te d'Azur, Observatoire de la C{\^o}te d'Azur, CNRS, Laboratoire Lagrange, Bd de l'Observatoire, CS 34229, 06304 Nice cedex 4, France\label{uca}
\and Observational Astrophysics, Division of Astronomy and Space Physics, Department of Physics and Astronomy, Uppsala University, Box 516, 75120 Uppsala, Sweden\label{uuppsala-oa}
\and Instituto de Astrof{\'i}sica de Canarias, V{\'i}a L{\'a}ctea s/n, E-38205 La Laguna, Tenerife, Spain\label{iac}
\and Departamento de Astrof{\'i}sica, Universidad de La Laguna, E-38205 La Laguna, Tenerife, Spain\label{ulaguna}
\and Nicolaus Copernicus Astronomical Center, Polish Academy of Sciences, ul. Bartycka 18, 00-716, Warsaw, Poland\label{ncac}
\and ROB - Royal Observatory of Belgium, Ringlaan 3, B-1180 Brussels, Belgium\label{obelgium}
\and INAF - Osservatorio di Astrofisica e Scienza dello Spazio, via P. Gobetti 93/3, 40129 Bologna, Italy\label{obologna}
\and Institut d'Astronomie et d'Astrophysique, Universit{\'e} Libre de Bruxelles, CP 226, Boulevard du Triomphe, B-1050 Bruxelles, Belgium\label{ulbruxelles}
\and Division of Astrophysics, Department of Physics, Lund University, Box 43, SE-22100 Lund, Sweden\label{ulund}
\and INAF - Osservatorio Astrofisico di Catania, Via S. Sofia 78, 95123 Catania, Italy\label{ocatania}
\and INAF - Osservatorio Asronomico di Palermo, Piazza del Parlamento, 1 90134 Palermo, Italy\label{opalermo}
\and Department of Astronomy, Stockholm University, AlbaNova University Center, SE-106 91 Stockholm, Sweden\label{ustockholm}
\and Max-Planck-Institut fur Astronomie, K{\"o}nigstuhl 17, D-69117 Heidelberg, Germany\label{mpia}
\and Niels Bohr International Academy, Niels Bohr Institute, Blegdamsvej 17, DK-2100 Copenhagen {\O}, Denmark\label{nbia}
\and GEPI, Observatoire de Paris, Universit{\'e} PSL, CNRS,  5 Place Jules Janssen, 92190 Meudon, France\label{oparis-m}
\and Departamento de F{\'i}sica de la Tierra y Astrof{\'i}sica \& IPARCOS-UCM (Instituto de F{\'i}sica de Partículas y del Cosmos de la UCM), Facultad de Ciencias F{\'i}sicas, Universidad Complutense de Madrid, E-28040 Madrid, Spain\label{ucmadrid}
\and Lund Observatory, Department of Astronomy and Theoretical Physics, Lund, Sweden \label{olund}
\and Institute of Theoretical Physics and Astronomy, Vilnius University, Sauletekio av. 3, LT-10257 Vilnius, Lithuania\label{uvilnius}
\and Centro de Astrobiologia (CSIS-INTA), Departamento di astrofisica, campus ESAC, Camino bajo de castillo 28  692 Villanueva de la Ca{\~n}ada, Madrid, Spain.\label{cda-vc}
\and Department of Chemistry and Physics, Saint Mary’s College, Notre Dame, IN 46556, USA\label{notredame}
\and Astronomical  Observatory,  Institute  of  Theoretical  Physics  and  Astronomy,  Vilnius  University,  Sauletekio  av.  3,  10257  Vilnius, Lithuania\label{uvilnius-ao}
\and Observat{\'o}rio Nacional - MCTI (ON), Rua Gal. Jos{\'e} Cristino 77, S{\~a}o Crist{\'o}v{\~a}o, 20921-400, Rio de Janeiro, Brazil\label{on-mcti}
\and Space Sciences, Technologies, and Astrophysics Research (STAR) Institute, Universit{\'e} de Li{\`e}ge, Quartier Agora, B{\^a}t B5c, All{\'e}e du 6 ao{\^u}t, 19c, 4000 Li{\`e}ge, Belgium\label{uliege}
\and Astrophysics Group, Keele University, Keele, Staffordshire ST5 5BG, United Kingdom\label{ukeele}
\and European Space Agency (ESA), European Space Research and Technology Centre (ESTEC), Keplerlaan 1, 2201 AZ Noordwijk, The Netherlands\label{esa}
\and Department of Physics and Astronomy, University of Padova, v. dell'Osservatorio 2, 35122, Padova, Italy\label{upadova}
\and INAF - Osservatorio Astronomico di Roma, Via Frascati 33, I00040 Monte Porzio Catone (Roma), Italy\label{oroma}
\and Nucleo Milenio ERIS \& Instituto de Estudios Astrofisicos, Universidad Diego Portales, Ej{\'e}rcito 441, Santiago, Chile\label{udp}
\and INAF - Osservatorio Astronomico di Padova, Vicolo dell’Osservatorio 5, I-35122, Padova, Italy\label{opadova}
\and Departamento de F{\'i}sica Aplicada, Facultad de Ciencias, Universidad de Alicante, 03690 San Vicente del Raspeig, Alicante, Spain\label{ualicante}
}

   \date{}

 
  \abstract
  {
  The Gaia-ESO Survey is a public spectroscopic survey that has targeted $\gtrsim10^5$ stars covering all major components of the Milky Way from the end of 2011 to 2018, delivering its public final release in May 2022. Unlike other spectroscopic surveys, Gaia-ESO is the only survey that observed stars across all spectral types with dedicated, specialised analyses: from O ($T_\mathrm{eff} \sim 30,000-52,000$~K) all the way to K-M ($\gtrsim$3,500~K). The physics throughout these stellar regimes varies significantly, which has previously prohibited any detailed comparisons between stars of significantly different type. In the final data release (internal data release 6) of the Gaia-ESO Survey, we provide the final database containing a large number of products such as radial velocities, stellar parameters and elemental abundances, rotational velocity, and also, e.g.,  activity and accretion indicators in young stars and membership probability in star clusters 
  for more than 114,000 stars. The spectral analysis is coordinated by a number of Working Groups (WGs) within the Survey, which specialise in the various stellar samples. Common targets are analysed across WGs to allow for comparisons (and calibrations) amongst instrumental setups and spectral types. Here we describe the procedures employed to ensure all Survey results are placed on a common scale to arrive at a single set of recommended results for all Survey collaborators to use. We also present some general quality and consistency checks performed over all Survey results. }

   \maketitle
%
\section{Introduction}
The launch of the European Space Agency's astrometric \textit{Gaia} mission in 2013 \citep[e.g.][]{perryman01,prusti,gaiadr2,gaiadr3} has prompted a new wave of Galactic studies. \textit{Gaia} is delivering precise distances and kinematics, photometry and spectrophotometry for more than 1.5 billion stars as well as radial velocities and chemical abundances for the brighter stars in the sample \citep[See further][]{gaiadr3,recio-blanco2022}. A variety of ground-based spectroscopic surveys have been carried out since the 2010's to collect complementary stellar parameters, elemental abundances and radial velocities, which, when combined with \textit{Gaia} astrometry, have the power to revolutionise our view of the Milky Way. Spectroscopy breaks the degeneracy between foreground extinction and stellar temperature to which the \textit{Gaia} Blue Photometer and Red Photometer Prism (BP/RP) spectrophotometry data alone are susceptible \citep{bailer-jones11}. The last decade saw the advent of such spectroscopic surveys in anticipation of \textit{Gaia}'s exquisite astrometry. Surveys such as the Radial Velocity Experiment  \citep[RAVE][]{rave}, GALactic Archeology with HERMES \citep[GALAH][]{desilva15}, the Apache Point Observatory Galactic Evolution Experiment \citep[APOGEE][]{Majewski17}, and the Gaia-ESO Survey \citep{randich22, gilmore22} are complete or close to complete, and the next generation of multi-fibre spectroscopic surveys are underway -- such as WEAVE \citep{weave2022, weave2012}, already taking observations, and 4MOST \citep{4most2014} and MOONS \citep{moons2020}, which are dedicating significant effort to operations in advance of observations, and SDSS-V, under development -- all using multi-object spectroscopy to provide detailed chemical and kinematic information for statistically significant samples of Milky Way stars.

In order to be able to draw accurate conclusions about Galactic structure, formation and evolution, we need a set of consistent measurements of stellar properties across the Hertzsprung-Russell (H-R) diagram, sampling all major components of the Milky Way. The Gaia-ESO Survey (hereafter, GES) has done this for the first time at high resolution on an 8-m class telescope, from the southern hemisphere. GES is a large pan-European effort, employing the VLT FLAMES instrument \citep{pasquini02} to obtain high quality spectra of $\sim 10^5$ stars across the H-R diagram. GES is producing stellar atmospheric parameters, elemental abundances and radial velocities for all stellar populations, which span the Galaxy from the halo to star forming regions, sampling the thin and thick discs, the bulge and open and globular clusters. 

The large number of spectra harvested by spectroscopic surveys in the current era requires an automated analysis procedure. Typically, one pipeline is developed and applied to all of the spectra within a survey. It is well known, however, that the various spectral analysis methods suffer from strong systematics, due in part to factors such as the choice of atmosphere models, or the atomic and molecular transitions employed. Within the Gaia-ESO Survey, considerable effort has been invested to improve the quality of the input line lists \citep{ruffoni14,heiter15b, heiter21}.

The Gaia-ESO Survey has a unique analysis structure \citep{gilmore22, randich22}. While the model atmospheres and the atomic/molecular data are fixed, the data are analysed by multiple different analysis teams hereafter referred to as "nodes". Each node runs a pipeline that generally employs a different method from other nodes, executed by experienced spectroscopists that are familiar with the pipeline. In this regard, the Gaia-ESO Survey is at a unique advantage to all other spectroscopic surveys: \emph{almost every spectroscopic analysis method ever considered is included in the Survey}, allowing us to make the first objective comparison between analysis methods, characterise the level of the systematics present in stellar spectroscopy, characterise the random and systematic uncertainty contributions for all measurements, and provide a robust ensemble measurement of stellar parameters and elemental abundances for the Survey.
 
Within GES, a major focus has been placed on producing stellar parameters that are both internally self-consistent and externally calibrated, with respect to a well-determined calibration sample of benchmark stars \citep{jofre14, blanco-cuaresma14}. The GES spectra not only cover a wide range of stellar populations (and thus parameter space) and are analysed with a variety of pipelines, but they are also taken with a variety of instrumental configurations designed to cover the characteristic spectral features of each stellar spectral type. 

The effort to transform such an inhomogeneous set of data and results onto a single, self-consistent scale is non-trivial. Essential to this process is the availability of a comprehensive set of calibrators across the H-R diagram. These calibrators include globular and open clusters spanning a wide range in metallicity as well as the \textit{Gaia} benchmark stars. The  design of the observational calibration programme for GES is described in \citet{pancino17}. Additionally, to facilitate exploitation of all current and future spectroscopic surveys we need a practical cross-survey calibration strategy with other Southern and Northern surveys. This requires both the analysis of a common set of calibration targets and the placing of the stellar parameter and abundance results on a consistent physical scale. GES takes an important step towards a cross-survey calibration by defining this scale.

In this paper, we present the strategy used to homogenise the GES stellar parameters, elemental abundances and radial velocities and discuss the challenges faced in attempting to define a self-consistent, externally-calibrated scale for such a broad parameter range and for such a wide variety of analysis pipelines and methods.
The structure of the paper is the following: \\
Section~\ref{sec_2} presents the observations carried out for the GES project and the distribution of the tasks among the different Working Groups. In Section~\ref{sec_3}, the multi-method, multi-pipeline design of the GES analysis is described along with the homogenisation workflow. Section \ref{sec:par} presents the set of quality checks and tools used to provide the users with a  global set of stellar parameters that will be used to compute the elemental abundances. 
In Section~\ref{sec:abu}, we describe how  the abundance homogenisation has been performed. Section~\ref{sec:QC} gathers the sequence of quality checks used to validate the homogenisation process element by element. Section~\ref{sec:rv_vsini} reports the results of the radial velocity determination of the GES sample and the homogenisation process. In Section~\ref{sec:errors} we report the determination of  the errors for the stellar parameters and the abundances. Section~\ref{sec:flags} makes a short report on the propagation of the technical and 
peculiar sets of flags defined by WG14 that can be used to trace the analysis of the spectra and their quality. 
The last section (Sec.~\ref{sec:discussion}) summarises the results of the GES survey and presents broad comparisons with other spectroscopic surveys.


\section{Observations and spectral analysis workflow}
\label{sec_2}
Gaia-ESO was awarded 300 nights as an ESO Public Spectroscopic Survey on the VLT, with an additional 40 nights later granted to compensate for bad weather and technical downtime. Observations were taken between December 2011 and January 2018 using the FLAMES spectrograph \citep{Pasquini2002} in multi-object spectroscopy mode with the GIRAFFE\footnote{\url{https://www.eso.org/sci/facilities/paranal/instruments/flames/inst/Giraffe.html}} ($\sim$ 140 fibres) and UVES \citep{dekker00} (8 fibres for U580\footnote{One fibre of the eight was broken}, 6 for U520) instruments. The wavelength ranges of the instrumental set-ups used are listed in Table 1. The GIRAFFE spectra are reduced using the dedicated Cambridge Astronomical Survey Unit (CASU) pipeline \citep{gilmore22}. The UVES spectra are reduced using a modified version of the ESO pipeline \citep{modigliani04}. 
GES observing blocks are split into two or more exposures and individual spectra are stacked to produce nightly stacked spectra for each field. When all observations from a particular field are complete (certain fields are repeated across nights with arbitrary separation), all spectra for an object are stacked to produce a final stacked spectrum known as a ``singlespec'' for each object or ``CNAME'' (the GES object name based on its coordinates, equivalent to the ESO ``OBJECT''). Where spectra are available in the ESO archive for the GES calibrators and objects in cluster fields in the GES instrumental setups, these have been retrieved, reduced with the GES pipelines and added to the GES dataset.
Radial velocities (RVs) are determined for all individual and stacked spectra \citep{sacco14, gilmore22}.

\begin{figure}
  \resizebox{\hsize}{!}{\includegraphics{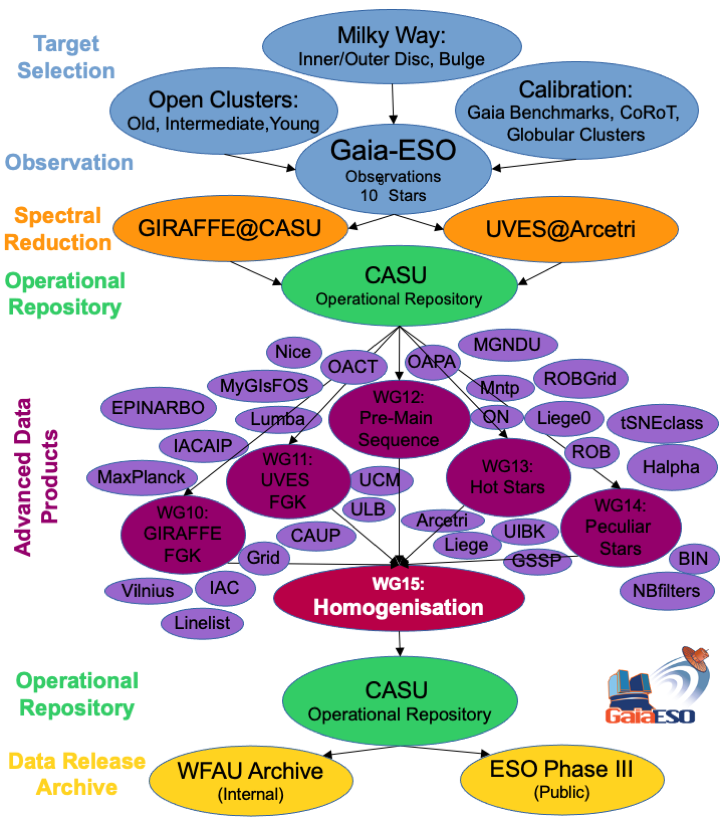}}
  \caption{Gaia-ESO data flow diagram for iDR6 showing the key stages from target selection to data release via the archives.} 
  \label{fig:idr6_dataflow_ppt}
\end{figure}

\begin{figure*}
\centering
 \includegraphics[width=0.775\linewidth]{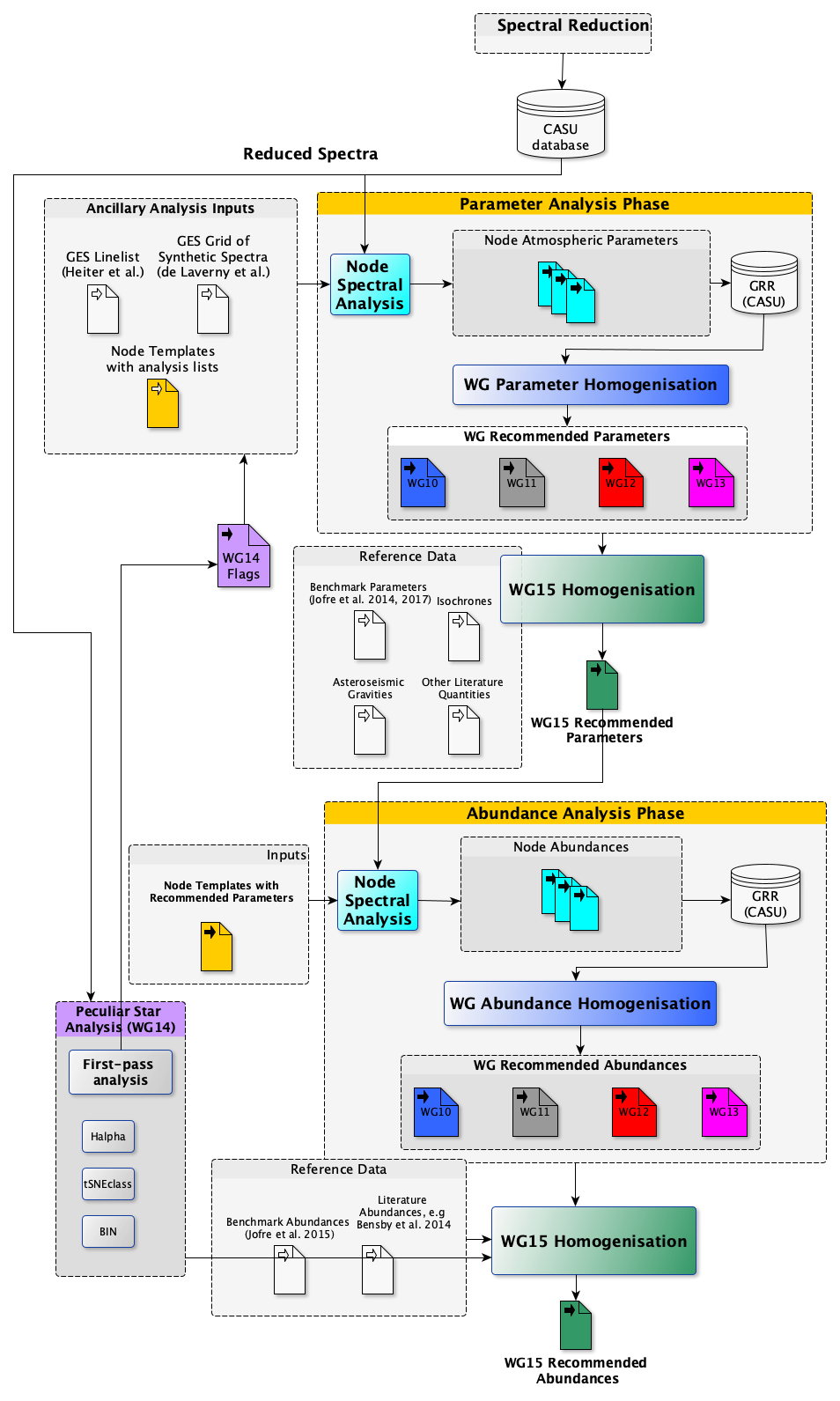}
  \caption{Gaia-ESO data processing diagram for iDR6 showing the complexity of the interfaces between the reduction, parameter, abundance and homogenisation processes. The GRR is the GES Results Repository at CASU.} 
  \label{fig:idr6_dataflow_yed}
\end{figure*}

GES internal Data Release (iDR) cycles consist of the following general procedure which is illustrated in Figs.~\ref{fig:idr6_dataflow_ppt} and \ref{fig:idr6_dataflow_yed}. 
In Fig.~\ref{fig:idr6_dataflow_ppt}, the general flow is described: 
targets are selected under the three programmes (Open Clusters, Milky Way, Calibrations) which are observed as necessary using UVES and GIRAFFE  \citep[see][for the target selection in the three categories, respectively]{randich22, gilmore22, pancino17};  raw spectra from a selected time period are reduced and released to the spectral analysis teams from the operational database at CASU in a standard FITS format \citep{wells81} with radial velocities and useful ancillary information such as observing parameters and photometry attached in FITS extensions (the spectral metadata). The teams analyse the data and return catalogues of their results which are then homogenised to produce a final catalogue of recommended results. Six data analysis cycles (iDRs) were completed as part of GES \citep[See][]{randich22}.  
 In Fig.~\ref{fig:idr6_dataflow_yed}, the internal analysis and homogenisation steps are highlighted: the first phase of the determination of the stellar parameters, followed by a homogenisation per WG and a general homogenisation operated by WG15; the second phase of determination of abundances, using the homogenised stellar parameters as input, and the definition of the final set of abundances passing through the WG homogenisation and the final WG15 validation. All steps are supported by the use of calibrators. 

The structure of the analysis teams in GES is described briefly here. More detail is provided in \cite{gilmore22}. 

There are four spectral analysis WGs dedicated to the analysis of different samples of stars within the GES consortium. Multiple analysis nodes operate their specific pipelines on the data within each WG. A further WG is dedicated to the characterisation of outliers:
\begin{itemize}
\item[1.] {\sc WG10}: this {\sc WG} analyses the GIRAFFE spectra of FGK stars both in the Milky Way (MW) and in open and globular clusters.
Four different GIRAFFE setups are analysed by {\sc WG10}: {\em i)} HR10-HR21 for  MW field stars; {\em ii)} HR15N for FGKM stars in open clusters;
{\em iii)} HR9B for earlier type stars in open clusters; {\em iv)} HR21 for  MW bulge stars.
\item[2.] {\sc WG11}: this {\sc WG} analyses  the UVES spectra of FGK stars both in the Milky Way and in open and globular clusters.
Two setups are used: U580 for late-type stars, and U520 for early-type stars.
\item[3.] {\sc WG12}:  this {\sc WG} analyses  the UVES and GIRAFFE spectra of main- and pre-main sequence stars (PMS) in young open clusters
with U580 for UVES and HR15N for GIRAFFE.
\item[4.] {\sc WG13}:  this {\sc WG} analyses the UVES and GIRAFFE spectra of stars in young clusters containing OBA-type stars.
For UVES the setup is U520, while for GIRAFFE HR3, HR4, HR5B, HR6, HR9B and HR14A are used.
\item[5.] {\sc WG14}:  this {\sc WG} works to identify and characterise outlier stars from the whole Survey. These are stars presenting peculiarities which endanger the determination of their stellar parameters and abundances by standard routines. The stars are tagged using a specially developed library of flags to allow filtering of the dataset during the homogenisation and scientific analyses.
\end{itemize}

The characteristics of each setup in UVES\footnote{\url{https://www.eso.org/sci/facilities/paranal/instruments/uves/doc/ESO\_411892\_User\_Manual\_P109.pdf}} and GIRAFFE\footnote{\url{https://www.eso.org/sci/facilities/paranal/instruments/flames/inst/specs1.html}} in which targets were observed for Gaia-ESO, and the working groups that analysed each setup are given in Table~\ref{tab:epsetups}.

\begin{table*}
\begin{center}
\caption{Table of instrumental set-ups used by the analysis Working Groups within GES. The GIRAFFE instrument was refocused in February 2015. The R$_\mathrm{old}$ values refer to pre-refocusing values and R gives the new values, post-refocusing.} 
\begin{tabular}{llccccl}
\hline\hline
Instrument & Setup & $\lambda_{\mathrm{min}}$ (\AA) & $\lambda_{\mathrm{max}}$ (\AA) & R$_\mathrm{old}$ & R  & WG\\
\hline 
UVES & 520 & 4140 & 6210 & 47000 & 47000  & 11,12,13,14 \\
UVES & 580 & 4760 & 6840 & 47000 & 47000 & 11,12,13,14 \\
GIRAFFE & HR3 & 4033 & 4201 & 24800 & 31400 & 13,14 \\
GIRAFFE & HR4 & 4188 & 4392 & 32500 & 35550 & 13,14 \\
GIRAFFE & HR5A & 4340 & 4587 & 18470 & 20250 & 13,14 \\
GIRAFFE & HR6 & 4538 & 4759 & 20350 & 24300 &  13,14 \\
GIRAFFE & HR9B & 5143 & 5356 & 25900 & 31750  & 10,13,14 \\
GIRAFFE & HR10 & 5339 & 5619 & 19800 & 21500 & 10,12,14 \\
GIRAFFE & HR14A & 6308 & 6701 & 17740 & 18000 & 13,14 \\
GIRAFFE & HR15N & 6470 & 6790 & 17000 & 19200 & 10,12,14 \\
GIRAFFE & HR21 & 8484 & 9001 & 16200 & 18000 & 10,12,14 \\
\hline \hline
\end{tabular}
\label{tab:epsetups}\\
\end{center}
\end{table*}

Each WG lead collates the results from their nodes and performs an initial homogenisation to put the parameters on a consistent scale for the WG. The interim results including node results files and WG Recommended results are delivered to the CASU operational database pending final homogenisation and delivery to the internal GES Science Archive at WFAU. The final homogenised results form the basis of the public Gaia-ESO catalogues delivered to the ESO archive.

A detailed description of the contents of each internal Data Release and each public data release via ESO is contained in \citet{randich22}.

\subsection{Calibration samples}

GES makes use of the set of calibrators selected by the Calibration and Standards Working Group (WG5), which is responsible for the observational calibration strategy for GES \citep{pancino17}. By using the calibrators in a uniform way, current precise knowledge for select samples is extended to much larger samples of stars, covering a wider parameter space. 

\begin{figure*}
    \centering
    \begin{minipage}{.48\textwidth}
        \centering
        \includegraphics[width=1\linewidth,trim=4cm 1.5cm 4cm 1.0cm,clip]{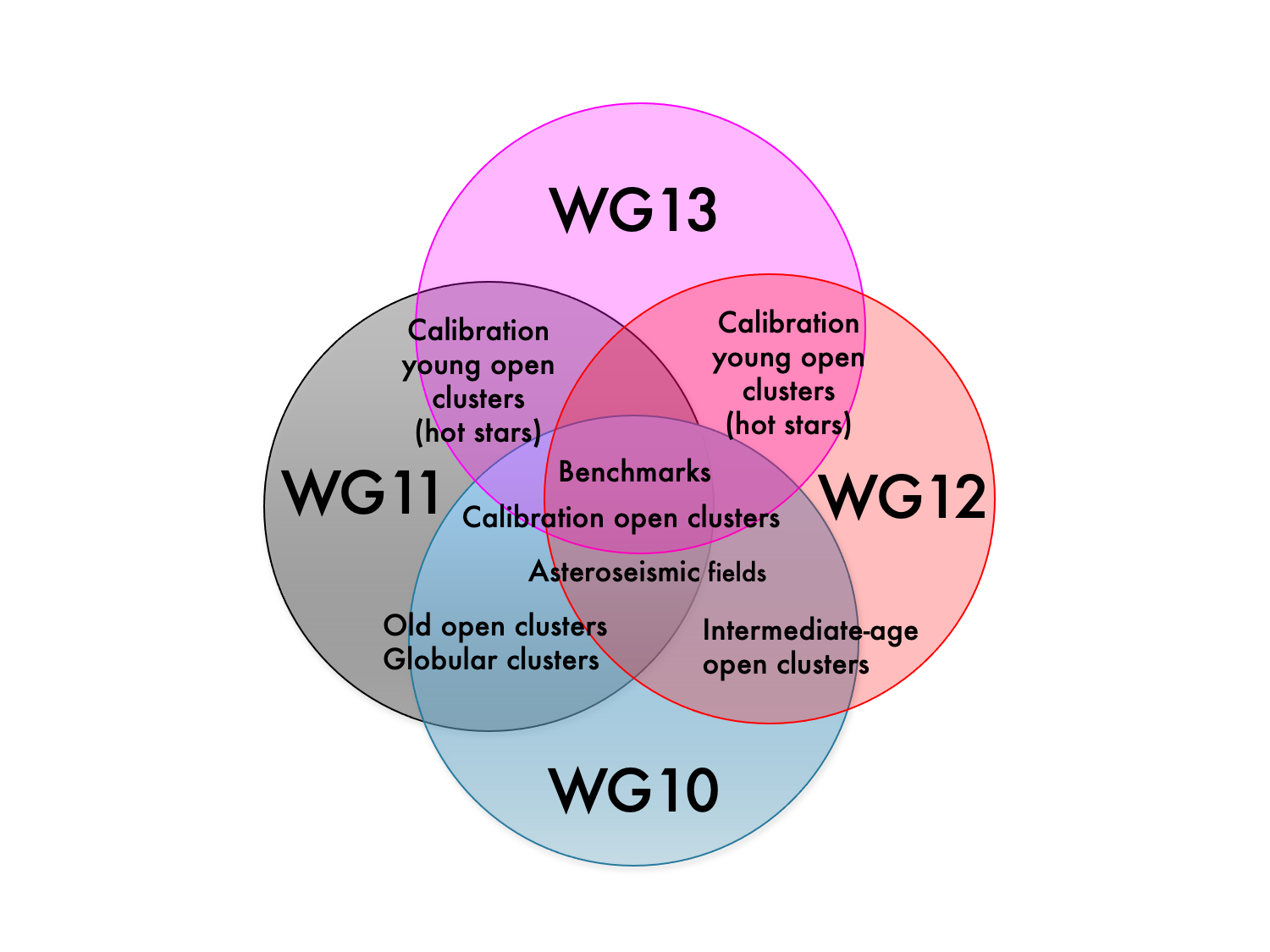}
        \caption{Venn diagram describing the role of calibration samples in the interconnection between the WGs.}\label{Fig:venn_calibrators}
    \end{minipage}%
    \hfill
    \begin{minipage}{.48\textwidth}
        \centering
        \includegraphics[width=1\linewidth,trim=4cm 1.5cm 4cm 1.0cm,clip]{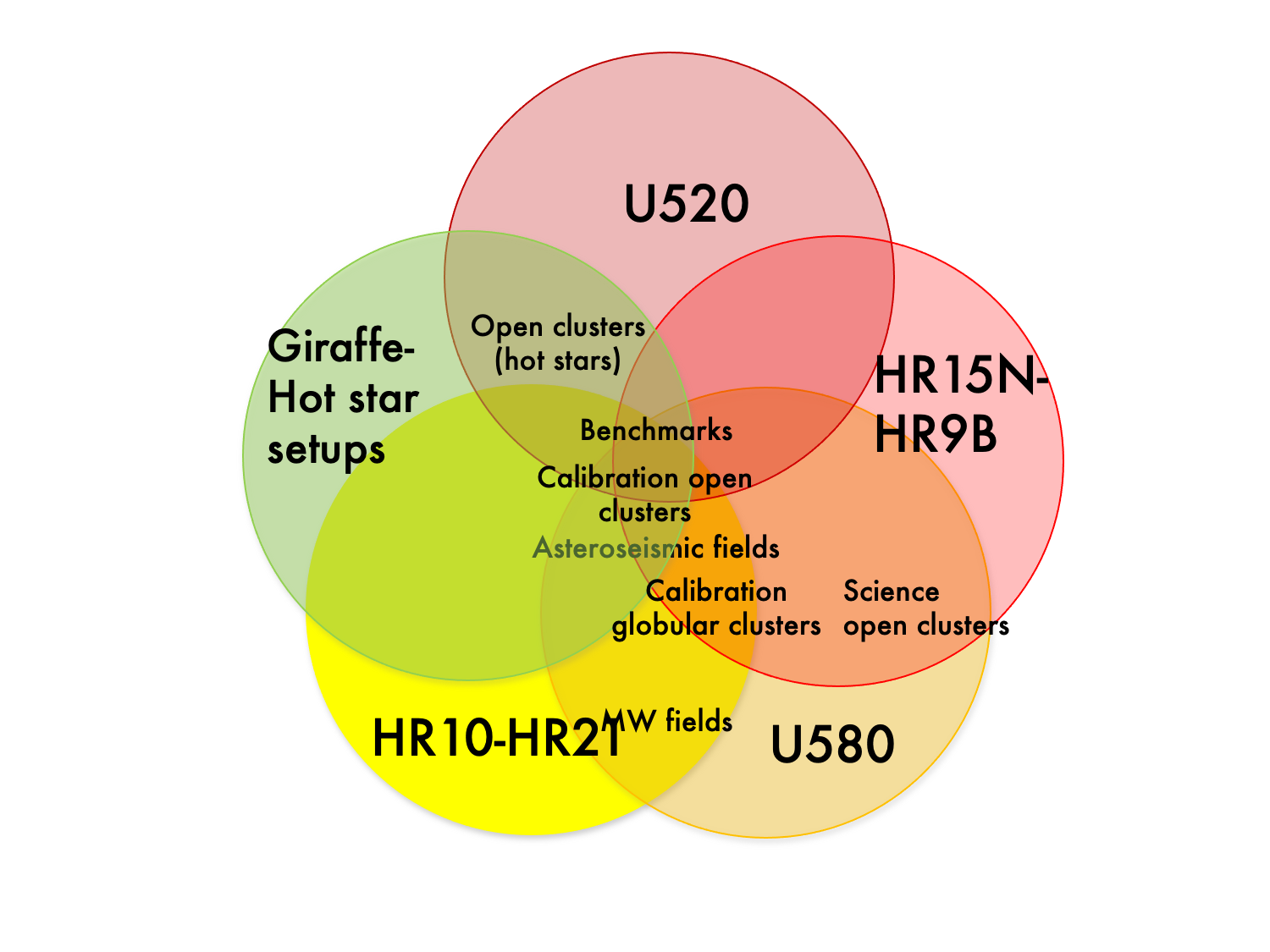}
        \caption{Venn diagram describing the setups with which the calibration samples have been observed (see also Table~\ref{tab:epsetups}).} \label{Fig:venn_setups}
    \end{minipage}
\end{figure*}

The calibration programme cannot take a one-size-fits-all approach due to the different specialisations of the various WGs. To allow an inter-calibration of the work of the different WGs, GES has been designed to have several samples in common amongst the analysis WGs,  graphically represented in the Venn diagram in Fig.~\ref{Fig:venn_calibrators}. In Fig.~\ref{Fig:venn_setups} we present a Venn diagram of the different setups used for the calibrator samples: in particular, benchmark stars are observed in all combinations of setups, to be analysed by all WGs, whereas other samples of calibrators are observed with the setups more suited for their analysis, e.g. young calibrator open clusters containing hot stars with U520 and a combination of GIRAFFE setups or calibrator globular clusters with U580 and HR10-HR21. The benchmarks are used to tie the GES results to a well-determined external scale. 

The {\em Gaia} Benchmark Stars were selected to comply with a variety of restrictions such that they serve for reference, as described in detail in \citet{heiter15}, and summarised here. Firstly, each star should have a measurement of its angular diameter, parallax and bolometric flux. This allows the effective temperature to be determined using the Stefan-Boltzmann relation, i.e. independent of the assumptions of spectroscopy. Secondly, the stars should adequately sample the parameter space for stellar populations in the Milky Way. This means that the sample is built to include dwarfs and giants, and have a spread in metallicity. Thirdly, the stars need to be located near the equator, such that they are observed from both hemispheres. The parameters are determined with the following procedure. First, the effective temperature is determined using fundamental relations \citep{heiter15}. Then, surface gravity is determined using temperature, bolometric flux, parallax and mass from a stellar track . Finally, $T_{\rm eff}${} and $\log g${} are fixed to the values obtained above and chemical abundances, including metallicity, are derived from high-Signal-to-Noise-Ratio (high-SNR) and high-resolution optical spectra \citep{blanco-cuaresma14} by various spectral analysis methods \citep{jofre14, jofre15}. This leads to a sample of 34 stars with accurate temperatures with a precision of about 100~K,  surface gravities and abundances with precisions of 0.1~dex. The latest catalogue can be found in \citet{jofre18}. 

One of the main motivations to assemble the {\em Gaia} Benchmark Star dataset was the calibration of GES. Therefore, substantial spectral analysis of the sample has been performed following the spirit of GES, namely using a combination of analysis methods (and adopting many of the GES WG11 methods), using the same line list and atmosphere models, and looking for new candidates to match the needs of GES.

The original set of {\em Gaia} benchmarks was expanded for the final release of GES to include stars at lower metallicity and at higher and lower temperatures (warm and cool benchmarks) to better cover the parameter space of the different science samples \citep[which are described in, e.g.][]{Stonkut16}. A set of hot stars, with well-known stellar parameters, was included for WG~13  and  several cool M-dwarf stars for WG~12. A sample of metal-poor candidates was proposed by \citet{hawkins16}, and a workshop to understand specific differences of the analysis methods in GES using the benchmarks was organised \citep{jofre17}. 

\begin{figure}
  \resizebox{\hsize}{!}{\includegraphics{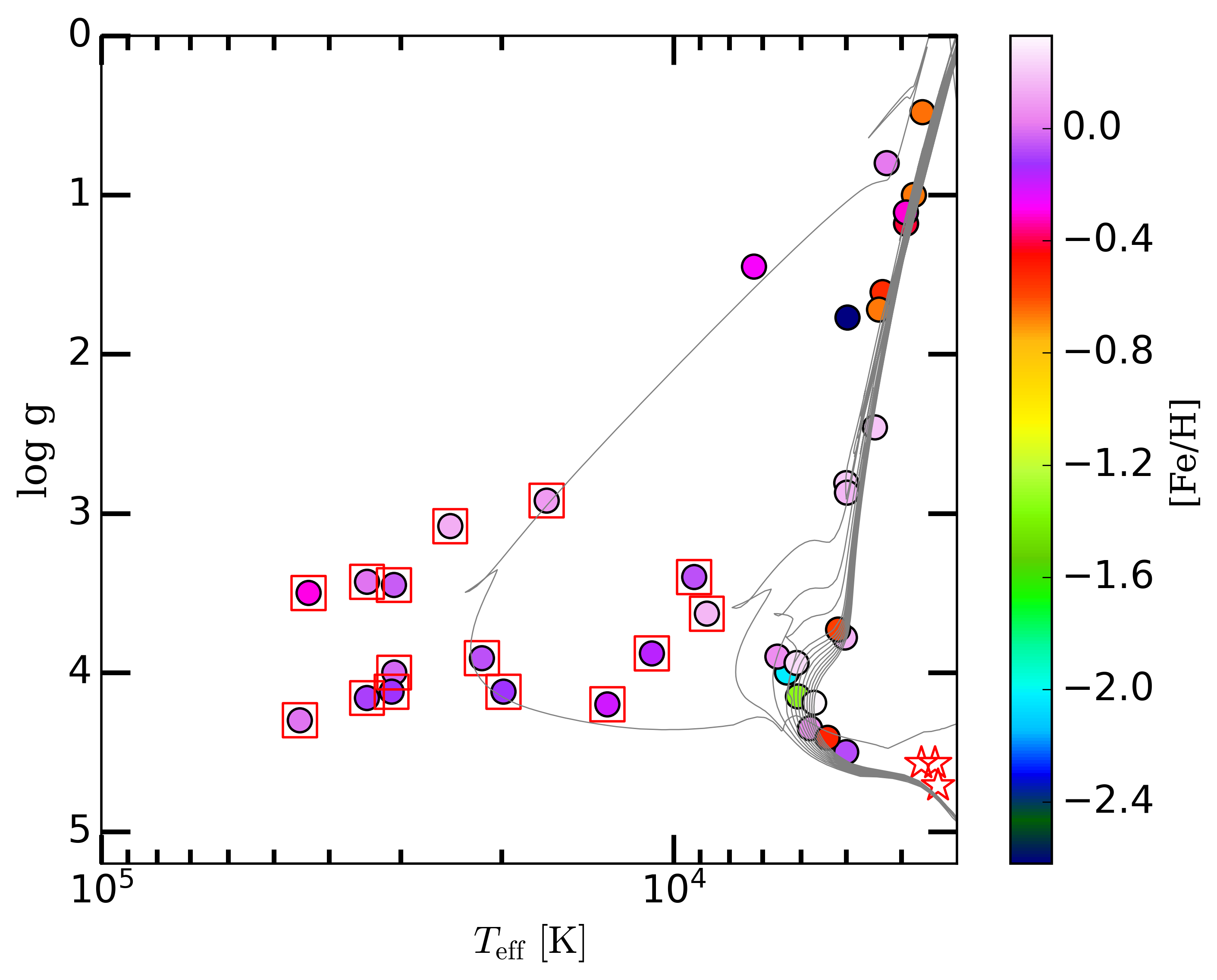}}
  \caption{Kiel diagram of the benchmark stars observed in GES. Stellar parameters are from the final data release. The symbols are colour-coded by [Fe/H]. PARSEC isochrones at solar metallicity with ages from 0.2 to 13 Gyr are shown with grey curves. Warm benchmarks are marked with squares, while cool ones are marked with stars.}
  \label{fig_benchmarks}
\end{figure}

The Kiel diagram of the final sample of benchmark stars, available in the final release iDR6 is presented in Fig.~\ref{fig_benchmarks}. The sample of the benchmark stars, divided into warm benchmarks (GE\_SD\_BW), FGK benchmarks (GE\_SD\_BM or AR\_SD\_BM) and cool benchmarks (GE\_SD\_BC), covers the parameter space mapped by the various WGs of GES.

Another two main classes of calibrator used in GES are well-studied open and globular clusters. Calibration using clusters is especially important for the WGs operating on stars at the edges or outside of the FGK range  and to test the method on groups of stars that have the same ages, distances, and metallicities, but different masses and evolutionary phases. The calibration open clusters are observed in setups matching the Milky Way field setups and those of the globular clusters, in addition to the setups used for the open cluster science. 
The literature metallicities of the  final set of calibrator clusters for iDR6 are listed in Table~\ref{tab:cal_clusters}.

\begin{table}
\begin{center}
\caption{Literature metallicities of globular and open clusters adopted as calibrators.} 
\begin{tabular}{lccl}
\hline\hline
Name & [Fe/H] & type  & Reference\\
\hline
M15         &   $-2.37$   & GC  & \citet{harris96} \\             
NGC4590     &   $-2.23$   & GC  & \citet{harris96} \\     
NGC4372     &   $-2.17$	& GC  & \citet{harris96} \\           
NGC4833     &   $-1.92$  & GC  & \citet{pancino2017gc} \\     	      
M2          &   $-1.47$	& GC  & \citet{pancino2017gc} \\     			
NGC1904     &   $-1.51$   & GC  & \citet{pancino2017gc} \\     				
NGC6752     &   $-1.48$   & GC  & \citet{pancino2017gc} \\     				
M12         &   $-1.37$   & GC  & \citet{harris96} \\     				
NGC1261     &   $-1.27$   & GC  & \citet{harris96} \\     				
NGC362      &   $-1.12$   & GC  & \citet{pancino2017gc} \\     				
NGC1851     &   $-1.07$   & GC  & \citet{pancino2017gc} \\     				
NGC2808     &   $-1.03$   & GC  & \citet{pancino2017gc} \\     				
NGC104      &   $-0.71$   & GC  & \citet{pancino2017gc} \\     				
NGC5927     &   $-0.39$   & GC  & \citet{pancino2017gc} \\  
NGC6553     &   $-0.25$   & GC  & \citet{harris96} \\ 
\hline
NGC2243	    &   $-0.48$	& OC  & \citet{pancino17} \\ 
Berkeley32	&   $-0.21$	& OC  & \citet{friel10}  \\ 
Melotte71	&   $-0.27$	& OC  & \citet{pancino17} \\ 
NGC2420	    &   $-0.05$	& OC  & \citet{pancino17} \\ 
M67         &   $+0.03$   & OC  & \citet{randich06} \\ 
NGC3532	    &   $+0.00$	& OC  & \citet{pancino17} \\ 
NGC2477	    &   $+0.07$	& OC  & \citet{pancino17} \\ 
NGC6705	    &   $+0.12$	& OC  & \citet{pancino17} \\ 
NGC6253	    &   $+0.34$	& OC  & \citet{pancino17} \\ 
\hline \hline
\end{tabular}
\label{tab:cal_clusters}\\
\end{center}
\end{table}

A relative newcomer to calibration sets for stellar surveys is samples for which asteroseismic measurements are available upon which $\log g${} is determined. GES included in its calibration plan the observation of targets from two key asteroseismic missions: CoRoT and K2 \citep{pancino17}.

\begin{figure*}
  \resizebox{\hsize}{!}{\includegraphics{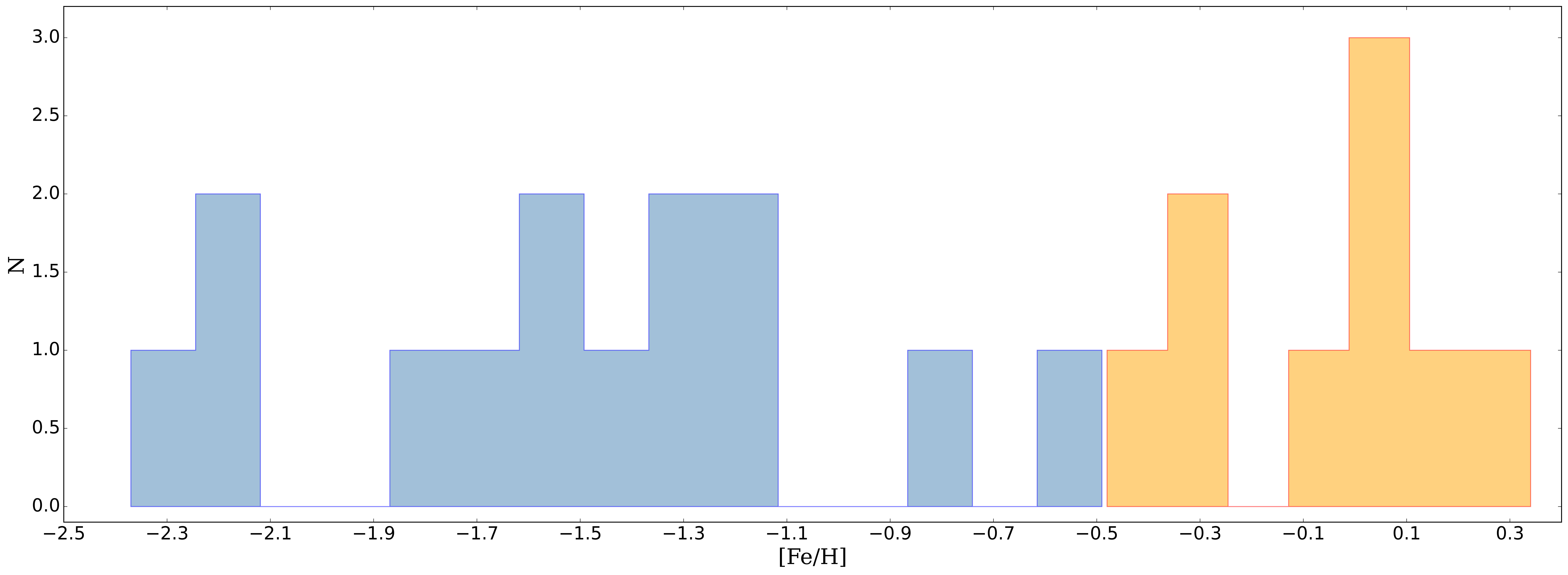}}
  \caption{Histogram of the metallicity covered by the calibrator clusters (in blue, globular clusters, in orange open clusters).}
  \label{fig:gc_oc_calibrators}
\end{figure*}

\section{Homogenisation Procedure}
\label{sec_3}

\begin{figure*}
  \resizebox{\hsize}{!}{\includegraphics[trim=0cm 3cm 3cm 4cm,clip]{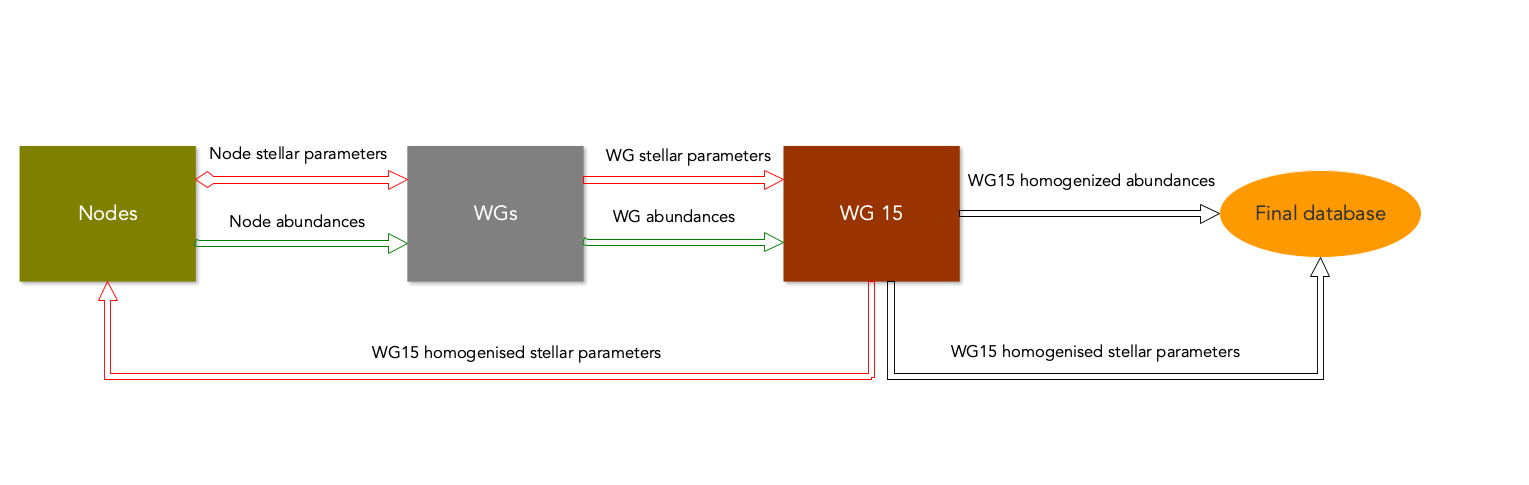}}
  \caption{Schematic view of the GES analysis approach: the arrows in red mark the first cycle of analysis dedicated to the determination of the stellar parameters; the arrows in green indicate the analysis of abundances (with homogenised parameters), while in black we have the  process of homogenisation and preparation of the final catalogue.  }
  \label{fig:ges_schem}
\end{figure*}

The multi-method, multi-pipeline design of the GES analysis, implemented through the analysis node and WG structure outlined above, means that multiple results are delivered for many GES stars. This includes both parallel analyses of the same stellar samples within a WG, and the analysis of common calibration samples across WGs.  To provide a final consistent set of results, the role of WG15 was to homogenise the recommended results from WGs 10--13 on to a common scale. The main product of WG15 is a catalogue of recommended astrophysical parameters, elemental abundances, radial and rotational velocities, other specific quantities and flags per star (or per CNAME). 
A schematic view of the GES analysis approach is presented in Fig.~\ref{fig:ges_schem}:  the analysis process starts from the nodes, which transmit their results to the WG. The first step, indicated by red arrows, denotes the determination of stellar parameters. Once homogenised by WG15, the stellar parameters are transmitted back to those nodes which determine the abundances. These are then homogenised by the WGs, and finally combined, together with the stellar parameters, by WG15 in the final database. 

WG15 was led by P. Fran\c cois [from DR1 to DR6] and over the course of the survey the working group members were A. Hourihane [DR1-DR6],  C. Worley [DR1-DR6], L. Magrini [DR2-DR6], A. Gonneau [DR5-DR6] and A. Casey [DR3-DR5].

\subsection{Quality and format checks}  

The first step of the data analysis is to perform sanity checks on the files provided to WG15 from the four analysis WGs. A first pass is performed by a dedicated automated tool (the {\sc FITSChecker} which was developed and maintained over the lifetime of the survey by C. Worley, A. Casey, D. Murphy and A. Gonneau). This tool flags issues in the file formats, data statistics and data completeness in a report that is sent to the submitter. The submitter must resolve any issues and resubmit the file until it is accepted by the tool as adhering to the GES data model. The GES data model is described in two technical documents governing the spectral formats from the spectral reduction pipelines\footnote{Gaia-ESO Spectral Data and Formats, July 2013} and the analysis results catalogues from the nodes and WGs\footnote{Gaia-ESO Survey Fixed Format FITS Template iDR5 Stellar Parameters \& Chemical Abundances, May 2016}. WG15 members also carried out a visual inspection of the data statistics summarised in the {\sc FITSChecker} report to identify any remaining spurious values and outliers. These were raised with the relevant WG leads for resolution.

\subsection{The homogenisation flow: from stellar parameters to elemental abundances} 
 In this Section the homogenisation flow is summarised, for which the quality checks performed on each step are described in Secs.~\ref{sec:par}, \ref{sec:abu} and \ref{sec:rv}. 
The homogenisation workflow starts with the application of an algorithm that defines a set of rules to obtain the best set of stellar parameters in the case of multiple observations with different setups. This choice is not only based on spectral resolution or SNR, but also on which type of observation is best suited to the type of star and which WG uses the most appropriate methods, e.g. WG13 for hot stars, or WG12 for cool stars. 

On the one hand, the internal analysis processes of WG10, WG11 and WG12 have been fully consistent in the last data release and have provided stellar parameters on the same scale, thanks to continuous interaction between the WG leads and the WG15 team \citep[see][for details on the analysis of each WG]{smiljanic14, lanzafame15, worley2023}. In particular, in the last data release the analysis of the U580 and U520 spectra assigned to WG12 was performed by all WG11 nodes and it was included in the homogenisation workflow of WG11, ensuring a consistent treatment of the data. Similarly, the WG12 GIRAFFE spectra (HR15N) were homogenised with the same code as the WG10 results. 

Details about the mapping of the WG10 and WG12 results on to WG11 are given in \citet{worley2023}. On the other hand, the WG13 results are located in a different region of the parameter space, and obtained with different methods. Therefore, they are treated separately and not homogenised with the results of the other WGs. 

\begin{figure*}
  \resizebox{\hsize}{!}{\includegraphics{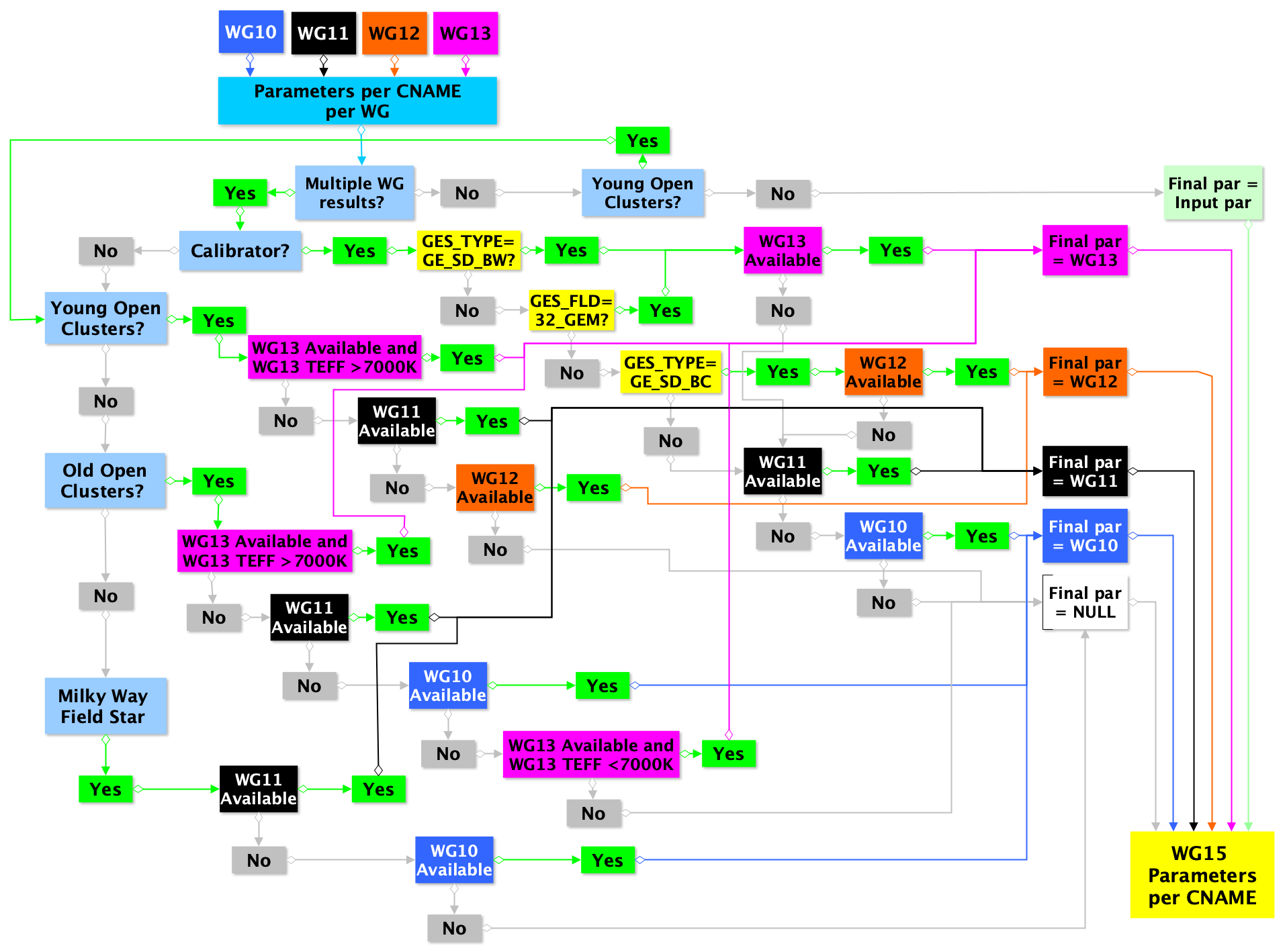}}
  \caption{Schematic view of the GES WG15 Parameter Homogenisation Work-Flow.}
  \label{fig:wg15_param_wrkflw}
\end{figure*}

For these reasons, the WG15 algorithm does not apply further corrections (offsets or linear relations) to stellar parameters coming from the various different WGs. 
In the end, the homogenisation algorithm allows us to have a single set of parameters for each CNAME, uniquely chosen following the flow represented in Figure~\ref{fig:wg15_param_wrkflw}.

In the case where a given target is analysed by more than one WG, the results from WG11, as the analysis of the high-resolution dataset, are taken as the priority. In the case of hot stars, essentially in young clusters (age<100 Myr), the parameters of WG13, if available and if $T_{\rm eff} > 7000$~K, are chosen.  Finally, in the case of cool stars ($T_{\rm eff} < 7000$~K) in young clusters (age < 100 Myr), the results of WG12 are preferred.

Once WG15 has produced a single set of stellar parameters per CNAME, 
the homogenisation cycle returns them to the nodes in WGs 10-12, which  carry out an abundance analysis.  
For both the parameter and abundance phases, nodes were provided with a list of verifications to carry out on the key calibration sets for their own quality control prior to submission. With the relevant sets of node files, the WG leads performed their own internal homogenisation of the parameters in the parameter phase, and the abundances in the abundance phase, to provide a set of recommended results for each CNAME specific to the WG datasets. 

At the end of each phase, WG15 performed a cross-WG homogenisation to produce the final recommended set of parameters for the parameter phase, and then abundances for the abundance phase. The homogenised WG14 flags are also included    \citep[see][for details on the analysis of each WG]{vaneck2022}. A first-pass analysis of the flags becomes available as a reference set during the parameter homogenisation. 

The radial velocities are homogenised by WG15 in parallel, utilising stars analysed in common across different instrumental set-ups for calibration, as is done for parameter and abundance homogenisation. The procedure followed for radial and rotational velocity homogenisation is described in Section~\ref{sec:rv_vsini}.

\section{Stellar parameter homogenisation: quality checks}
\label{sec:par}
 The aim of the homogenisation process is to provide a recommended set of stellar parameters consistent with each other, regardless of the setup used (medium or high resolution and the covered spectral range).  
To this purpose, the GES strategy acts with several different tools, including: {\em i)} A set of well-defined calibrators, the benchmark stars, described above, covering the whole parameter space ($T_{\rm eff}$, $\log g$, $\mathrm{[Fe/H]}$);  {\em ii)} a sample of targets observed with different setups, and whose stellar parameters are derived by  different WGs;  {\em iii)} the Kiel diagrams of stars in the Milky Way fields having metallicity in a given, restricted, metallicity interval to be compared with the corresponding theoretical isochrones (we use Kiel diagrams rather than H-R diagrams as we have an estimate of the $\log g${} rather than the luminosity of each object);  {\em iv)} member stars in open and globular clusters, which share the same age and metallicity, and can be considered simple stellar populations, at least to a first approximation, and thus their stellar parameters are directly comparable with the corresponding isochrones; {\em v)} a sample of asteroseismic targets observed in the K2 and CoRot fields. 

The various sub-samples are used directly as calibrators to map the results onto the reference ones (in the internal WG procedures, see \citet{worley2023}), or as  final checks of the WG results and on the final set of global parameters by WG15. 

\subsection{Benchmark stars} 
The Gaia Benchmark Stars, described above, are used as a reference set during the WG homogenisation to define the parameter scale. The benchmark sample has been expanded from the initial set to better cover the parameter space needed for the global homogenisation of the Survey results. The sample available in iDR6 contains 42 stars in total, 21 FGK stars, 16 warm benchmarks (OBA stars) and 5 cool benchmarks (M stars).
 
As part of the quality checks on the parameters, several diagnostic plots are used. For example, the difference between the $T_{\rm eff}$, $\log g${} and $\mathrm{[Fe/H]}$ determined by each of the Working Groups for the benchmarks and the reference value is plotted with the result selected by the WG15 algorithm highlighted, to ensure the appropriate quality of the results selected according to the rules. 
 
\begin{figure*}
  \centering
  \includegraphics[width=1\linewidth]{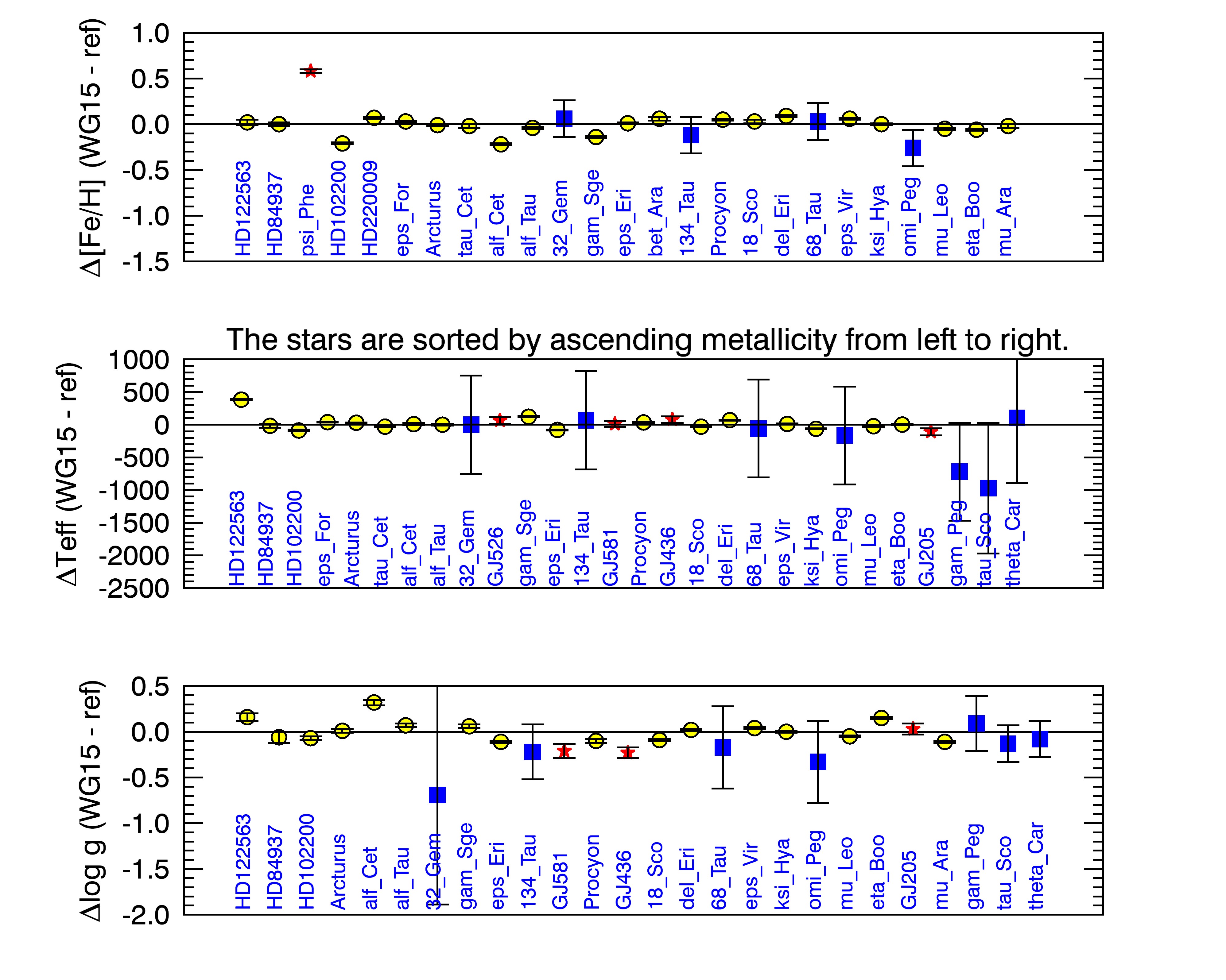}
\caption{WG15 parameters versus reference parameters as described in \citet{pancino17}. Warm benchmarks (OBA stars) are marked with blue squares, while the cool benchmarks (M stars) are marked with red stars. The stars are presented with ascending [Fe/H] along the abscissa. }
\label{fig:WG15_benchmark_param}
 \end{figure*}

The homogenised WG15 results are shown in Fig.~\ref{fig:WG15_benchmark_param} . The plots of Delta Parameter per benchmark are ordered by benchmark reference metallicity. For better visualisation of the results, the x-axis contains the list of benchmarks.
The results  demonstrate good agreement across the parameter space of the GES results with the literature ones. Two stars ($\tau$ Sco and $\gamma$ Peg) show a  temperature difference larger than 500~K, but these two stars are warm stars with temperatures above 22,000~K. A gravity difference of $-0.68$~dex is found for 32~Gem; however, it is within the error estimates. This star is an A9III-type star for which the gravity is difficult to estimate. Meanwhile, the GES parameters for the Sun are determined from archival spectra contained in the FLAMES solar atlas and reduced with the GES pipelines, and homogenised to values of $T_{\rm eff}$ of $5751\pm11$~K, $\log g$ of $4.35\pm0.02$~dex and $\mathrm{[Fe/H]}$ of $0.02\pm0.02$~dex. These are consistent with ($\mathrm{[Fe/H]}$) or close to ($T_{\rm eff}$, $\log g$) the literature values for the Sun from \citet{heiter15}.

\subsection{Milky Way field} 
\begin{figure*}
\begin{subfigure}{.5\textwidth}
 \centering
 \includegraphics[width=1\linewidth]{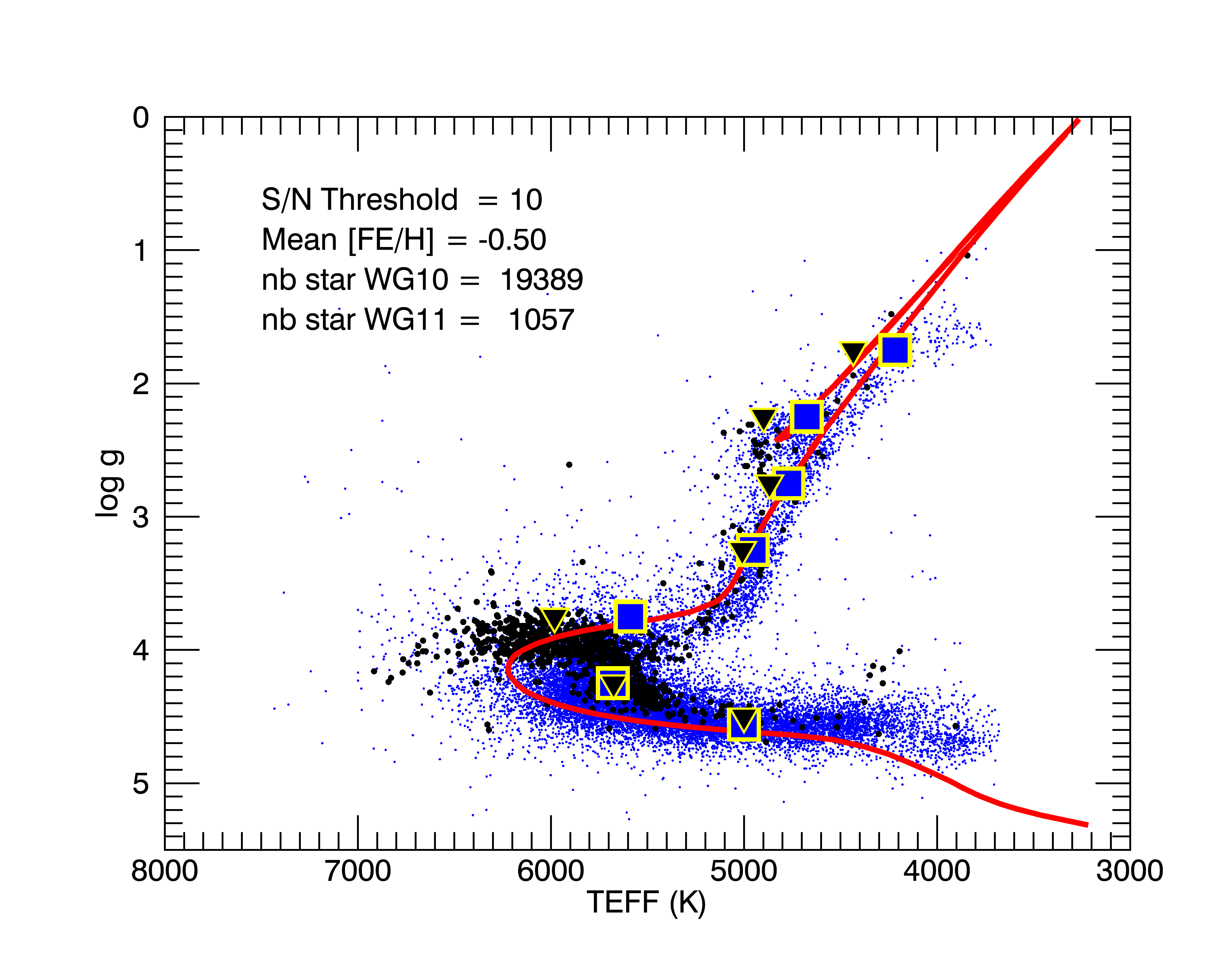}
\end{subfigure}%
\begin{subfigure}{.5\textwidth}
  \centering
   \includegraphics[width=1\linewidth]{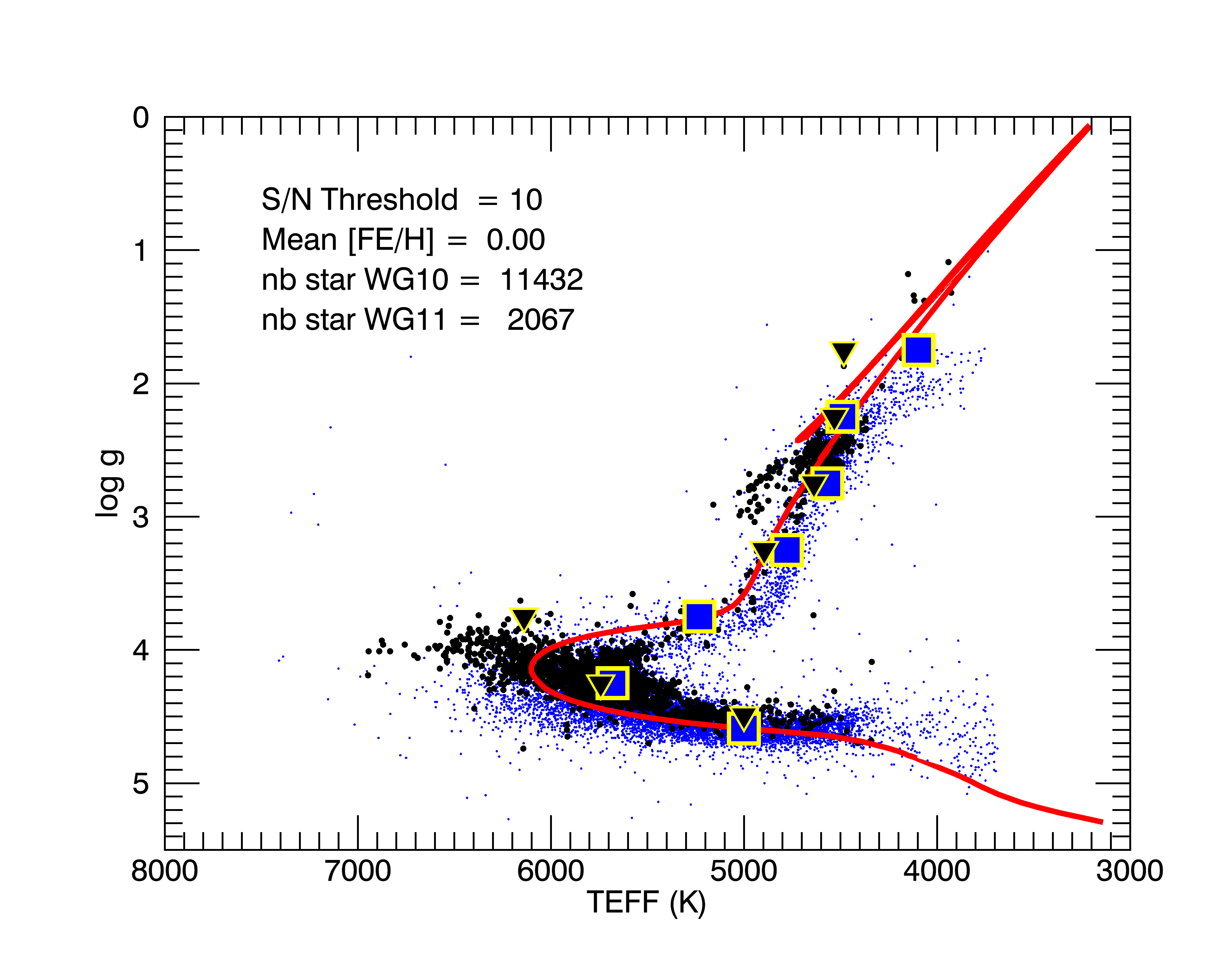}
\end{subfigure}
\caption{GES iDR6 Kiel diagrams for Milky Way field stars for WG10 (the blue points) and WG11 (the black points) for two metallicity bins [-0.75:-0.25], [-0.25:+0.25] and S/N > 10. Isochrones are from PARSEC \citep{bressan12} for z=0.01, z=0.02 and age=5.7 Gyr. The black triangles (resp. blue rectangles) are median
values for the WG10 (resp. WG11) stars in $\log g$ bins of 1~dex for the giant branch and in $T_{\rm eff}${} bins of 500~K for the main sequence stars.}
 \label{fig:mwhrd}
\end{figure*}

The Milky Way (MW) fields contain both UVES (WG11) and GIRAFFE (WG10) spectra. Target selection is described in \citet{Stonkut16}.  With the assumption that the two sets of spectra sample the same stellar population, we can compare
the distributions  in the  $T_{\rm eff}$-$\log g${} plane for the stars for both samples and check for offsets in $T_{\rm eff}${} and/or $\log g${} (see Figure~\ref{fig:mwhrd} for the bin centred at solar metallicity and for the one at $[Fe/H] = -0.5$).
The isochrones are only representative and they do not correspond to a specific age-metallicity relation fitted to the sample. Their
parameters are compatible with the metallicity bin used in each plot. They are plotted as
representative of the shape and the location of the Main Sequence and the giant branch. The large black triangles (resp. blue rectangles) are median
values for the WG11 (resp. WG10) stars in $\log g$ bins of 1~dex for the giant branch and in $T_{\rm eff}${} bins of 500~K for the main sequence stars.
 
\subsection{Calibration open and globular clusters}
\begin{figure*}
  \resizebox{\hsize}{!}{\includegraphics{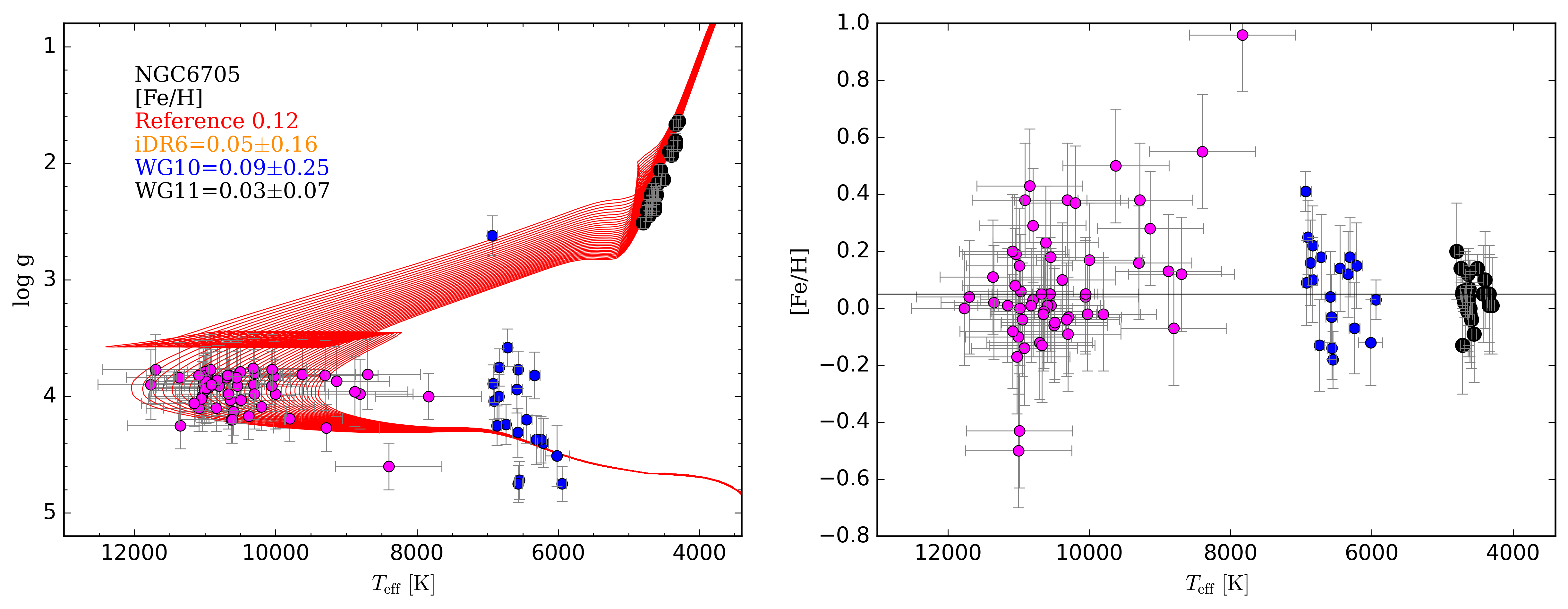}}
  \caption{The stellar parameters of the members of NGC6705: on the left, the Kiel diagram showing the PARSEC isochrones in the age and metallicity range of NGC6705 (solar metallicity, age from 200 to 500 Myr, steps of 50 Myr) and the data obtained by WG10 (blue), WG11 (black) and WG13 (magenta); on the right, [Fe/H] vs $T_{\rm eff}$ for the same stars. The black horizontal line is the mean metallicity for NGC6705 in \citet{randich22}. In the plot, we include only stars with E\_LOGG < 0.35. 
  } 
  \label{Fig:ngc6705}
\end{figure*}

\begin{figure*}
  \resizebox{\hsize}{!}{\includegraphics{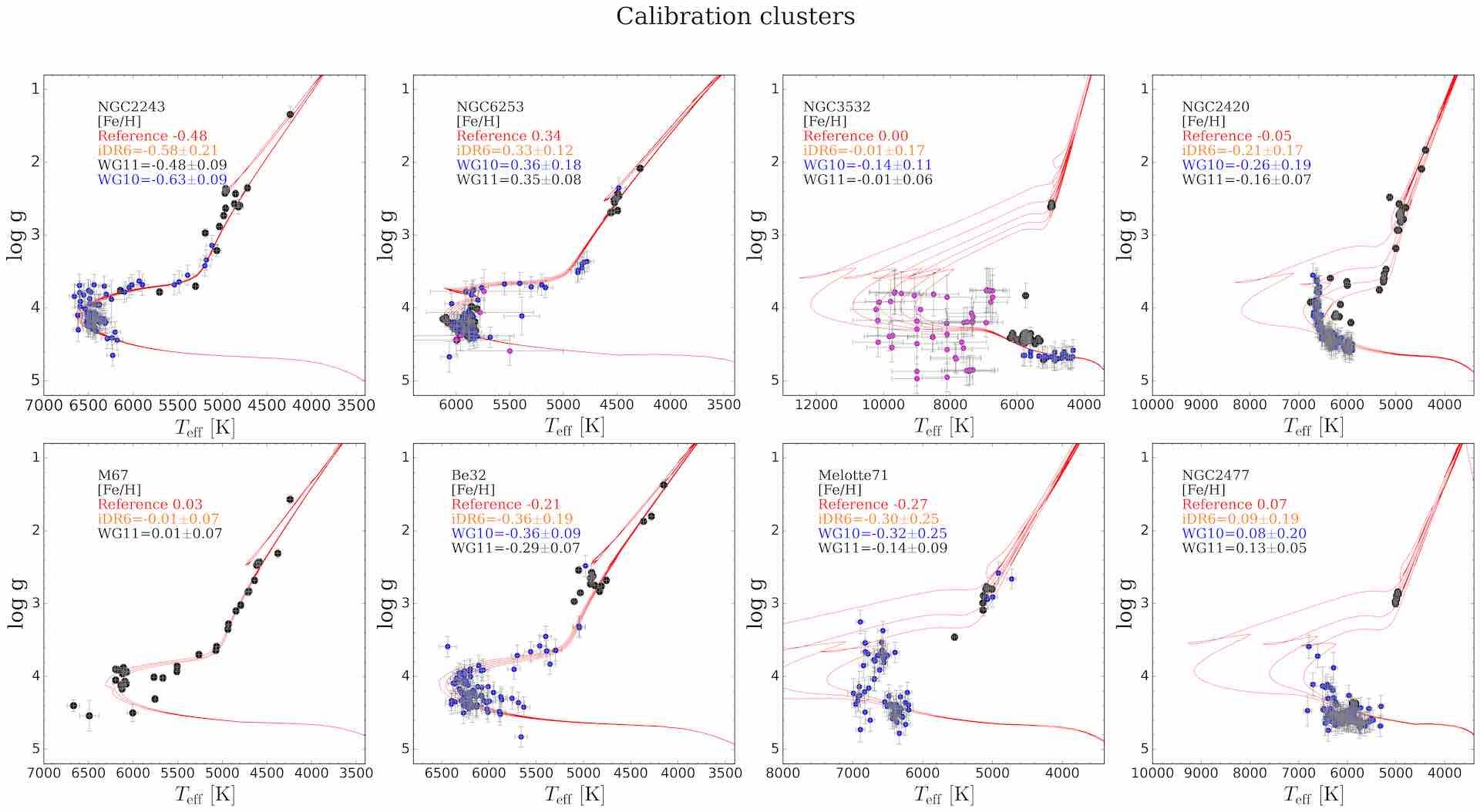}}
  \caption{Kiel diagram of the  calibration open clusters with homogenised data from  WG10 (blue), WG11 (black) and WG13 (magenta) and the PARSEC isochrones for the age and metallicity of the clusters. In the plot, we include only stars with E\_LOGG $<$ 0.35. } 
  \label{fig_all_calib_ocs}
\end{figure*}

As described in \citet{pancino17}, clusters form an essential part of the calibration set and provide a way to compare simple stellar populations of known ages with the outputs of theoretical models. In addition, they allow us to map stars analysed by different WGs on to a common scale when no actual stars in common are present between WGs (which is an unavoidable issue when dealing with WGs which work on, e.g., hot, massive cluster stars and cool pre-main sequence stars). Specific clusters have been observed purposely for calibration. In addition to the calibration clusters, we also make use of clusters observed as science targets, in particular those in which a large number of stars were observed. 
Clusters were used to validate the results in various ways. 
First, we verified the good agreement in the $\log g$-$T_{\rm eff}$ plane with the isochrones corresponding to the age and metallicity of the various clusters. 
In addition, using the membership published in \citet{jackson22J}, we selected members of open clusters with a probability of membership > 0.99. Membership analysis for Melotte\,71 and Br32 is not available in \citet{jackson22J}, so we plot in the figure the members selected on the basis of their radial velocity. 
We computed the average metallicity of their members examined by the different WGs and we used them to identify possible offsets due to the analysis process.
 An example is shown in Figure~\ref{Fig:ngc6705}, in which we plot the stellar parameters of the members of NGC6705, one of the nine calibration open clusters,  obtained by three different WGs and homogenised by WG15. We also compare the metallicity obtained from the different samples of stars, with the mean metallicity of the cluster provided  in \citet{randich22}.
In this figure, we show the results of WG10 obtained with the combination of the two setups HR10 and HR21, which are used to observe the MW fields and for calibration purposes in calibration clusters and in benchmark stars.  Although the data obtained at lower resolution from WG10 and for hot stars from WG13 show a larger scatter than the one from WG11 data, there is good agreement with the isochrones for the stellar parameters, and with the average metallicity of the cluster as derived by different WGs. 
In Fig.~\ref{fig_all_calib_ocs}, we show the Kiel diagrams of the other  calibration open clusters, adding  an illustrative isochrone for each cluster. The PARSEC isochrones are selected at the cluster metallicity and age, considering a typical uncertainty in the age determination as given in \citet{cantat20}.  
Overall the agreement in the parameters along the isochrone is very good for the results obtained by the different WGs, with an excellent tracking of the cluster sequence.

\begin{figure*}
 \centering
\includegraphics[width=0.95\linewidth]{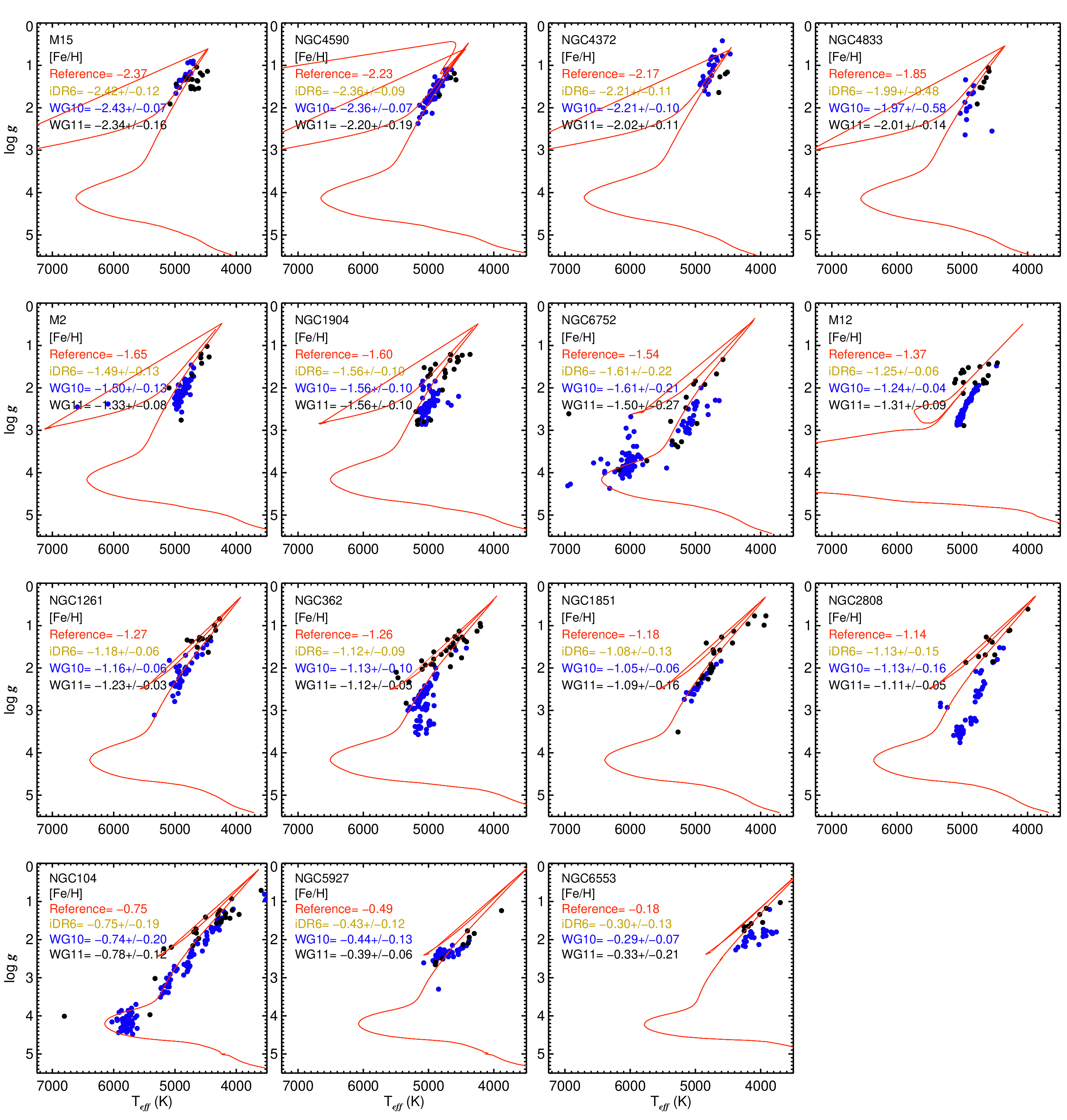}
 \caption{Kiel diagram for each of the 15 globular clusters in iDR6 shown as increasing in metallicity from left to right. An isochrone at the reference values of age and metallicity for each globular cluster is displayed. The results are separated by WG: WG11 in black and WG10 in blue. }
\label{fig:idr6gcs}
 \end{figure*}

There are stars from 15 globular clusters in GES iDR6, selected such that the globular clusters span a wide range in metallicity (see Fig.~\ref{fig:gc_oc_calibrators} for their metallicity distribution). These were observed with HR10, HR21 and U580, the setups used for the Milky Way fields. Figure~\ref{fig:idr6gcs} shows the Kiel diagram for each GC. 

The stars shown for each globular are those defined as cluster members using Gaia DR3 proper motions and GES radial velocities as described in \citep{worley2023}.

An illustrative isochrone at the reference values for age and metallicity for each globular cluster is shown in Figure~\ref{fig:idr6gcs}. The WG11 (black) and WG10 (blue) results are shown with median metallicity calculated for each and for iDR6 in total. The reference metallicity is also provided.

Overall there is very good agreement of the GES globular cluster stellar parameters with the isochrones. By distinguishing between WG11 and WG10 by colour, the agreement between the two WGs along the stellar evolution sequences confirms the consistency between the two sets of results. However, the WG10 results for two globular clusters, M12 and NGC2808, show a non-trivial disagreement with the respective isochrone, indicating further detailed study is warranted but this is outside the scope of this paper. 

\begin{figure}
 \centering
\includegraphics[width=0.9\linewidth]{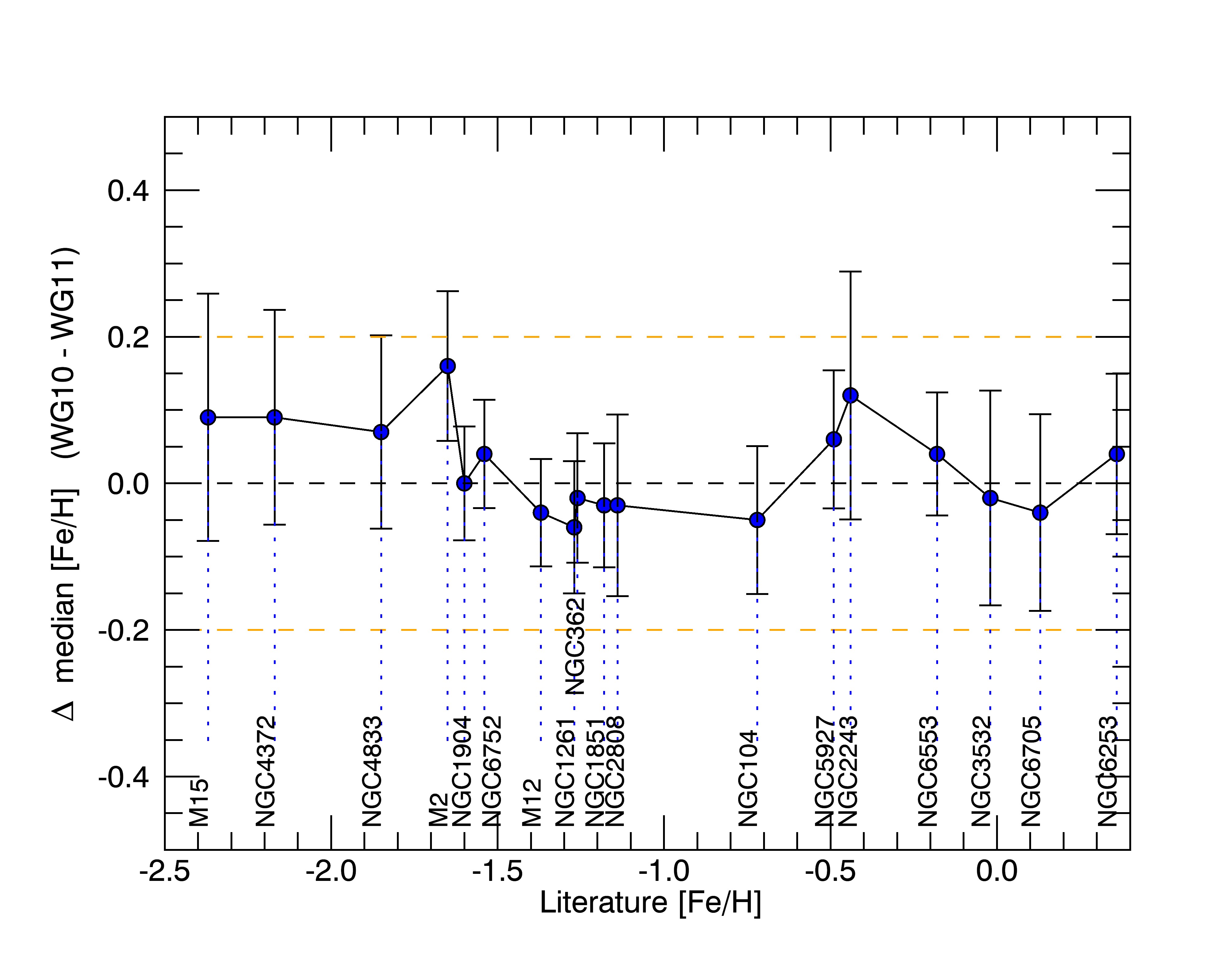}
 \caption{[Fe/H] bias amongst the GES GCs and calibrating OCs.}
\label{fig:idr6gcsbias}
 \end{figure}

Figure~\ref{fig:idr6gcsbias} shows the offset between the WG10 and WG11 mean [Fe/H] for the calibrating globular and open clusters. For each cluster the median difference between WG10 and WG11 is represented as blue circles. Orange dotted lines show offset limits at  $\pm 0.2$~dex. The figure shows how good the agreement between the results of WG10 and WG11 is.
M2 is the only cluster with a bias of the order of 0.15~dex.  
We measured a mean bias of $0.04\pm 0.07$~dex between the WG10 and the WG11 [Fe/H] median values.   

\subsubsection{Asteroseismic Calibrations}
\begin{figure}
 \centering
\includegraphics[width=1.0\linewidth]{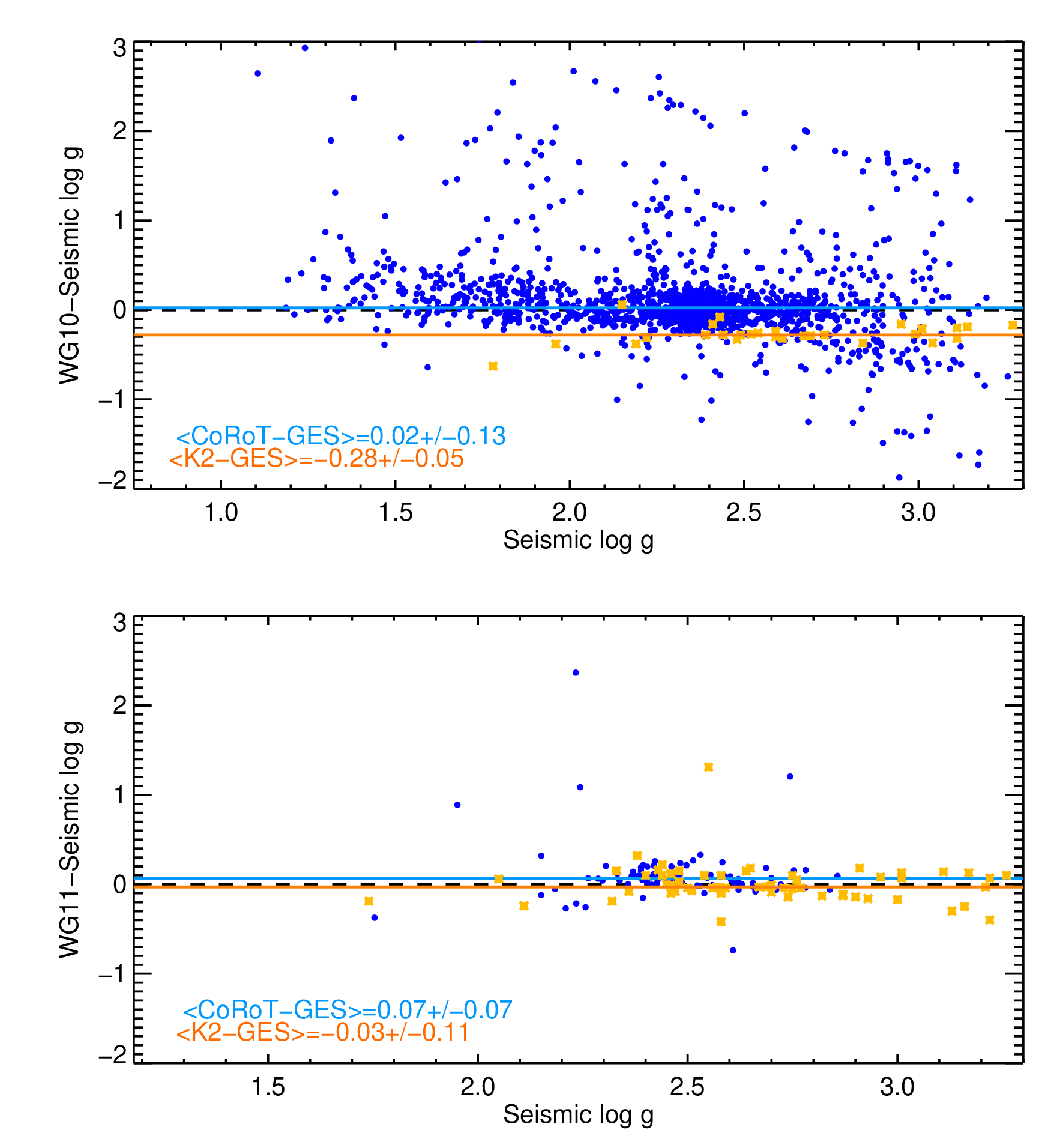}
 \caption{Comparison between WG10 (top) and WG11 (bottom) $\log g$ values and the seismic $\log g$ from CoRoT and K2. The median and MAD are also given.}
\label{fig:logg_k2crt}
 \end{figure}

The agreement with the $\log g$ of the asteroseismic samples of CoRoT and K2 as described was investigated by WG15.  For WG10, 1,512 CNAMEs had both CoRoT and GES $\log g$, while 28 CNAMEs had both K2 and GES $\log g$. For WG11, 86 CNAMEs had both CoRoT and GES $\log g$, while 62 CNAMEs had both K2 and GES $\log g$. The differences of these values with respect to the seismic $\log g$ values are shown in Figure~\ref{fig:logg_k2crt}, and the median difference and standard deviation in each set are provided.

In three of the four datasets there is good agreement between the seismic and GES $\log g$ values. However, the WG10 $\log g$ values of the K2 sample are over-estimated by 0.28$\pm$12 dex. While the agreement of the WG10 values with CoRoT is good there is a large scatter ($-0.07\pm0.51$). The nature of the WG10 spectra as shorter wavelength ranges and lower resolution compared to the WG11 spectra most likely contributes to both effects as noted in \cite{worley2020}. More extensive calibration samples combining spectroscopy and asteroseismology are needed to explore and refine this approach, which is being pursued by upcoming surveys.

\subsubsection{Stars in common between Working Groups}
\begin{figure*}
 \centering
\includegraphics[width=0.9\linewidth]{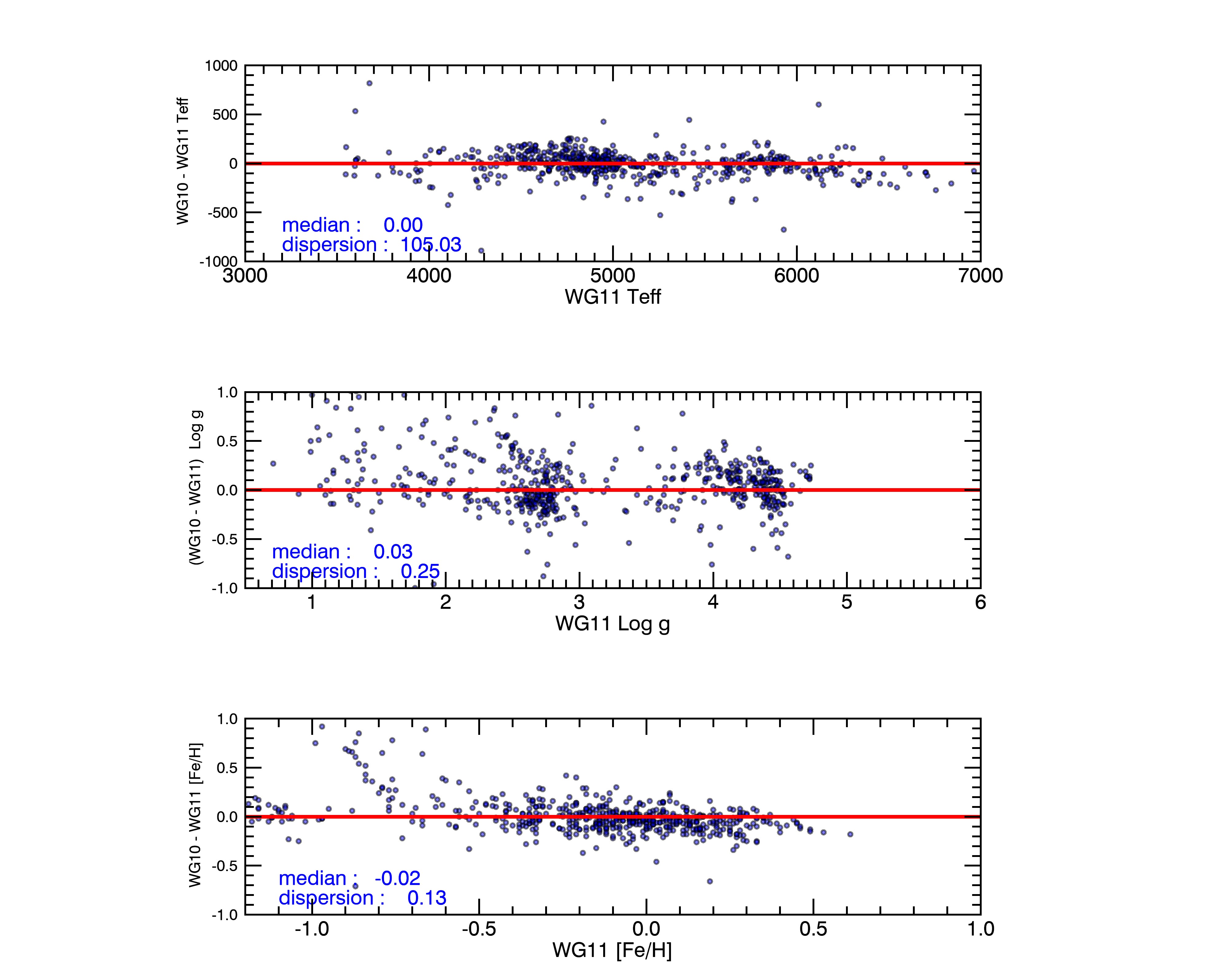}
 \caption{Comparison of $T_{\rm eff}$, $\log g$  and [Fe/H] between WG11 and WG10 for stars in common.}
\label{fig:wg11wg10_common}
\end{figure*}

The GES strategy consists of having a number of targets observed in several setups, and consequently analysed by several WGs. These targets make it possible to verify the consistency of the results obtained by the various WGs. 
There are about 708 spectra in common between WG11 and WG10. Most of them are
benchmark stars and stars in clusters. There is also a small number of stars in the MW field.
In Figure~\ref{fig:wg11wg10_common}  we show the comparison between  the difference WG10$-$WG11 versus WG10   results for $T_{\rm eff}$, $\log g$ and [Fe/H]. 
The results are in very good agreement,  with a low median difference for the different parameters. 

\begin{figure*}
 \centering
\includegraphics[width=0.9\linewidth]{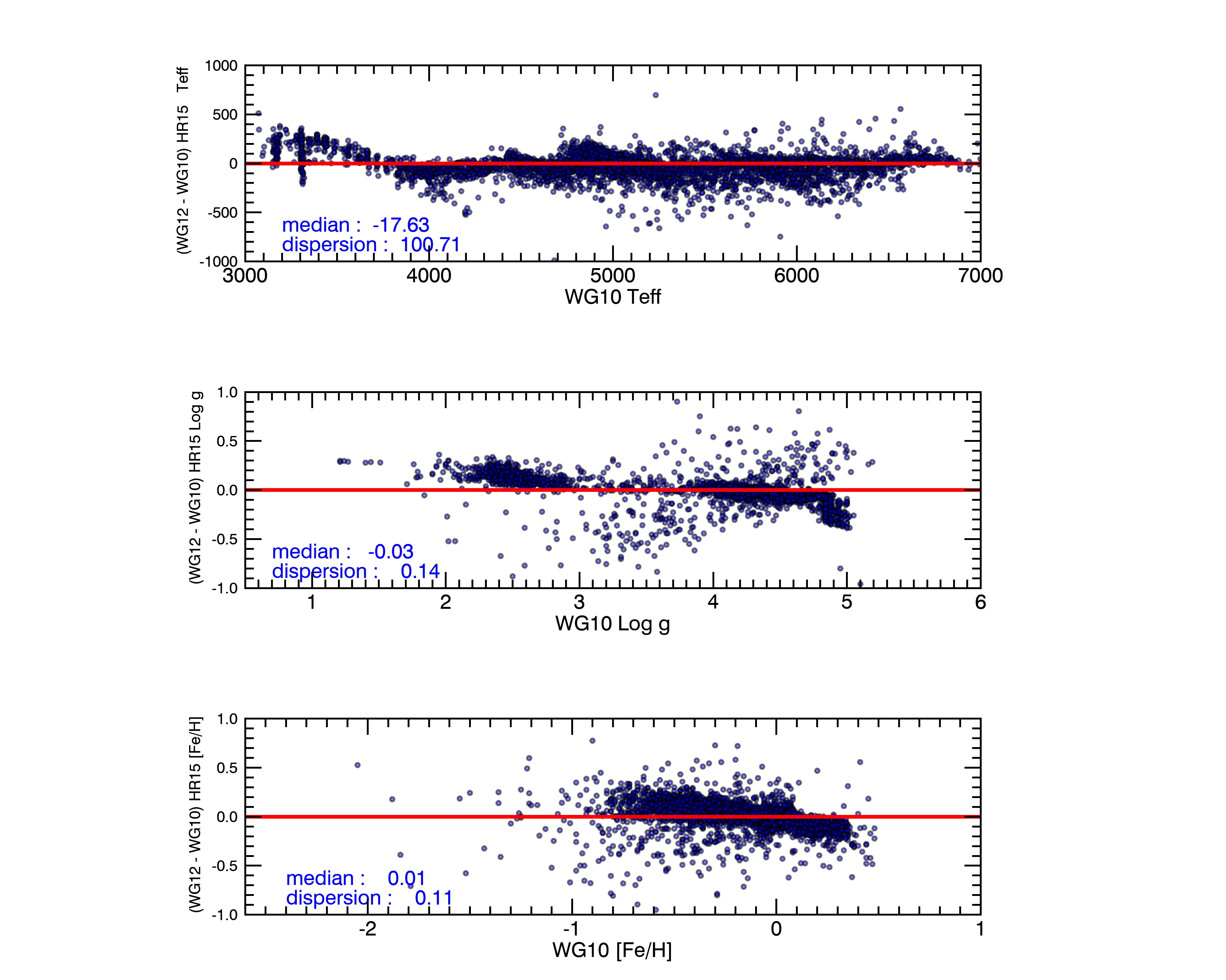}
 \caption{Comparison of $T_{\rm eff}$, $\log g$  and [Fe/H]  between WG12 and WG10 for stars in common in the setup HR15. Median values and dispersion are given in each subpanel. }
\label{fig:wg12wg10}
 \end{figure*}

There are 5,171 spectra in common between WG12 and WG10 for the setup HR15. 
Most of these
are stars in clusters analysed by both working groups, then there are the usual calibrators,
mainly benchmark stars.
In Figure~\ref{fig:wg12wg10}  we show the comparison between the WG12 and the WG10 results for $T_{\rm eff}$, $\log g$
and [Fe/H].
The results are in  good agreement, as shown by the median difference and the dispersion. No bias correction is needed.

\begin{figure*}
 \centering
\includegraphics[width=0.9\linewidth]{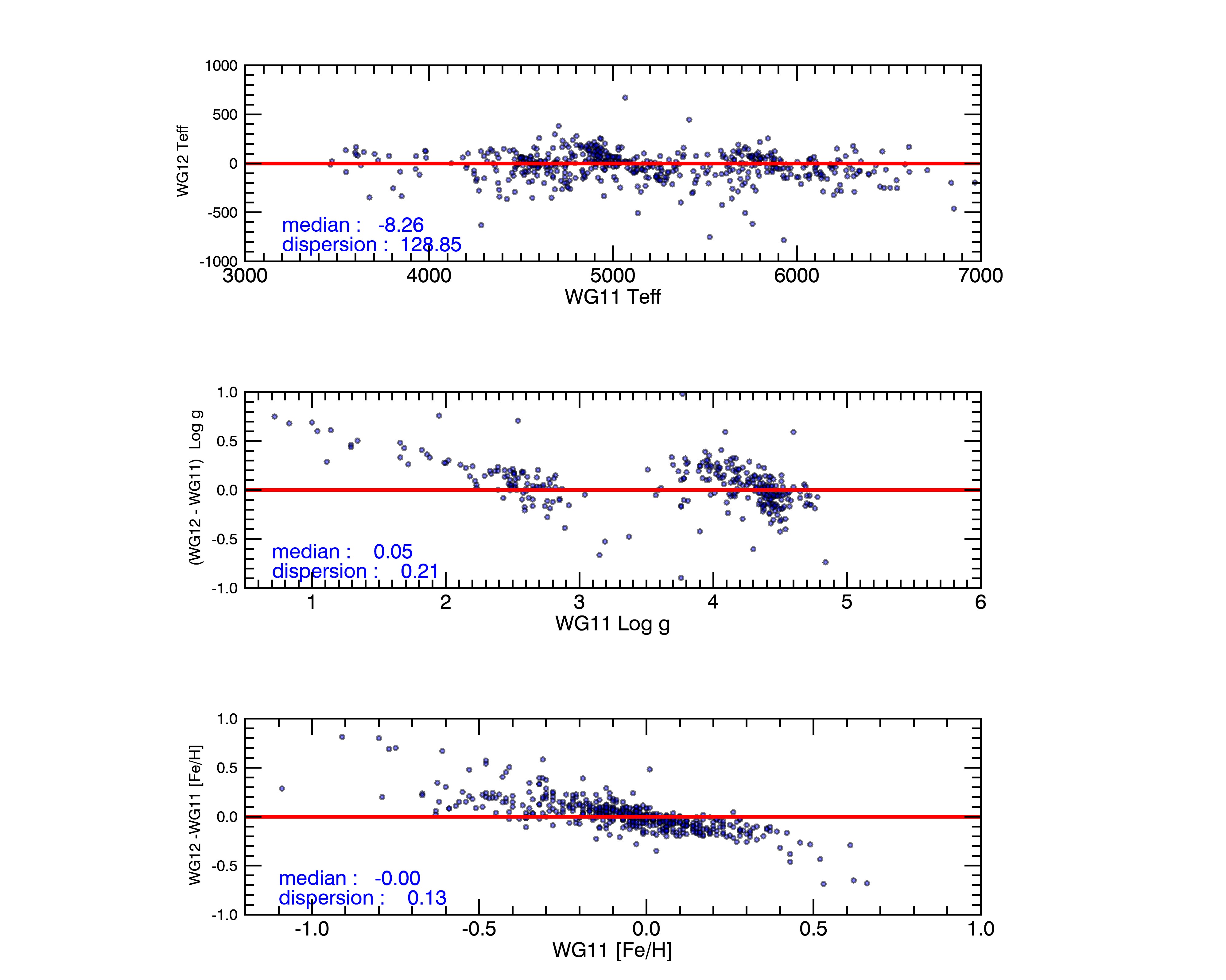}
 \caption{Comparison of stellar parameters between WG12 and WG11 for stars in common.}
\label{fig:wg12wg11}
 \end{figure*}

There are more than 600 spectra in common between WG11 and WG12. Most of them are
stars in clusters analysed by both working groups. Also there are the usual calibrators, mainly
benchmark stars.
In Figure~\ref{fig:wg12wg11}  we show the comparison between WG11 and WG12 results for $T_{\rm eff}$, $\log g${}
and [Fe/H].
The differences between WG12 and WG11  are small indicating   very good agreement. There is no systematic offset to apply given the very low median difference compared to the dispersion. However, there is a trend in the [Fe/H] difference with an increase of the difference when the metallicity decreases. A substantial difference is also found for the difference in gravity at low $\log g$, below $\log g$ $\simeq$ 1 dex.  

\section{Abundance homogenisation}
\label{sec:abu}
In the following section, we discuss the homogenisation of the abundances derived by WG10 and WG11. 
We do not include in the discussion the abundances of WG12 and WG13 for the following reasons: 
on the one hand, in the final release the abundances of WG12 obtained at high resolution with UVES were treated consistently with those of WG~11, following the same analysis flow. They therefore became part of the WG11 sample of abundances; on the other hand, medium-resolution observations of WG12 with the HR15N setup essentially allow us to measure Li abundance. Li is homogeneously determined for the entire survey by a single node, the analysis of which is described in detail in \citet{franciosini22}. 

Finally, the abundances of WG13 are obtained for stars in different regions of the parameter space: the derived abundances are for different  elements and ionisation states with respect to those obtained in FGK stars,  often strongly influenced by diffusion and non-LTE effects, thus not directly comparable with those of cooler stars, even if they belong to the same cluster. 

We refer to \citet{blomme2022} for a complete description of the process of analysis and homogenisation of WG13 spectra. 

\begin{figure}
  \resizebox{\hsize}{!}{\includegraphics{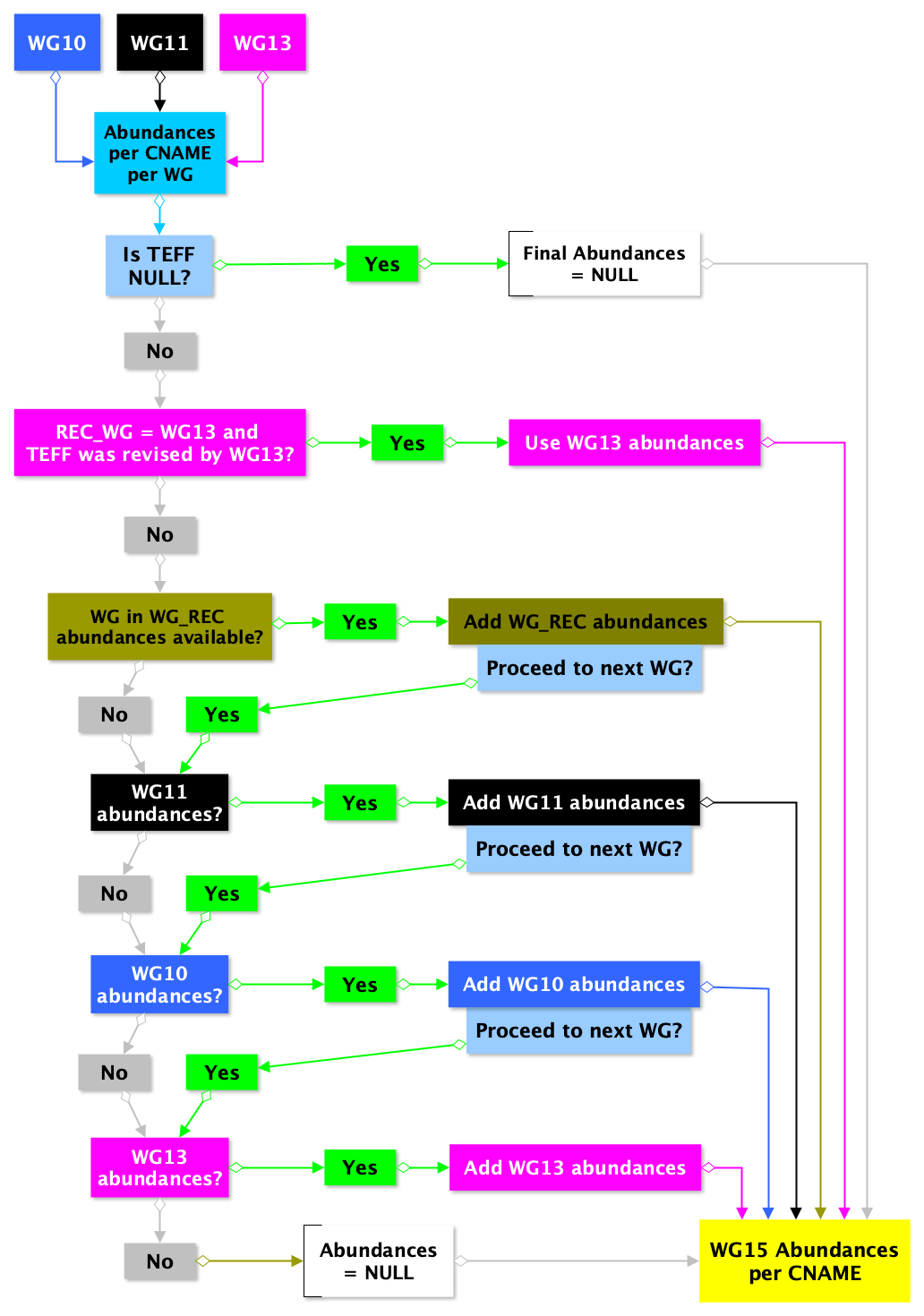}}
  \caption{Schematic view of the GES WG15 Abundance Homogenisation Workflow.}
  \label{fig:wg15_abun_wrkflw}
\end{figure}

Figure~\ref{fig:wg15_abun_wrkflw} illustrates the rules applied in order to homogenise the elemental abundances from WG10, WG11 (including the analysis of the WG12 UVES spectra) and WG13.

Here, we focus on the WG10 and WG11 abundances, whose elemental abundances for the different setups are computed starting from the homogenised stellar parameters. 
As shown in the analysis of the abundances derived from the different nodes in WG11 \citep{smiljanic14}, this does not guarantee that they are automatically perfectly consistent. 
The process of mapping the results of WG10 onto those of WG11, described in detail in \citet{worley2023}, alleviated any eventual discrepancy between the results of the two WGs. 
As the large wavelength range and the high resolution of the UVES
spectrograph permits a more precise determination of the stellar parameters and abundances than do the GIRAFFE spectra, the abundances obtained from the UVES spectra are taken as a reference for both the process of homogenisation and for defining the final checks.  
The task of WG15 in this final data release cycle is therefore limited to a final set of checks for consistency and homogeneity of the results, using the various tools and calibrators available. In what follows, we describe the main quality checks performed on the elemental abundances. 

\subsection{Light elements with abundances from single Nodes}
\label{sec_licno}
The abundances of light elements Li, C, N and O do not enter into the abundance homogenisation cycle since they are derived by single nodes: the Arcetri node for Li and the Vilnius node for CNO \citep{tau15}. 

The lithium abundance is measured from the doublet lines at 670.8~nm in the U580 and HR15N setups.  At the HR15N resolution, the doublet is blended with the nearby FeI line at 670.74~nm, while the two components are separated in UVES. In the final release, the Li abundances were derived in a homogeneous way by means of the equivalent widths (EWs) using a set of curves of growth \citep{franciosini22}  specifically derived for GES.

In the case of GIRAFFE, where only the total blended Li$+$Fe EW can be measured, the Li-only EW was first computed by applying a correction for the Fe blend. When the line is not visible or  barely visible, an upper limit to the EW, equal to the uncertainty, or to the measured EW if higher, is given.

The abundances of C, N and O are derived from molecular bands and atomic lines, with a simultaneous fit of the three abundances. 
The  $^{12}$C$^{14}$N molecular bands 6470-6490 \AA~, the C$_2$ Swan (1,0) band head at 5135\AA, the C$_2$ Swan (0,1) band head at 5635.5\AA, and the forbidden [O~{\sc I}] line at 6300.31\AA~ are used.  The analyses are performed through spectral synthesis with the $Turbospectrum$ code \citep{plez2012}.
For the determination of the oxygen abundance,  the oscillator strengths of the two lines of Ni are taken into account \citep{Johansson03}. 

The carbon abundance is also derived from atomic lines (C~{\sc I}). In this case, the abundance is derived by several nodes, and homogenised with the same procedure as for the other elements. 

In the final database the abundances of C and N from molecular bands are indicated with C\_C2 and C\_CN.

\section{Quality checks on the elemental abundances}
\label{sec:QC}
\begin{figure*}
  \includegraphics[width=1\linewidth]{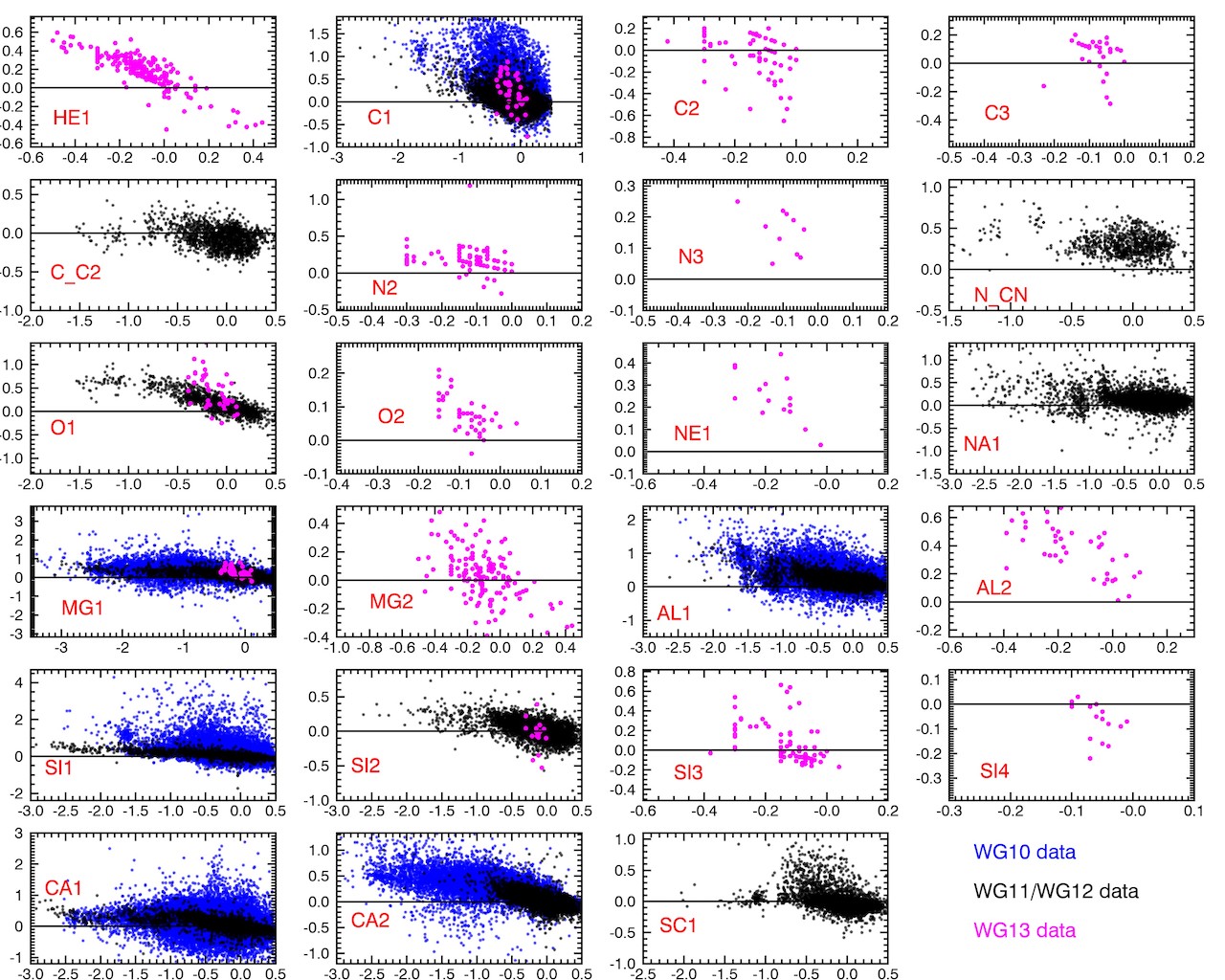}
\caption{Abundance ratios [X/H] versus [Fe/H] for all  the elements analysed in GES (except Li). Results for WG10 are  represented by blue circles, WG11 (and WG12) are shown as black circles and WG13 results are represented by magenta circles. C\_C2  are carbon abundances from C2 molecular bands, N\_CN  are nitrogen abundances from CN molecular bands}
 \label{fig:allabu1}
 \end{figure*}

\begin{figure*}
  \includegraphics[width=1\linewidth]{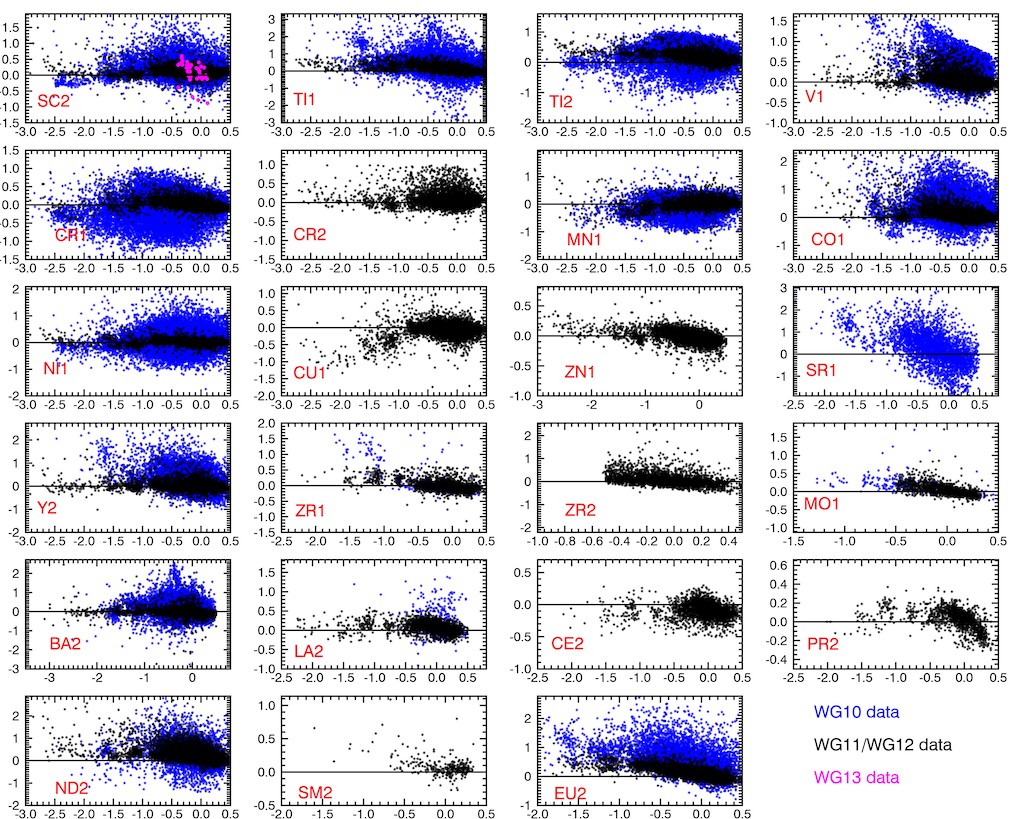}
\caption{Abundance ratios [X/H] versus [Fe/H] for all  the elements analysed in GES (except Li presented in detail in \citet{franciosini22}). Results for WG10 are  represented as blue circles, WG11 (and WG12) are shown as black circles and WG13 results are represented as magenta circles.}
 \label{fig:allabu2}
 \end{figure*}

In this section, we present some examples of the quality checks performed on the final abundance database. 
In Figs.~\ref{fig:allabu1} and \ref{fig:allabu2} the abundance ratios of all the elements analysed in GES (except Li) as a function of [Fe/H] are shown. The different colours represent the results from the different Working Groups. 

\subsection{Elemental abundances in benchmark stars}
\begin{figure*}
  \includegraphics[width=1\linewidth]{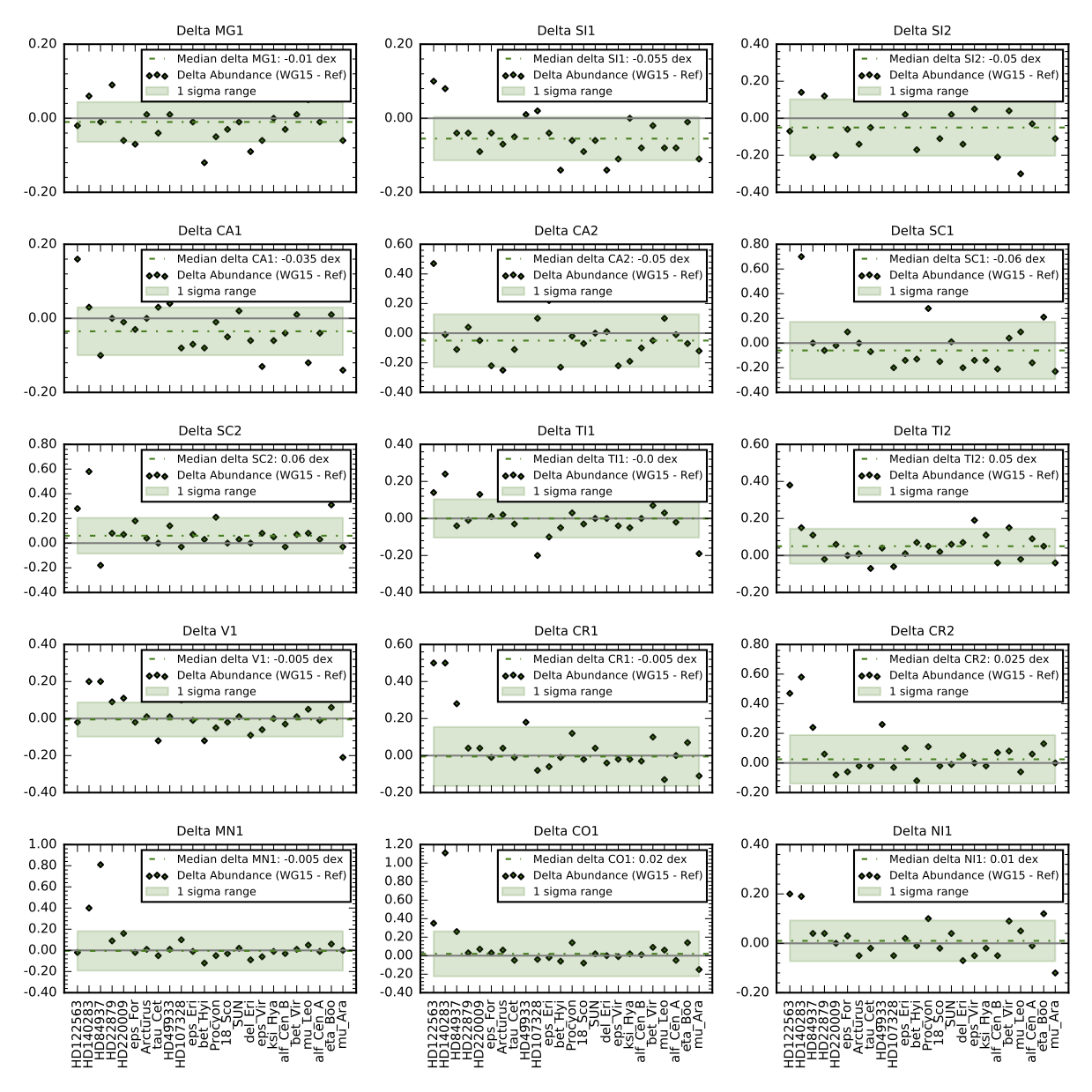}
\caption{Delta abundance  ratios [X/Fe]: WG15 minus reference abundances for benchmark stars from \cite{jofre15}.}
 \label{fig:bmdeltaabund1}
 \end{figure*}

Quality checks include a visual inspection of the WGs' and homogenised WG15 elemental abundance results in a variety of views, such as the abundance against $\mathrm{[Fe/H]}$ or per benchmark, with one plot per elemental species. For the elements where we have reference abundances available \citep[the ten elements, besides Fe, in][]{jofre15}, the plots of delta abundance with respect to the reference abundances are generated and viewed. Figure~\ref{fig:bmdeltaabund1} shows the delta abundance of the WG15 values minus the reference values per benchmark. The fifteen panels represent the ten elements with reference values, for some of which we have reported results for more than one ionisation species (such as CaI and CaII). The comparison to iDR5 was additionally checked and showed good consistency or improvement in the delta abundance results for iDR6. Results for the Sun are included in the delta abundance plots. See \citet{randich22} for more detail on the quality of the GES solar abundances.

By visually checking the WG results with the selected WG15 results highlighted, we were able to identify updates that could be made to the homogenisation rules, for example in cases where an abundance result was not available for particular benchmark stars in the preferred WG according to the homogenisation rules we implemented (e.g. WG11), but a result from another appropriate source was available (e.g. WG10). Whereas results from different WGs are not mixed for the stellar parameters (i.e. the parameters $T_{\rm eff}$, $\log g${} and $\mathrm{[Fe/H]}$ will always come from the same WG for a particular star), the rules for the abundance results are relaxed to allow mixing of the WG of origin of the results.

\subsection{Elemental abundances in field stars}
\begin{figure*}
  \includegraphics[width=1\linewidth]{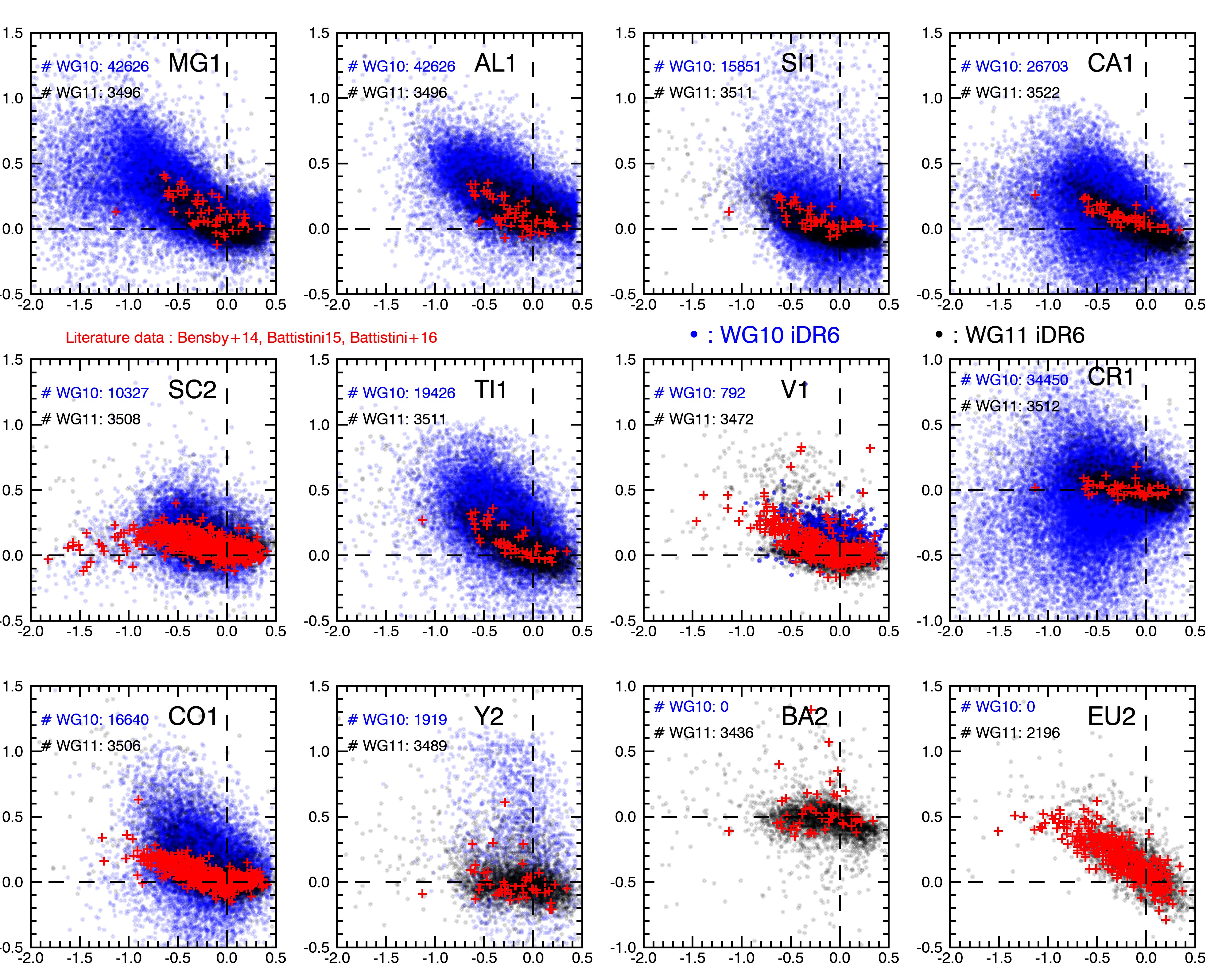}
\caption{WG10 and WG11 abundances  ratios [X/Fe] vs [Fe/H] for Milky Way stars : WG10 stars are represented by blue symbols and  WG11 stars by black symbols -  the red symbols refer to literature 
data \citep{bensby14,battistini15,battistini16}}.
 \label{fig:wg10wg11mwabu1}
 \end{figure*}

In Fig.~\ref{fig:wg10wg11mwabu1} we display the abundances in the [X/Fe] versus [Fe/H] planes of the elements in common between WG10 and WG11 and we compare them with a compilation of recent literature results. For WG10, we have selected only the results obtained from spectra with SNR $\ge 10$. 
The figures show not only the excellent agreement between WG10 and WG11 abundance trends, but they also 
demonstrate a very good match with the literature data coming from the  very high resolution (40,000-110,000), very high S/N (150-300)\footnote{Resolution and SNR values from \citet{bensby14}.} ratio spectra analyses from \citet{bensby14,battistini15,battistini16}. We also see clearly the larger  dispersion in the [X/Fe] ratios as a function of [Fe/H] found in the WG10 results compared to the WG11 results, naturally explained by the lower dispersion of the spectra used by WG10 and the smaller wavelength range.

\subsection{Elemental abundances in the combined cluster sample }

\begin{figure}
\begin{subfigure}{.5\textwidth}
  \centering
  \includegraphics[width=1\linewidth]{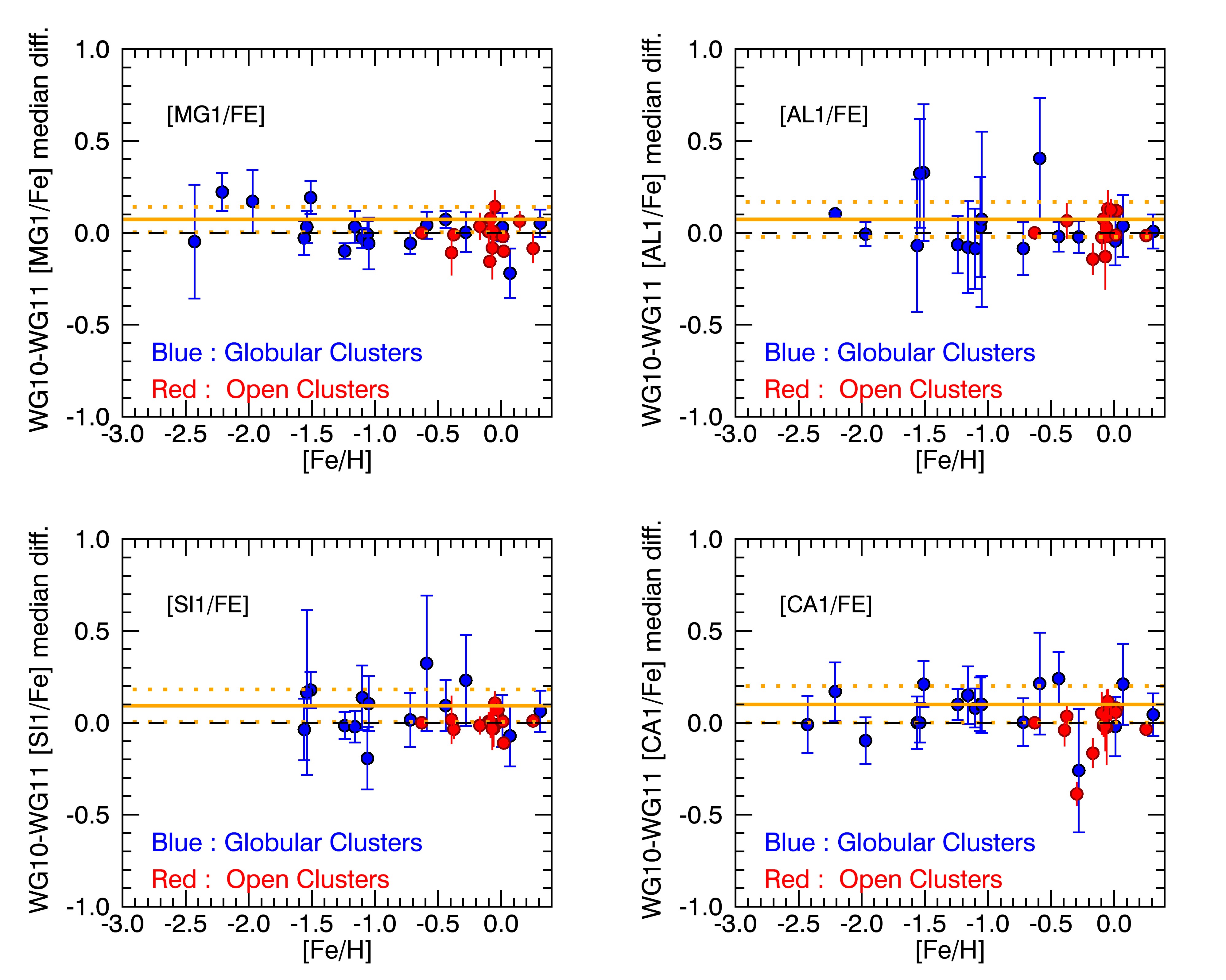}
\end{subfigure}\\
\begin{subfigure}{.5\textwidth}
  \centering
  \includegraphics[width=1\linewidth]{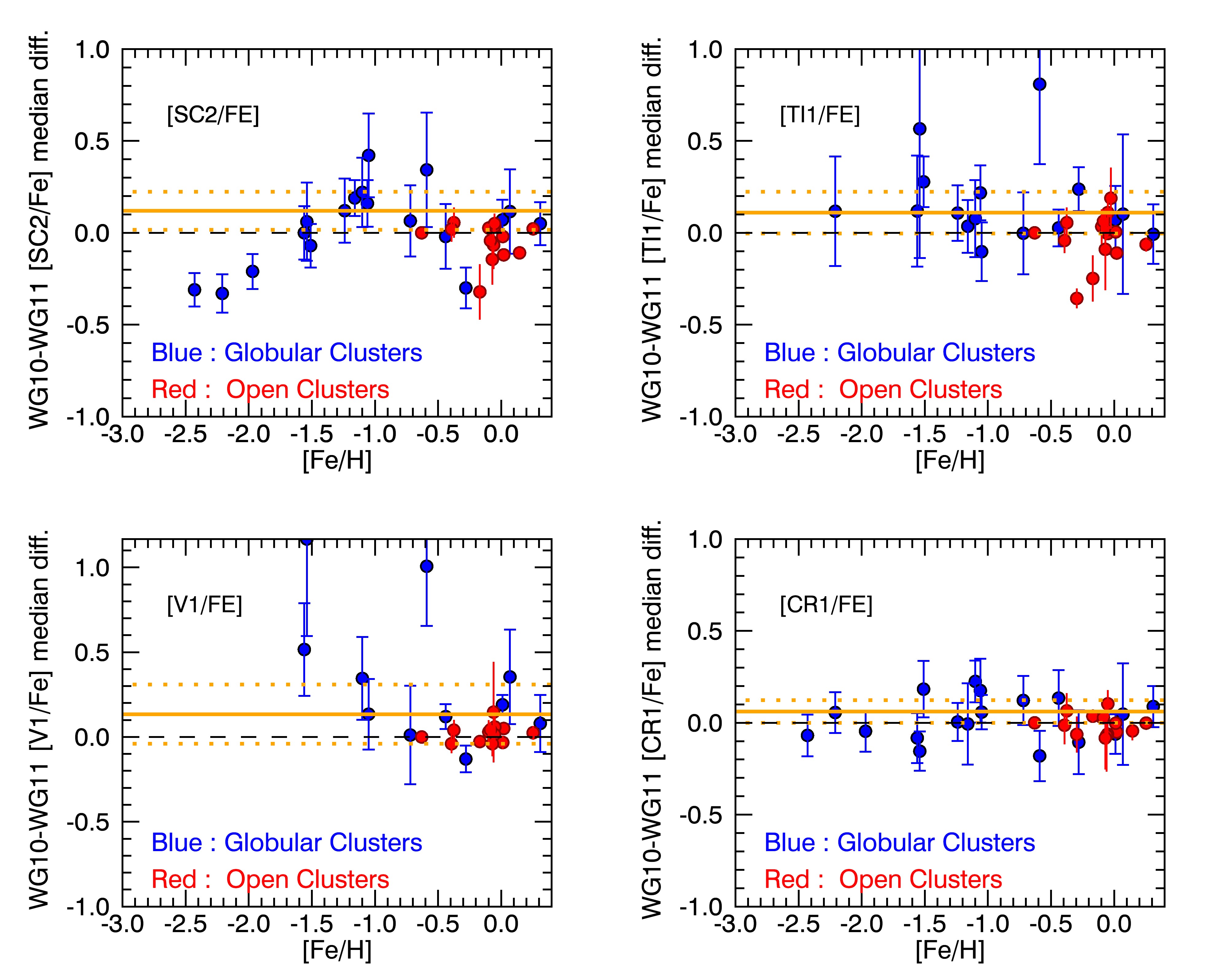}
\end{subfigure}\\
\begin{subfigure}{.5\textwidth}
  \centering
  \includegraphics[width=1\linewidth]{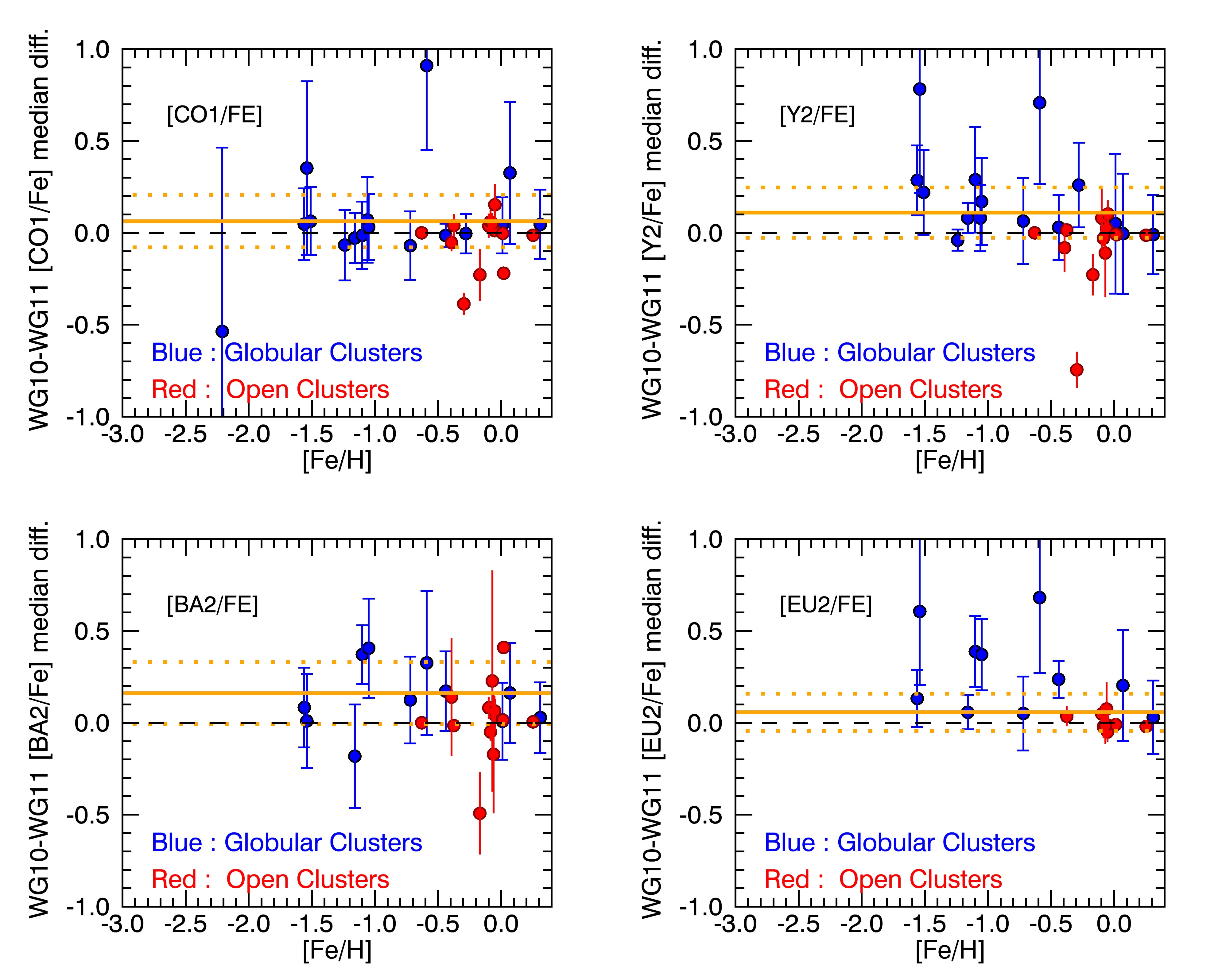}
\end{subfigure}\\
\caption{Open and globular clusters: Each figure represents the results for a different element. $\mathrm{[Fe/H]}$ is in abscissa. The ordinates are the values ``median WG11 abundances - median WG10 abundances''. The median global difference is shown as an orange line. Dotted orange lines indicate $\pm$1-$\sigma$ standard deviation from the median value.  }
\label{fig:wg10wg11ocgcabu}
\end{figure}

In Fig.~\ref{fig:wg10wg11ocgcabu}  the results for the open and globular clusters are combined.  For each cluster, the difference in the median abundances between WG10 and WG11 are computed and plotted. The orange line indicates the  median offset computed  for the sample of globular and open clusters. The dotted orange lines indicate the standard deviation  around the median value.  We  note that the offsets versus metallicity are in good agreement for both open and globular clusters.  
Discrepancies are found in some globular clusters for some elements that have only a few absorption lines measurable given the moderate resolution of the WG10 spectra.

\begin{figure*}
 \centering
\includegraphics[width=1.0\linewidth]{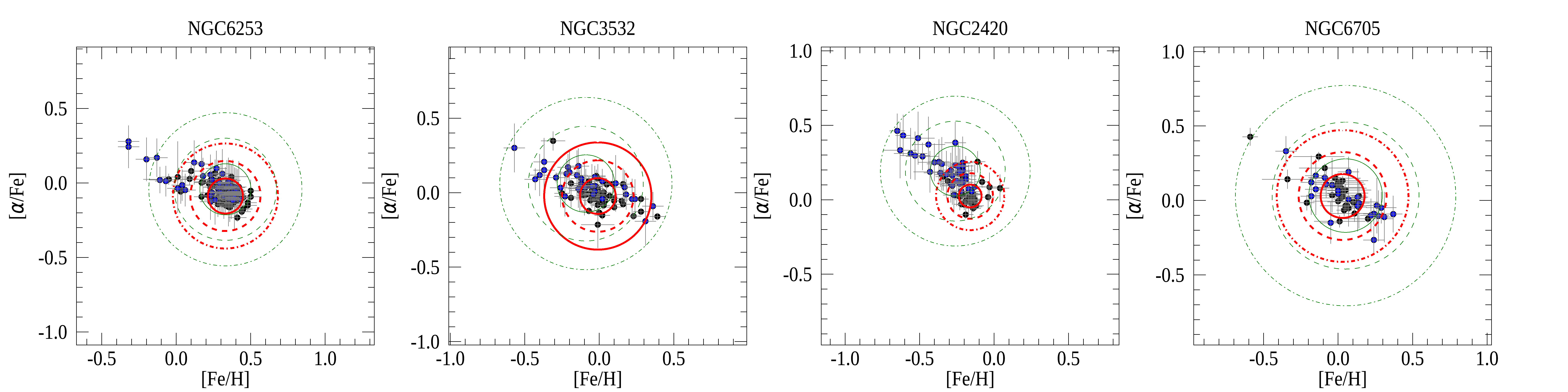}
 \caption{$\alpha$-element abundance over iron, computed averaging [Mg/Fe], [Si/Fe], [Ca/Fe] and [Ti/Fe] as a function of [Fe/H] in member stars of four calibration open clusters. In black the abundances from WG~11 and in blue from WG~10. The red circles mark the median WG~11 abundances within 1-$\sigma$, 2-$\sigma$, and 3-$\sigma$. The green circles indicate the same quantities but computed with WG~10 results. 
 }
\label{fig:oc_precision and accuracy}
\end{figure*}

\begin{figure*}
 \centering
\includegraphics[width=1.0\linewidth]{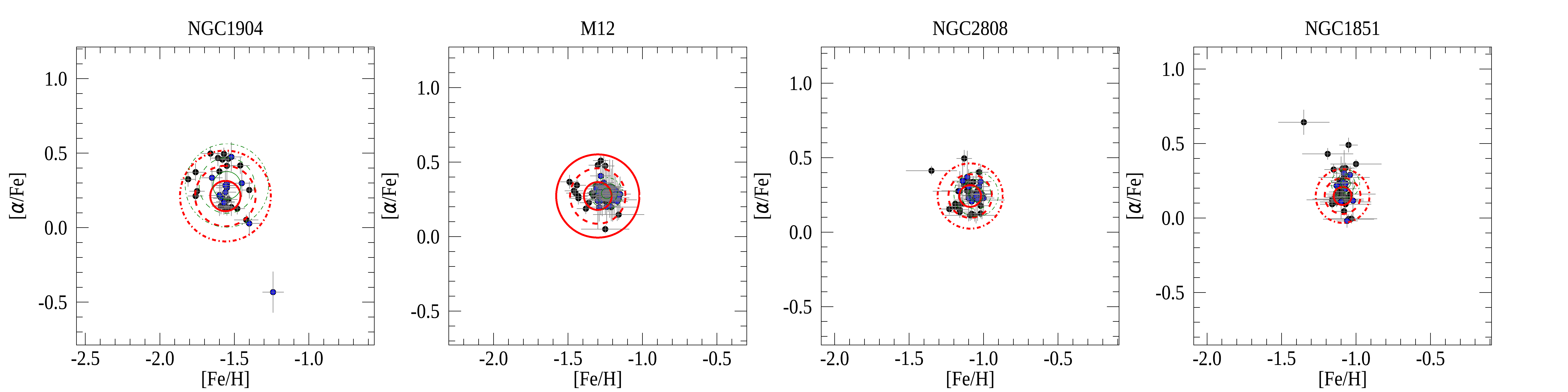}
 \caption{[$\alpha$/Fe] as a function of [Fe/H] in member stars of four calibration globular clusters. 
 Symbols and colours as in Fig.~\ref{fig:oc_precision and accuracy}.
 }
\label{fig:gc_precision and accuracy}
\end{figure*}

The abundances of members of  clusters derived by WG10 and WG11 can be used to estimate the final accuracy and precision of our homogenised abundances. 
In Figs.~\ref{fig:oc_precision and accuracy} and \ref{fig:gc_precision and accuracy}  we display, as an example, the average $\alpha$-element abundance over iron, computed averaging [Mg/Fe], [Si/Fe], [Ca/Fe] and [Ti/Fe] as a function of [Fe/H] in member stars of four calibration open clusters and in four calibration globular clusters. We computed the median abundances, separately, for member stars observed with the GIRAFFE setups HR10|HR21 and analysed by WG10 and for those observed with U580 and analysed by WG11. We removed from computing the median abundances the stars with E\_LOGG > 0.25 and with an uncertainty on [$\alpha$/Fe], e\_[$\alpha$/Fe] > 0.15. The members are selected to have a probability MEM3D > 0.95. 
Although U580 and HR10|HR21 are very different setups in terms of resolution and spectral range, the figures show that the homogenisation process produces values in very good agreement, in some cases within 1-$\sigma$, in slightly worse cases within 2-$\sigma$, as. e.g., for NGC~2420. The comparison indicates that WG10 results are on average accurate (similar median values as WG~11), but, as expected by their lower resolution and shorter spectral coverage, less precise. 

\section{Radial and rotational velocity homogenisation}\label{sec:rv_vsini}

\subsection{Radial Velocities}\label{sec:rv}
The \textit{Gaia} Radial Velocity Standards \citep[GRVS, ][]{soubiran13} were used to investigate the robustness of the Gaia-ESO radial velocities. While part of the calibration strategy, the 29 GRVS stars were observed in only five of the GES setups: HR10, HR15N, HR21, HR9B and U580. The radial velocities for all of the setups were calculated as described in \cite{gilmore22}. The radial velocities were calculated on the stacked spectra upon which the stellar parameters and abundances were also determined. For investigation of the radial velocity variations between individual or nightly stacked spectra, we refer to \citet{jackson15}. 

However, the remaining setups (HR3, HR4, HR5A, HR5B, HR6, HR14A, HR14B), HR15N and HR9B were part of the WG13 Hot Star analysis for which a radial velocity was calculated using some combination of these setups \citep{blomme2022}. This set of radial velocities was treated as a separate `setup' for the purposes of the below calibration procedure.

\begin{figure*}
 \centering
\includegraphics[width=1.0\linewidth]{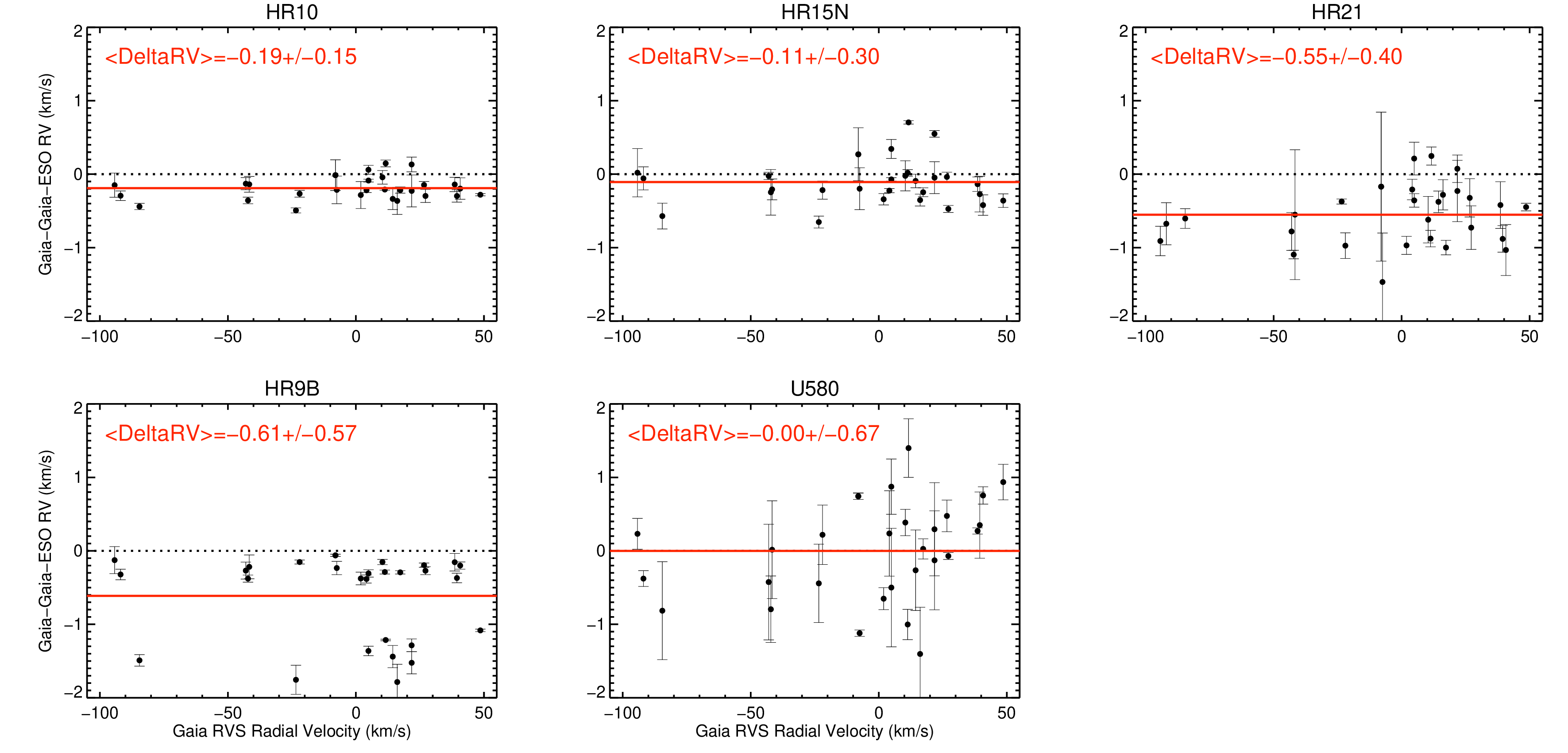}
 \caption{Difference of GRVS radial velocity to mean Gaia-ESO radial velocity calculated per GRVS per setup. The mean difference and standard deviation per setup are given.}
\label{fig:rvhomg_gaiarvs}
\end{figure*}

Figure~\ref{fig:rvhomg_gaiarvs} shows the difference between the GRVS radial velocity and radial velocity derived within Gaia-ESO for each GRVS for each relevant setup.

Note that often multiple spectra were obtained for each GRVS for each setup and so the mean of the radial velocities per GRVS was calculated per setup. If there was only one measurement the error taken was that associated with the value. If there were multiple measurements, the standard deviation of the measurements was taken as the error on the mean value. These are shown as error bars in Figure~\ref{fig:rvhomg_gaiarvs}.

Comparing the relative values between setups of the offset mean and standard deviation, and the standard deviations per GRVS per setup, HR10 shows the most robust agreement with the GRVS values. It was therefore selected as the baseline setup to which radial velocities for the other setups would be calibrated. The goal was not to then calibrate to the GRVS but rather to report the homogenised Gaia-ESO radial velocities.

As an internal calibration set the GRVS were limited in usefulness as they were only observed for 5 of the 12 setups used across the WGs. It was then necessary to construct a bootstrapping procedure to maximise the samples in common between setups in order to calibrate them onto HR10. Each setup was investigated against all other setups to see which had the most stars in common and also what the `in-common' set contained. In some cases it was just the Sun (e.g. HR5B and HR14B), so the calibrations for some setups are not particularly robust.

In general, there were four possible bootstrap procedures:
\begin{enumerate}
    \item Offset calculated directly with HR10
    \item Offset calculated directly with HR15N then bootstrapped to HR10
    \item Offset calculated directly with HR9B then bootstrapped to HR15N then bootstrapped to HR10
    \item Offset calculated directly with U580 then bootstrapped to HR15N then bootstrapped to HR10
\end{enumerate}

Table~\ref{tab:rv_bootstrap} gives the details on the bootstrap procedure for each setup and the resulting offset applied to calibrate the radial velocities of each setup to HR10.

\begin{table*}[htbp]
  \centering
  \caption{Bootstrap procedure applied per setup (X) for calibration of radial velocities to HR10. The offset, standard deviation and number of stars in common for the bootstrap to the initial setup (Y), and for the overall bootstrap (BS) to HR10 are given.}
    \begin{tabular}{lcl|ccc|cc}
    \hline\hline
     SETUP &  N &  SETUP & \multicolumn{2}{|l}{X-Y} & N & \multicolumn{2}{l}{X-BS-HR10} \\
     \multicolumn{1}{c}{X} & XxHR10 & \multicolumn{1}{c|}{Y} & Offset & STD & XxY & Offset & STD \\
    \hline
    HR15N & 3161  & HR10 & 0.09  & 0.21  & 3161  & 0.09  & 0.21 \\
    HR21  & 48133 & HR10 & 0.59  & 0.46  & 48133 & 0.59  & 0.46 \\
    HR9B  & 253   & HR15N & 0.08  & 1.30  & 558   & 0.17  & 1.32 \\
    U580  & 154   & HR15N & 0.21  & 0.53  & 695   & 0.30  & 0.57 \\
    U520  & 46    & U580 & -0.10 & 0.92  & 54    & 0.20  & 1.08 \\
    HR3   & 123   & HR9B & -0.13 & 2.75  & 334   & 0.04  & 3.11 \\
    HR4   & 116   & HR9B & -0.43 & 5.03  & 211   & -0.26 & 5.20 \\
    HR5A  & 131   & HR9B & -0.10 & 3.98  & 151   & 0.07  & 2.57 \\
    HR5B* & 1     & HR10 & 0.15  & 0.22  & 1     & 0.15  & 0.22 \\
    HR6   & 118   & HR9B & 0.00  & 7.03  & 123   & 0.17  & 2.74 \\
    HR14A & 145   & HR9B & 0.16  & 2.37  & 146   & 0.33  & 3.97 \\
    HR14B* & 1     & HR10 & 0.29  & 0.22  & 1     & 0.29  & 0.22 \\
    WG13Comb & 506     & HR10 & 0.24  & 3.94  & 507     & 0.24  & 3.94 \\
    \multicolumn{3}{l|}{* Only Sun in common} &  &       &       &       &  \\
   \hline\hline
     \end{tabular}%
  \label{tab:rv_bootstrap}%
\end{table*}%

Offsets were calculated between each of the setups and the zero-point of the GES RV scale, HR10. The offsets were then applied to put the other setups onto the HR10 scale. For WG13, RVs  based on a combination of WG13 setups
were calculated for particular clusters (NGC~3293, NGC~6705, Trumpler~14, NGC~6530, NGC~2244, NGC~3766, and NGC~6649) and an offset was calculated with respect to HR10.

\begin{figure}
 \centering
\includegraphics[width=1.0\linewidth]{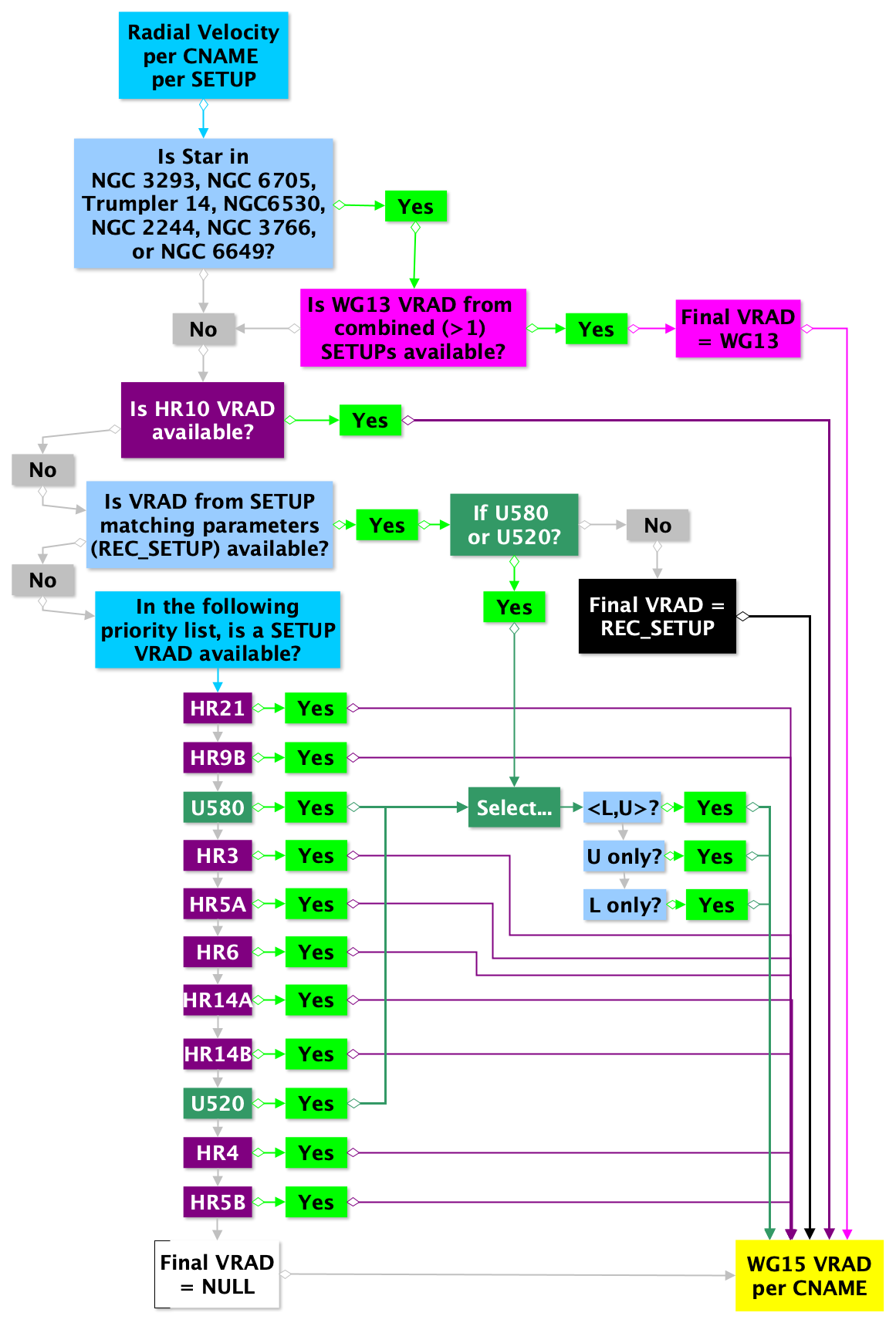}
 \caption{Schematic view of the GES WG15 Radial Velocity Homogenisation Work-Flow.}
\label{fig:rvhomg_wrkflw}
\end{figure}

Having assessed the baseline SETUP as HR10, and calculated offsets to put all RVs per SETUP onto the HR10 RV scale, the next stage was to assign an RV to each CNAME based on a set of rules. Figure~\ref{fig:rvhomg_wrkflw} illustrates the rules used to select an RV per CNAME. The offsets listed in Table~\ref{tab:rv_bootstrap} were applied when an RV other than from HR10 was selected.

The error is that associated with the selected value, except for the case when the value is the mean of the two values of the upper and lower arms of the UVES SETUP. In that case the error is calculated as the sum in quadrature of the errors on the two arms. The homogenised radial velocity is reported as VRAD in the final database.

\subsection{Rotational Velocities}\label{sec:vsini}
Rotational velocities were determined as part of the Arcetri UVES pipeline, the CASU GIRAFFE pipeline, by WG13 and by the OACT node. However, there was a re-calibration of the GIRAFFE instrument by ESO after the internal DR4 which changed the resolution such that the GIRAFFE radial velocity pipeline, which also derived rotational velocity \citep[See][]{gilmore22}, could not consistently determine \ensuremath{v\sin i} from the stacked spectra. Hence, after the internal DR4 no \ensuremath{v\sin i} were reported for the GIRAFFE spectra.

\begin{figure}
 \centering
\includegraphics[width=1.0\linewidth]{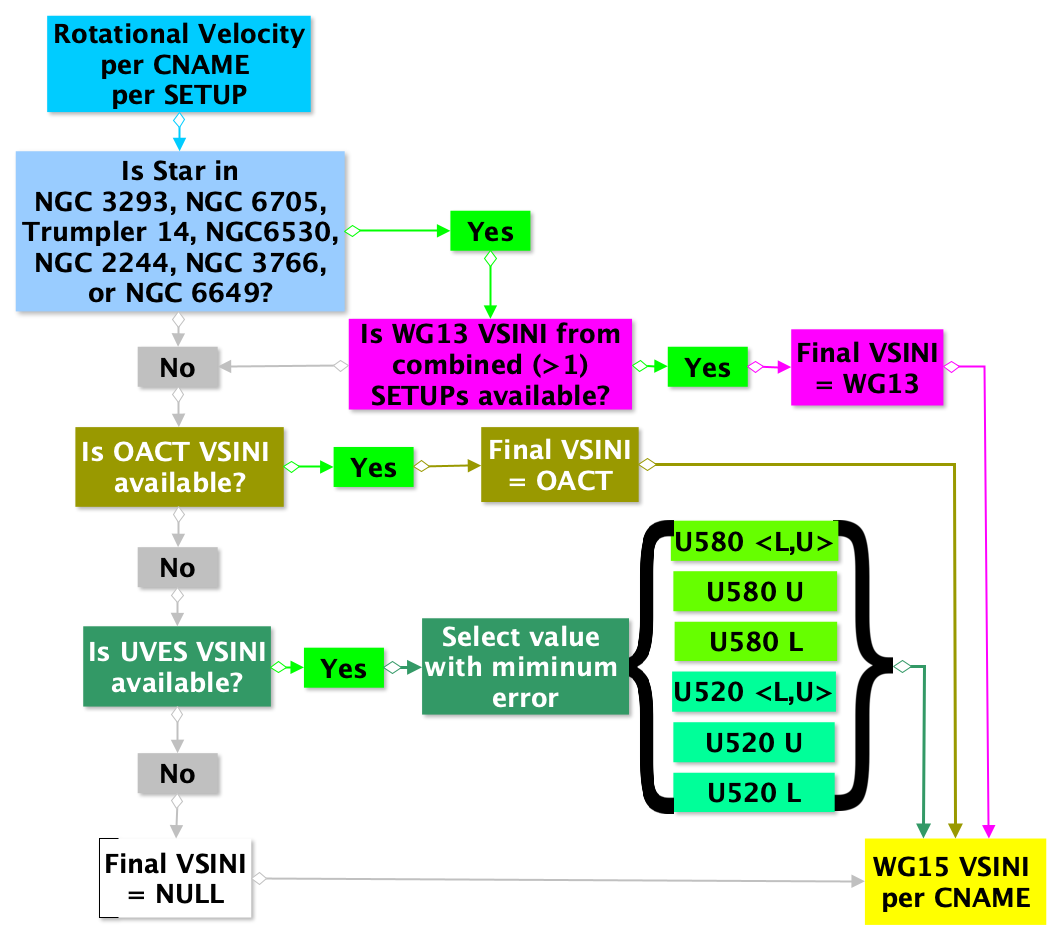}
 \caption{Schematic view of the GES WG15 Rotational Velocity Homogenisation Work-Flow.}
\label{fig:vsinihomg_wrkflw}
\end{figure}

Therefore the final GES catalogue reports \ensuremath{v\sin i} values only from the Arcetri UVES pipeline, WG13 or OACT. The rules governing the assignment of \ensuremath{v\sin i} are illustrated in Figure~\ref{fig:vsinihomg_wrkflw}.

The error is that associated with the selected value, except for the case when the value is the mean of the two values of the upper and lower arms of the UVES SETUP. In that case the error is calculated as the sum in quadrature of the errors on the two arms. 

\subsection{Signal-to-Noise Ratio}\label{sec:snr}
The Signal-to-Noise values reported in the final catalogue were selected to match the selection of the radial velocities. When a GIRAFFE RV was selected for the homogenised VRAD, the SNR from the specific GIRAFFE SETUP was selected. Similarly, when a UVES RV was selected, the SNR from the specific UVES setup was selected. The sequence is as shown in Figure~\ref{fig:rvhomg_wrkflw}.

However, often a combination of setups was used to calculate the VRAD. For instance, when combining UVES upper and lower arms, and the combination of setups used to calculate the WG13 VRADs. In these cases, the SNR values from the setups used in the calculated VRAD were summed in quadrature to provide the reported SNR.

\subsection{Other parameters}\label{sec:special_measurements}
In the final catalogue, there are some other parameters that do not enter into the homogenisation process. 

The photometric temperature from the infrared flux method \citep[TEFF\_IRFM, see][]{gonzalez2009} with its error are provided for more than 20,000 stars. This is derived as the weighted average of the IRFM values calculated on the 2MASS J, H, Ks bands.

Some specific parameters for young stars are provided by WG12, such as veiling (VEIL), parameters describing the mass accretion rate, such as the EW of the H$\alpha$ line (EW\_HA\_ACC) and  HA10, i.e. 10\% H$\alpha$, and parameters for the chromospheric activity obtained from the EW and flux of the H$\alpha$ emission line, EW\_HA\_CHR and FHA\_HA, and from the EW and flux the H$\beta$ one, EW\_HB\_CHR and FHB\_CHR. In a few cases also the mass accretion rate is provided (LOG\_MDOT\_ACC). All quantities are provided with their associated uncertainties, and are described in \citet{lanzafame15}. 

The GAMMA index is supplied for more than 20,000 stars. It is an alternative parameter to be used when it is not possible to properly estimate the surface gravity in stars observed with the GIRAFFE setup HR15N. For a complete description of the gravity and temperature indices, including GAMMA,  for HR15N we refer to \citet{damiani14}. 
In addition to Li abundance, described in Sec.~\ref{sec_licno}, the final catalogue provides  the EWs of Li lines: the measured equivalent width EW\_LI, with its associated error and indication of upper limit or measurement and the EW corrected for the contamination of the nearby Fe line, EWC\_LI, again with its error and upper limit indication. We refer to \citet{franciosini22} for a full description of Li EW measurement.  
Finally, the membership probability for stars in the field of  several open and globular clusters is provided in the MEM3D column. The membership analysis is described in \citet{jackson22J}. 

\section{The Stellar Parameter and Abundance Error Distributions}
\label{sec:errors}
For each measurement reported in the final GES catalogue, an associated uncertainty (e.g. in associated column 'E\_') is also reported. 

\begin{figure*}
 \centering
\includegraphics[width=1.0\linewidth]{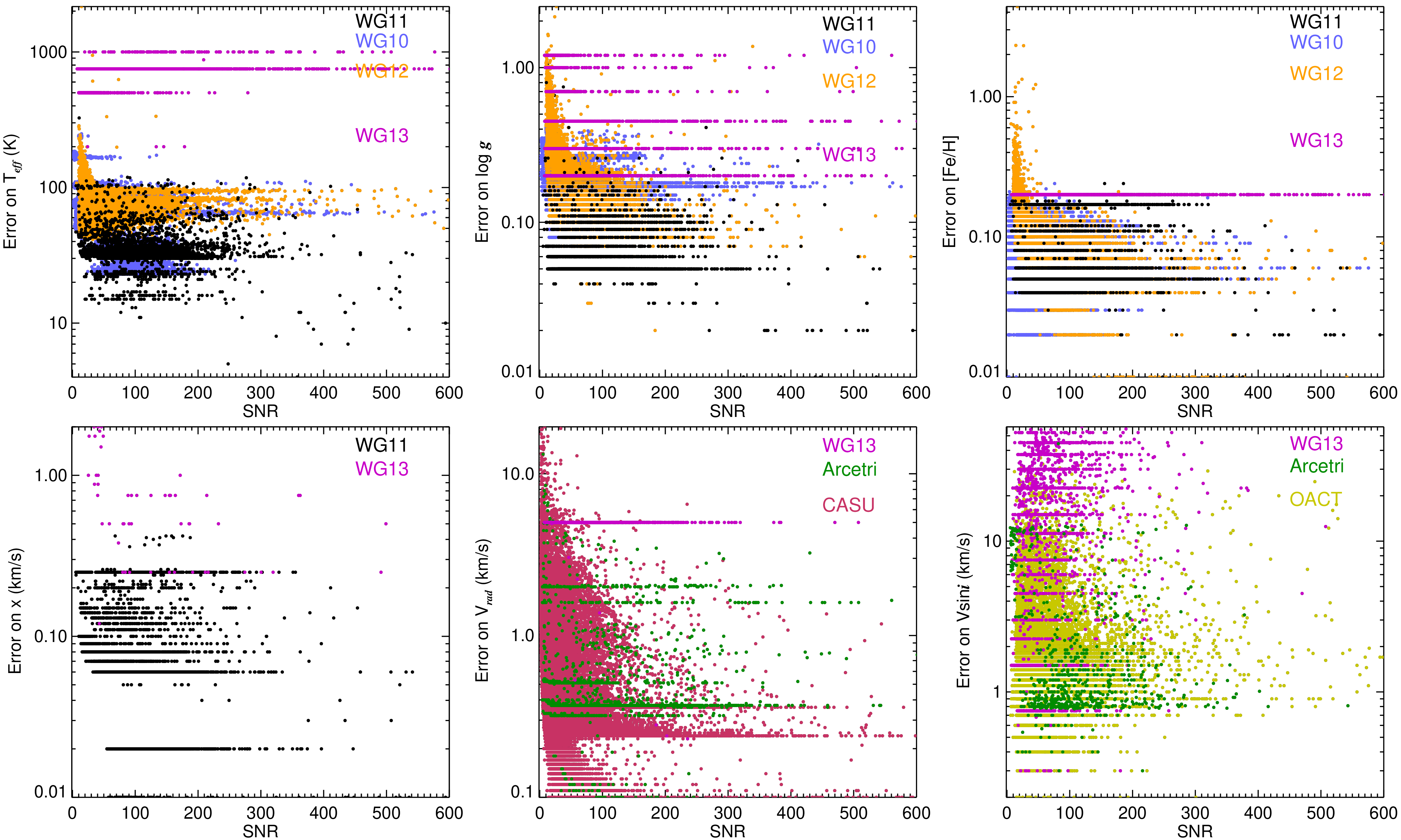}
 \caption{Error on $T_{\rm eff}${},~$\log g${}, ~$\mathrm{[Fe/H]}$,~ $\xi$,~$V_\mathrm{rad}$ and \ensuremath{v\sin i} against SNR. The provenances of the errors are indicated by colour (blue -- WG10, black -- WG11, orange -- WG12, magenta -- WG13, green -- Arcetri node, red -- CASU, yellow -- OACT node).}
\label{fig:errors_param}
\end{figure*}

Fig.~\ref{fig:errors_param} 
shows the reported error for each of $T_{\rm eff}${}, $\log g${}, $\mathrm{[Fe/H]}$, $\xi$, $V_\mathrm{rad}$ and \ensuremath{v\sin i} against SNR. The provenance of the errors are indicated by colour. For $T_{\rm eff}${}, $\log g${}, $\mathrm{[Fe/H]}$ and $\xi$ the provenance comes from WG10, WG11, WG12 or WG13. For $V_\mathrm{rad}$ the provenance is from the radial velocity pipelines, Arcetri for UVES or CASU for GIRAFFE respectively, or from further analysis by WG13 \citep{blomme2022}. See Section~\ref{sec:rv} for the provenance selection details. For \ensuremath{v\sin i} the provenance is from WG13, Arcetri or OACT. See Section~\ref{sec:vsini} for the provenance selection details.

\begin{figure*}
 \centering
\includegraphics[width=1.0\linewidth]{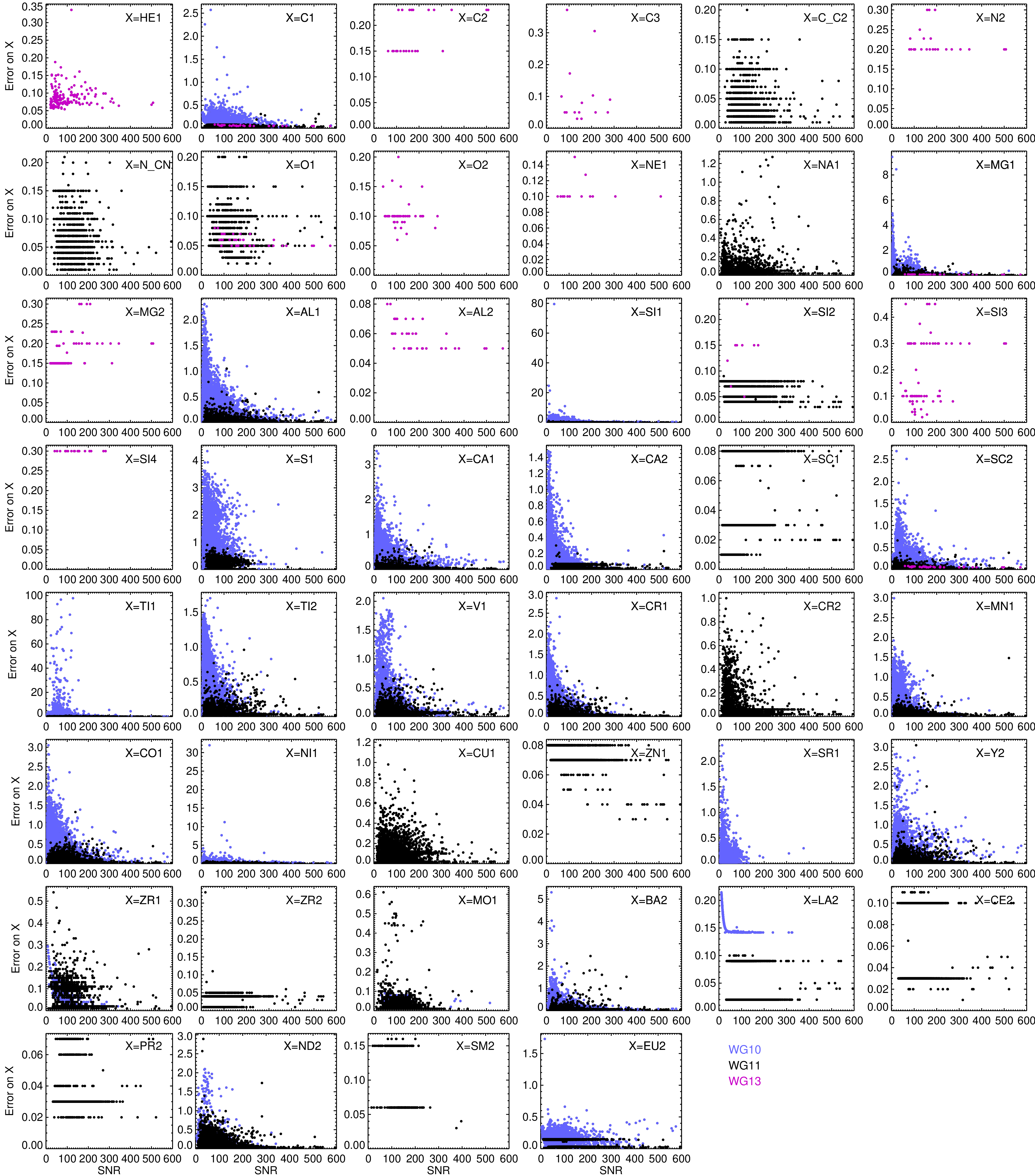}
 \caption{WG level error on each element abundance against SNR. The provenance of the errors are indicated by colour (blue -- WG10, black -- WG11, magenta -- WG13).}
\label{fig:errors_abun}
\end{figure*}

Similarly, Fig.~\ref{fig:errors_abun} shows the error for the element abundances reported by WG10, WG11 and WG13 (excluding L1 which is discussed in Section~\ref{sec:special_measurements}). 

The error model for each parameter and abundance is defined per provenance source, and no homogenisation of the errors occurs for the final catalogue. However, each provenance source, whether it is at the WG level or from either RV pipeline, provides an internally consistent error model reflective of the analysis at that level. See the descriptions in the associated papers for more details \citep{worley2020,gilmore22,blomme2022}. In general, as is shown in Fig.~\ref{fig:errors_abun}, the error models show decreasing error with increasing SNR as expected.

These error models are typically based on the measurements, not the errors provided by the node analysis. Therefore another column was provided for WGs to report an uncertainty based on the reported node analysis uncertainties, namely 'ENN\_'.

\begin{figure*}
 \centering
\includegraphics[width=1.0\linewidth]{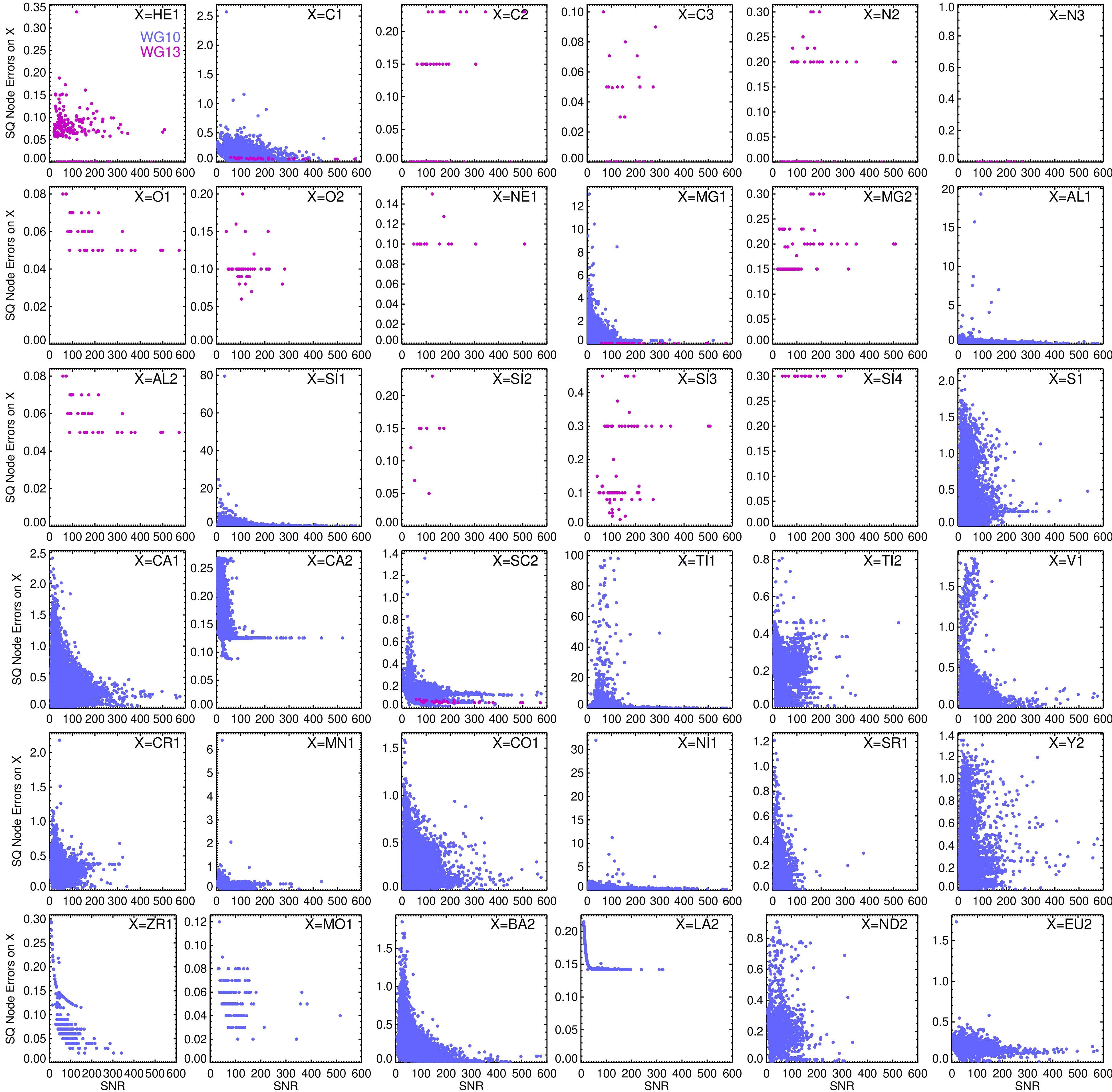}
 \caption{Errors based on the node analysis errors on abundances against SNR. The provenance of the errors are indicated by colour (blue -- WG10, magenta -- WG13).}
\label{fig:errorsnn_abun}
\end{figure*}

Fig.~\ref{fig:errorsnn_abun} shows errors based on the node errors for the element abundances for which these values are available.  

\section {Flags }
\label{sec:flags}

\subsection{Flag Homogenisation}
A sophisticated system of flags (detailed flags, hereafter) has been designed within the Gaia-ESO survey by WG14 \citep[see][ for details]{vaneck2022} to report and keep track of issues occurring during the analysis (TECH) and also to indicate physical peculiarities on a given target (PECULI).

The homogenisation is based on the dictionary of flags produced by  {\sc WG14}.
We compared the flags produced by the different {\sc WGs} and searched for possible conflicts.
In the case of differences in the confidence level flag, we took the highest confidence flag.
All the flags from the WG$\_$Recommended files are included with any duplicates removed.

The {\sc WG14} dictionary is available to the GES consortium on GES WG14 wiki page, and all the flags included in the final database are described in the document accompanying the public release in the ESO archive.

\subsection{WG15 Additional Flags}
\begin{figure*}
 \centering
\includegraphics[width=1.0\linewidth]{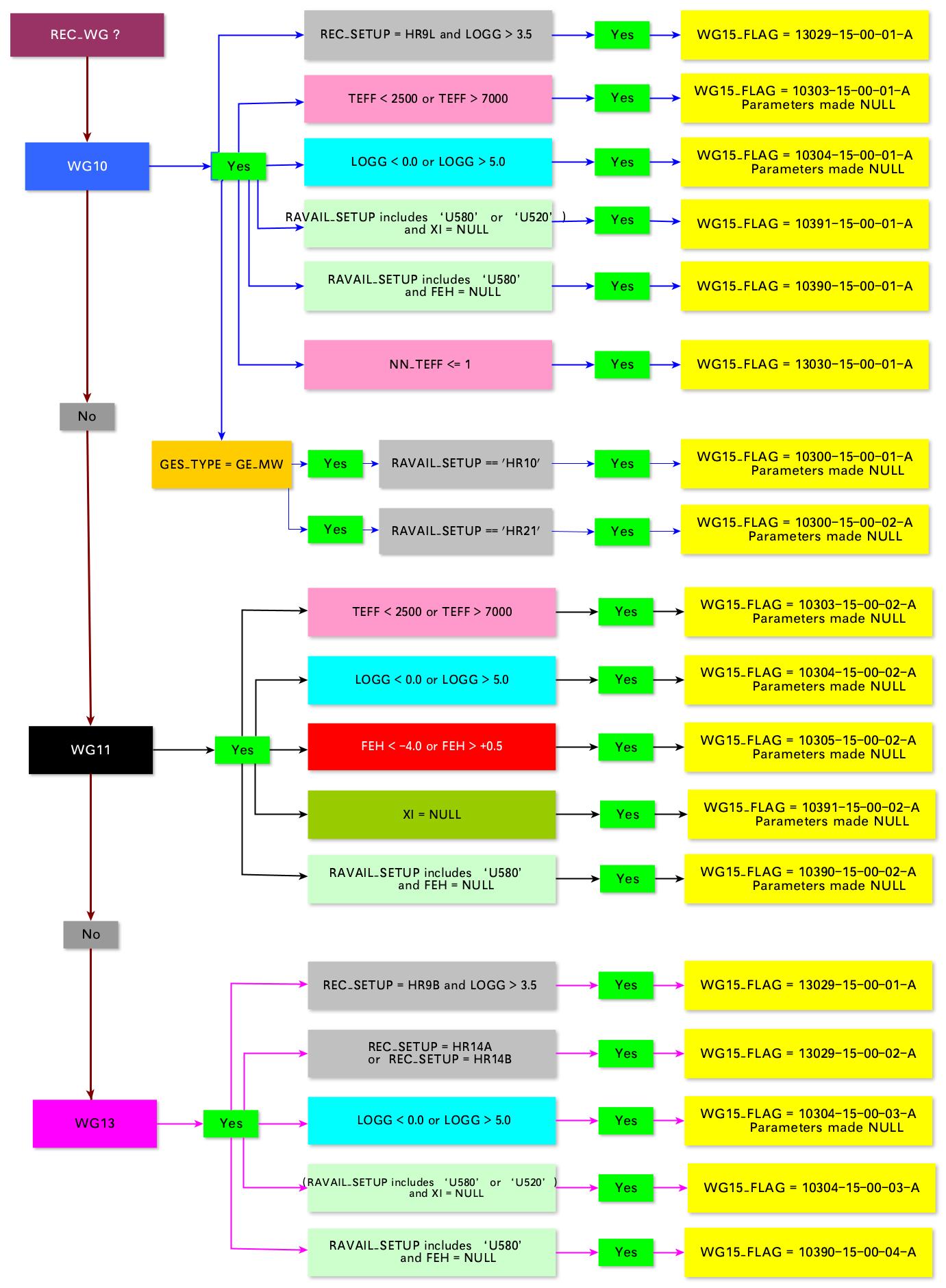}
 \caption{WG15 additional flags flow chart.}
\label{fig:wg14_flow}
\end{figure*}

Additional rules are added at the WG15 level, depending on the REC$\_$WG provenance previously assigned to each CNAME. This new set of rules aims to add WG15 flags or in some cases to remove stellar parameters. The detailed flow chart is shown in  Fig.~\ref{fig:wg14_flow}.

Parameters here mean the columns TEFF, LOGG, FEH, XI, MH, ALPHA$\_$FE and all associated number/error columns in the final database. Note that the flag suffixes do change between WGs, so they are not all identical even if they may at first appear to be. If multiple flags are activated, the WG15 flags are concatenated using ‘|’. The WG15 flags are then concatenated with the existing TECH column.

\subsection{Simplified Flags}
The TECH flags cover a broad range of topics (SNR, data reduction, determination \& quality of stellar parameters/abundances). The syntax of the flags allows one to quickly identify the issue (prefix), trace the originating working group (WG ID) and node (node ID) and, in some cases, have extra information (suffix). However, this system is too detailed for the end users aiming to quickly use the Gaia-ESO data. 

For iDR6, a system of simplified flags has thus been designed for the Gaia-ESO Survey.
These simplified flags must allow the end users to quickly filter the data to do their science.

They should allow the user to quickly reject objects with non-physical or highly suspicious results. They complete the information already carried by the error bars associated with the observables. Converting any flag to the simplified scheme will cause a loss of valuable information, and therefore, it is important to also release the detailed flags. The simplified flags consist of a small acronym (three letters) whose meaning is easily recoverable or can be easily guessed without looking at the documentation. They are coded with Booleans (False/True), each in an individual column, allowing the end users to easily sort upon them. The acronym and meaning of each flag are listed in the table below. A comment is also provided to specify when the flag is raised and to briefly illustrate the conversion from the detailed scheme to the simplified scheme. The default value of the simplified flag is False; in other words, only the value True carries information.

All TECH flags (except some `neutral' flags that are dropped during the conversion) have been translated into simplified flags (see next paragraph). On the other hand, only two simplified flags are defined to summarise the information carried by the most-used PECULI flags in order to quickly identify: (a) if the object is suspected to be a spectroscopic multiple (BIN), or (b) if emission lines are observed (EML).

Three simplified flags (SNR, SRP, SDS) deal with the intrinsic quality of the reduced spectra. The simplified flags pertaining to the stellar parameters (IPA, SSP, PSC) only deal with the effective temperature, the surface gravity, the metallicity and the microturbulence. Two simplified flags (NIA, SSA) give a general indication of the availability of abundance determinations (for any element but iron) for a given star. There is a dedicated simplified flag for the radial velocity (SRV), and for the rotational velocity (SRO).

It is not possible to have a limited set of simplified flags and at the same time to have a detailed assessment of each stellar parameter (and respectively, abundance). It means that the end-users have to perform some further checks (e.g. based on the detailed flags) to decide which abundances can be kept when an object has the flag "some suspicious abundances" raised. During the process of reducing the detailed flags to the simplified flags, a conservative approach was adopted, meaning that the problems might be less severe than indicated by the simplified flags. For example, the SSP (some suspicious parameters) or IPA (incomplete parameter) flags are sometimes raised when some, though not all, analysis nodes had uncertain parameters or abundances, and though other nodes might well have provided reliable results. Similarly, the flag SSA gives a general appreciation for the quality of abundance ratios attached to a given star. Given that, for instance, up to twenty chemical species are investigated in UVES observations, it is impossible for a unique simplified flag to give an accurate picture. Therefore, we advise that the flag SSA is used in a second step when outliers remain in the user's selection to identify objects for which a look at the detailed flags may be necessary. On the other hand, the simplified flags SNR, SRP, NIA may be used a priori to clean the user's sample.

The list of simplified flags and comments can be found in Appendix \ref{simplified}.

\section{Discussion and conclusion}
\label{sec:discussion}

In this section, we present some validation plots of the final recommended set of stellar parameters.

\subsection{The final Kiel diagram}

\begin{figure*}
 \centering
\includegraphics[width=1.0\linewidth]{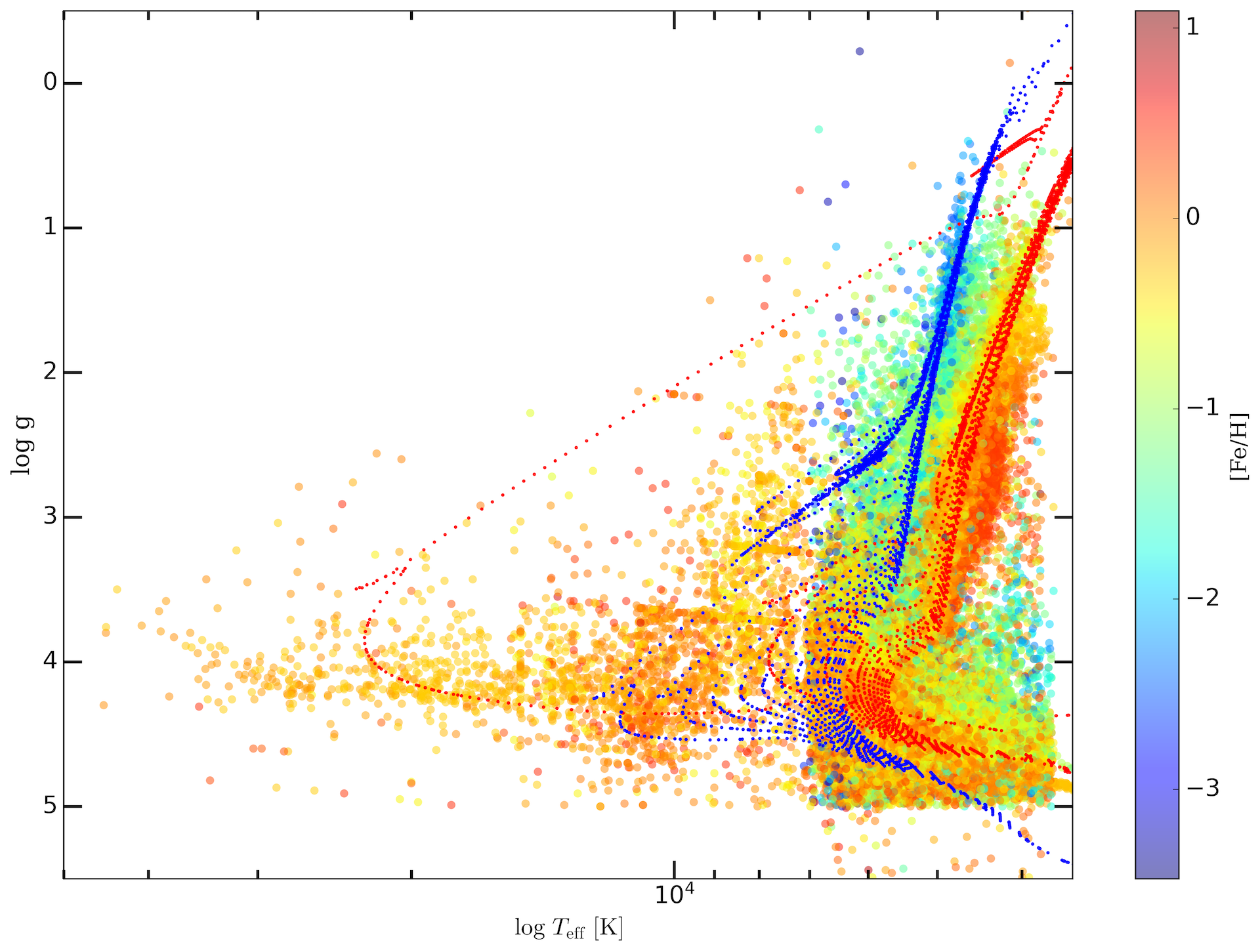}
 \caption{The Kiel diagram using the final recommended values for iDR6 (all stars). Two sets of  isochrones at solar metallicity and 0.01<Age[Gyr]<13 (in red) and at [Fe/H]$=-$2 and 1<Age[Gyr]<13 (in blue) are shown. }
\label{fig:idr6_kiel}
 \end{figure*}

In Fig.~\ref{fig:idr6_kiel}, we show the Kiel diagram of the entire latest release of GES (>114000 unique CNAMEs). The diagram shows the variety of spectral types analysed, a unique aspect of GES compared to other surveys: ranging from cool pre-main sequence (PMS) stars, to hot early-type stars and Red Giant Branch (RGB) stars, covering a metallicity range from $-2.5$ to 0.5~dex, from globular clusters to inner-disc open clusters. Comparison with two representative sets of isochrones at solar metallicity and $\mathrm{[Fe/H]} = -2$~dex indicates very good agreement, with a shift of RGB stars toward higher temperatures for the more metal-poor stars.
Also noticeable from the figure is the intrinsic difficulty in measuring the metallicities of cool PMS stars, whose spectra are dominated by molecular bands. 

\subsection{The [Mg/Fe] vs [Fe/H] diagram}

\begin{figure*}
 \centering
\includegraphics[width=1.0\linewidth]{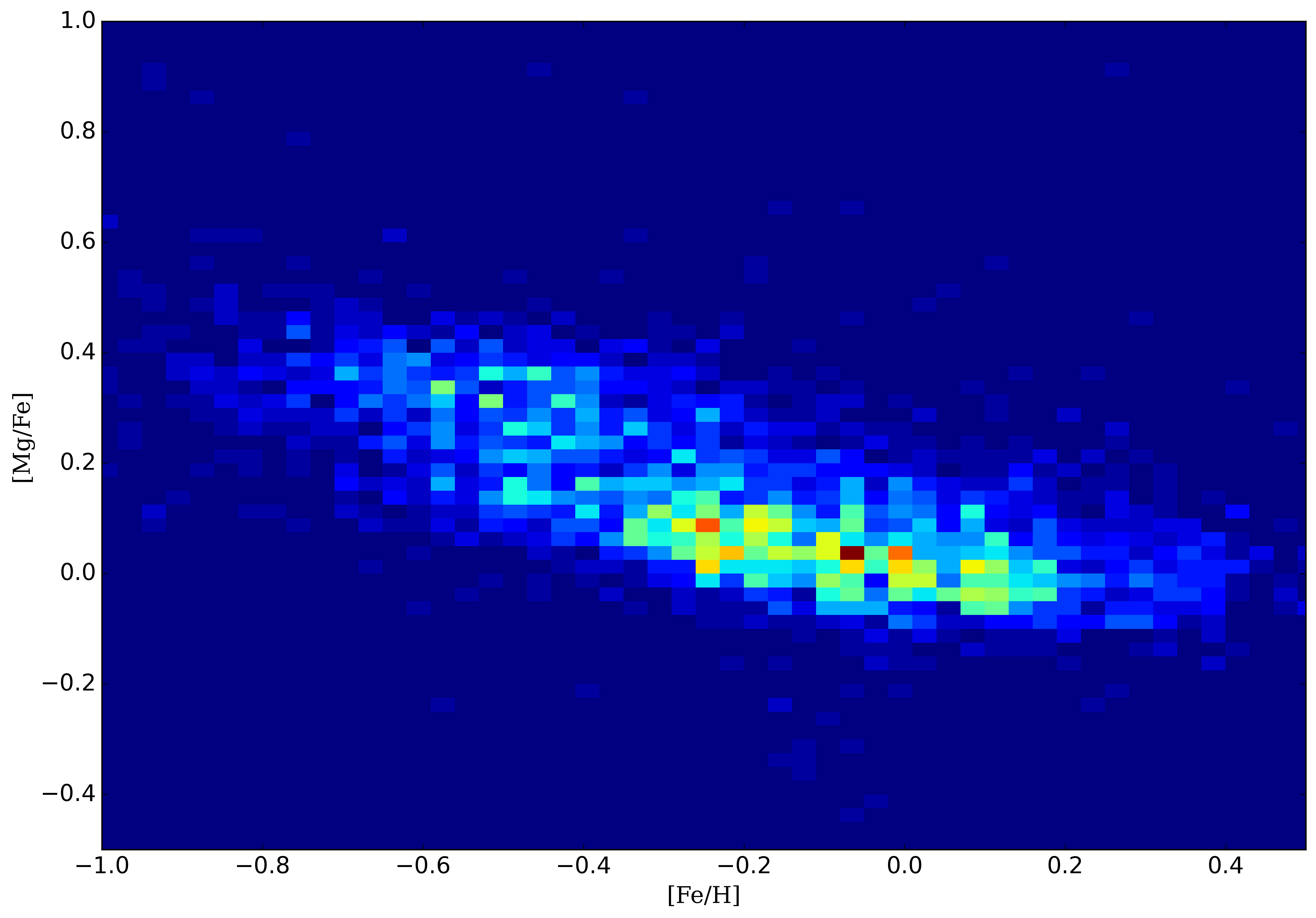}
 \caption{[Mg/Fe] vs [Fe/H] diagram for the MW field stars, plotted using the {\sc iDR6} final recommended values (selecting GES\_TYPE=GE\_MW stars and  SNR>50). }
\label{fig:idr6_alpha}
 \end{figure*}

In Fig.~\ref{fig:idr6_alpha}, we display the density plot of  [Mg/Fe] as a function of [Fe/H] for the MW field populations. 
This diagram is usually used to separate the thin disc population from the thick disc population since \citet{Wallerstein62}. The combined sample, including both UVES WG11 and GIRAFFE WG10 results,  indicates a gap at $\mathrm{[Mg/Fe]} \sim 0.2$ and $\mathrm{[Fe/H]} \sim -0.4$, in a location similar to the one discussed in \citet{recioblanco14} with data from a previous GES data release or in \citet{hayden15} with APOGEE data and in \citet{buder21} with GALAH data.  
In the illustrative Fig.~\ref{fig:idr6_alpha} we plot the entire GES MW sample, making a cut only in SNR. The exact location of the transition between thin and thick disc is a function of $R_{\rm GC}$ and of $z$, the height above the plane, thus it might vary as a function of the selected sample. 

\subsection{Comparison with GALAH and APOGEE surveys}
\begin{figure*}
\begin{subfigure}{.5\textwidth}
  \centering
  \includegraphics[width=1\linewidth]{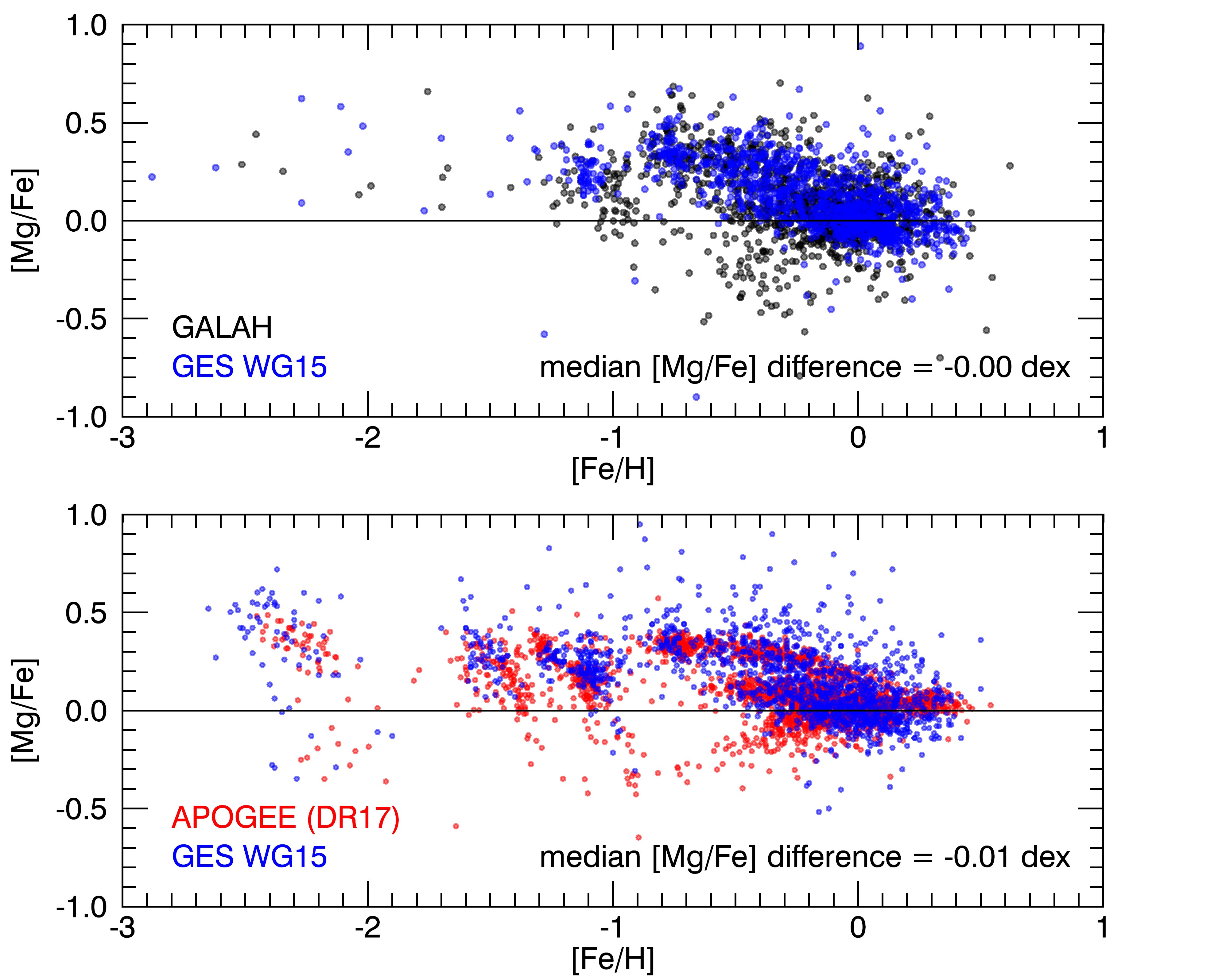}
\end{subfigure}%
\begin{subfigure}{.5\textwidth}
  \centering
  \includegraphics[width=1\linewidth]{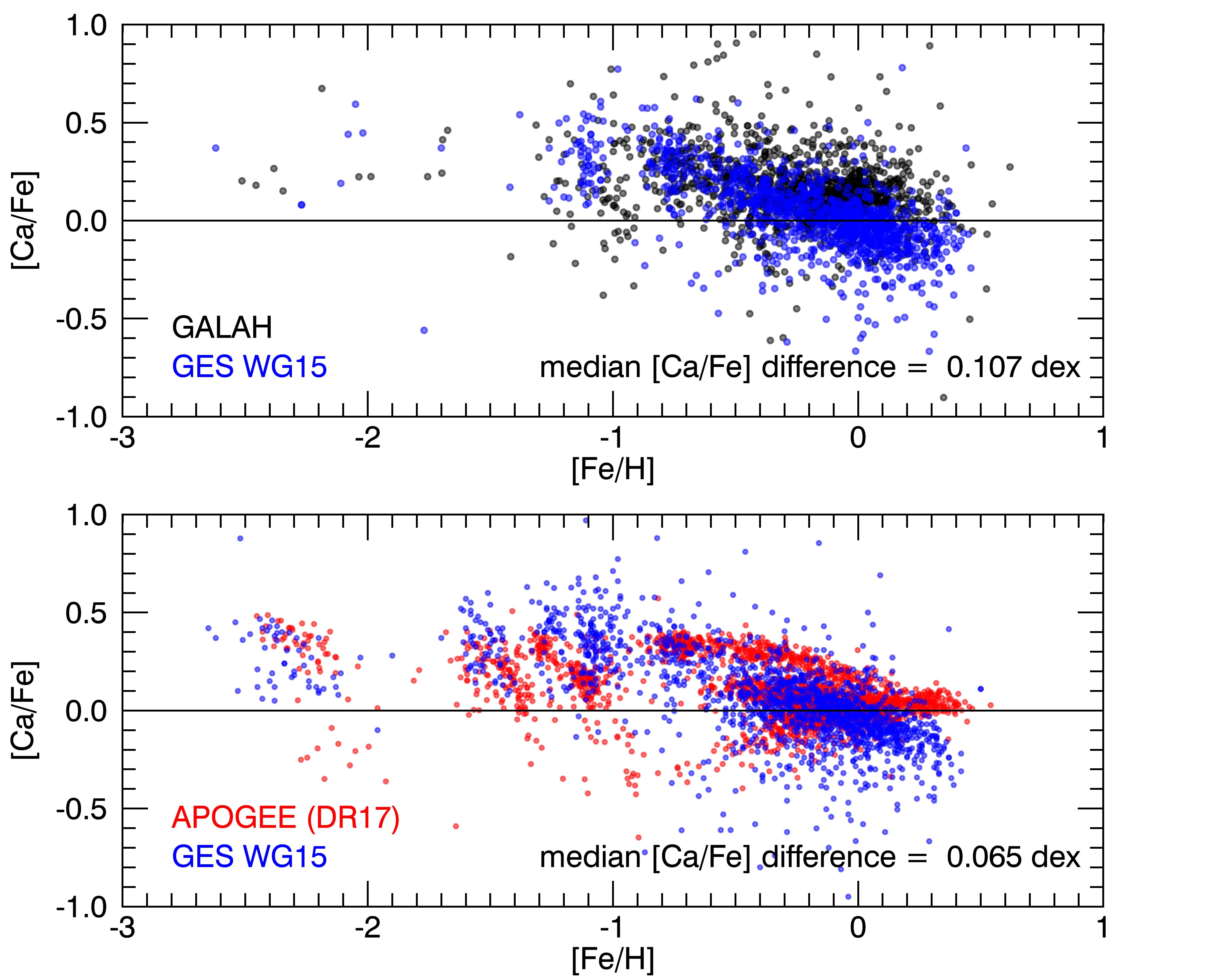}
\end{subfigure}\\
\begin{subfigure}{.5\textwidth}
  \centering
  \includegraphics[width=1\linewidth]{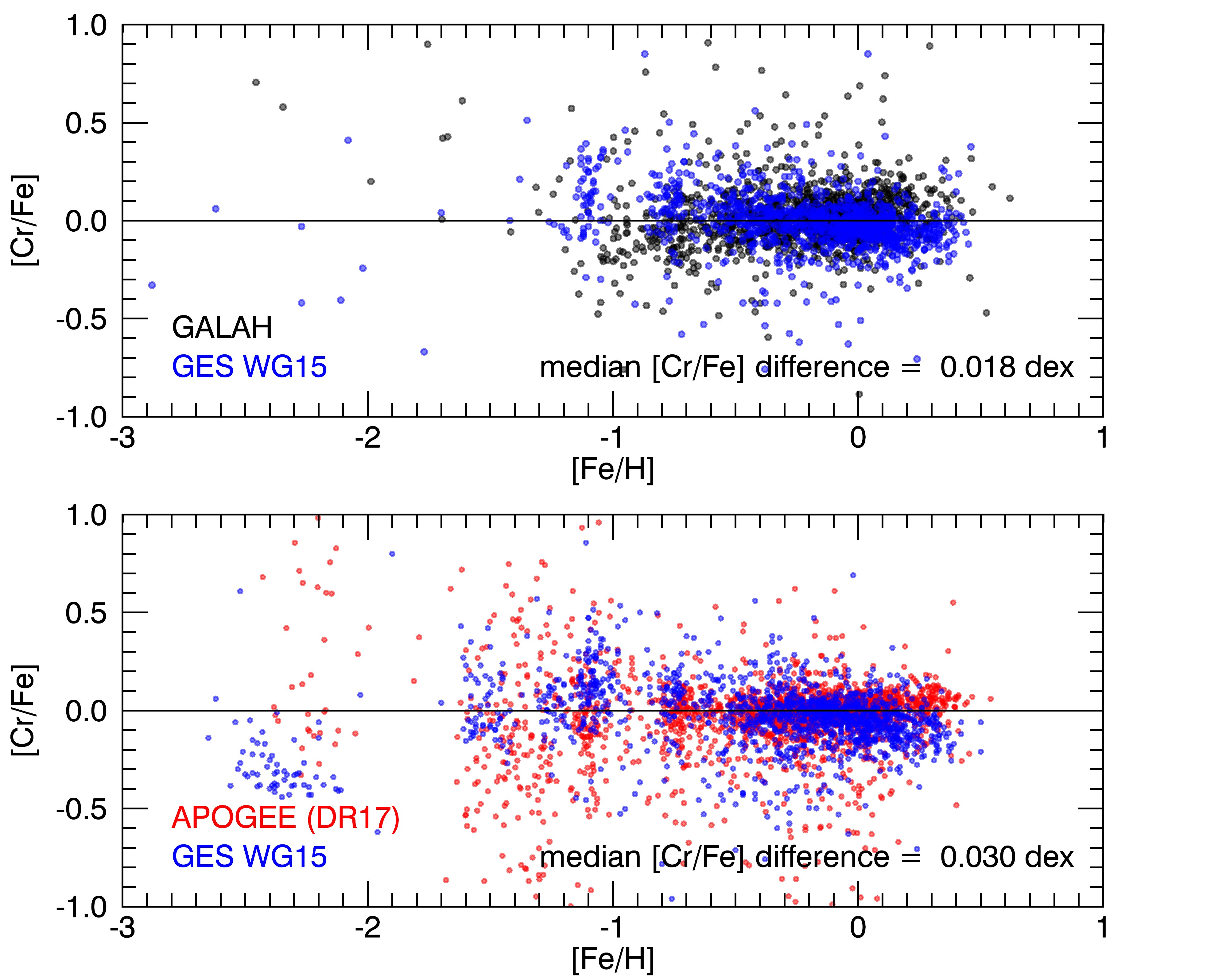}
\end{subfigure}%
\begin{subfigure}{.5\textwidth}
  \centering
  \includegraphics[width=1\linewidth]{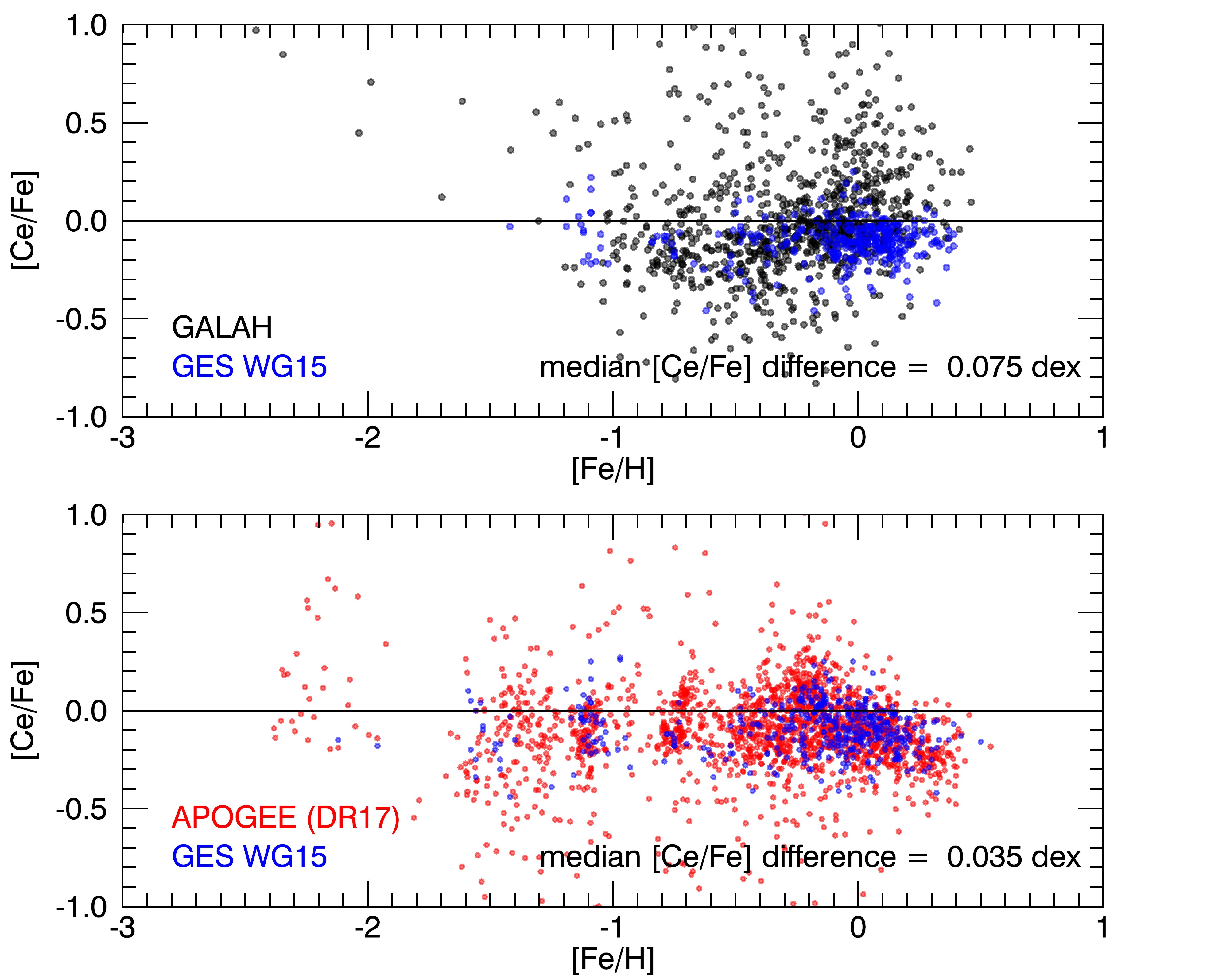}
\end{subfigure}\\
\caption{Comparison between WG15, GALAH and APOGEE abundance ratios for the sample of stars in common. WG15 data are represented as blue circles. Red symbols represent APOGEE data and black symbols are the data from the GALAH survey.  }
\label{fig:wg15_survey_comp}
 \end{figure*}

Fig.~\ref{fig:wg15_survey_comp} shows the comparison between GES and GALAH, and APOGEE, respectively. 
We selected the GES stars that are in common with each survey. The plots show typical comparisons for the $\alpha$-elements, Mg and Ca, an iron-peak element, Cr, and a neutron-capture element, Ce.  It may also be seen  that the dispersion found in the GES results is generally smaller than that found in the results of the other surveys -- a good example is for the element Cr. 
 Table~\ref{tab:Galah_APOGEE_WG15_median} presents the abundance [X/Fe] median differences and the associated dispersion  between GALAH, APOGEE and GES WG15 for the sample of stars in common. In most cases, the median difference is below 0.1~dex, demonstrating the excellent agreement between the surveys.

\begin{table}
\begin{center}
\caption{Median abundance [X/Fe] difference between GALAH, APOGEE and WG15 for the sample for stars in common. 
For the majority of the elements, the median difference is below 0.1~dex.} 
\begin{tabular}{lrrrr}
\hline\hline
     Element    &  \multicolumn{2}{c}{GALAH  $-$ WG15} &   \multicolumn{2}{c}{APOGEE  $-$ WG15}\\
\hline
     &   offset  &  std dev  &  offset & std dev  \\
\hline
C~I  &    0.16   &   0.35    &   0.01  &  0.31  \\
O~I   &  $-0.05$	 &   0.36    &  $-0.09$  &  0.15   \\
Na~I  &    0.01	 &    0.19   &  $-0.08$  &  0.34   \\
Mg~I  &  $-0.01$	 &    0.18   &  $-0.01$  &  0.16   \\
Al~I  &    0.04	 &    0.22   &  $-0.12$  &  0.17   \\
Si~I  &    0.04	 &    0.24   &  $-0.02$  &  0.22   \\
S~I   &           &           &   0.33  &  0.76   \\
Ca~I  &    0.11	 &    0.23   &   0.02  &  0.19   \\
Sc~I  &    0.10	 &    0.17   &         &	 	    \\   
Ti~I  &    0.03	 &    0.23   &  $-0.02$  &  0.22   \\
Ti~II &    $-0.05$	 &    0.25   &	     &	       \\
V~I   &    0.13	 &    0.35   &	     &	       \\
Cr~I &     0.02	 &    0.24   &   0.03  &  0.31   \\
Mn~I  &    0.00	 &    0.32   &   0.04  &  0.22   \\
Co~I  &    0.06	 &    0.42   &  $-0.01$  &  0.29   \\
Ni~I  &    0.04	 &    0.22   &   0.02  &  0.16   \\
Cu~I  &    0.09	 &    0.23   &	       &	       \\
Zn~I  &    0.04	 &    0.29   &         &	       \\
Y~II  &   $-0.03$	 &    0.39   &	      &	       \\
Zr~I  &    0.12	 &    0.52   &	      &	       \\
Mo~I  &    0.15	 &    0.37   &	      &	       \\
Ba~II  &    0.10	 &    0.36   &	      & 	      \\ 
La~II  &    0.06	 &    0.34   &	      &	       \\
Ce~II  &    0.07	 &    0.32   &   0.03 &  0.18   \\
Nd~II  &    0.17	 &    0.39   &	      &	       \\
Sm~II  &   $-0.05$	 &    0.44   &	      &         \\
Eu~II  &   $-0.05$	 &    0.22   &        &          \\

\hline \hline
\end{tabular}
\label{tab:Galah_APOGEE_WG15_median}\\
\end{center}
\end{table}

\subsection{Comparison with \textit{Gaia} radial velocities and calibrated metallicities}

\begin{figure}
 \centering
\includegraphics[width=1.0\linewidth]{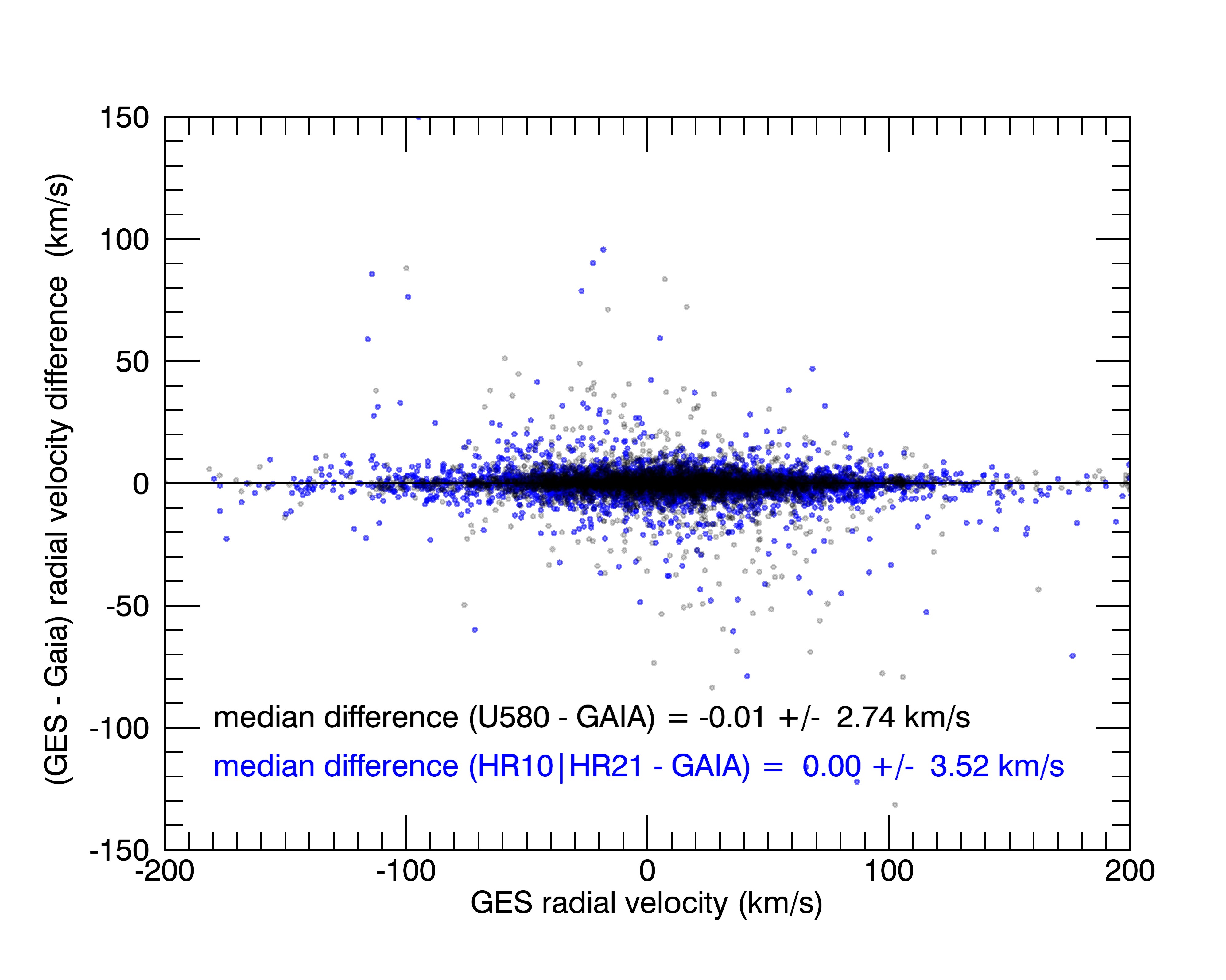}
 \caption{ Radial velocity difference between GES and \textit{Gaia} DR3 for the stars in common. The GES samples HR10|HR21 and U580  have been plotted in different colours.  }
\label{fig:ges_gaia_vrad}
 \end{figure}

In Fig.~\ref{fig:ges_gaia_vrad}, we plot the radial velocity difference (GES - Gaia DR3) as a function of the \textit{Gaia} radial velocity.
The plot has been made for two GES sub-samples in two different instrumental configurations (HR10 and HR21 GIRAFFE spectra) and U580 UVES spectra). The median difference is close to zero and the dispersion is respectively 2.74 and 3.52 for the two setups. The agreement between the GES and the \textit{Gaia} radial velocities is excellent for both setups. 

\begin{figure}
 \centering
\includegraphics[width=1.0\linewidth]{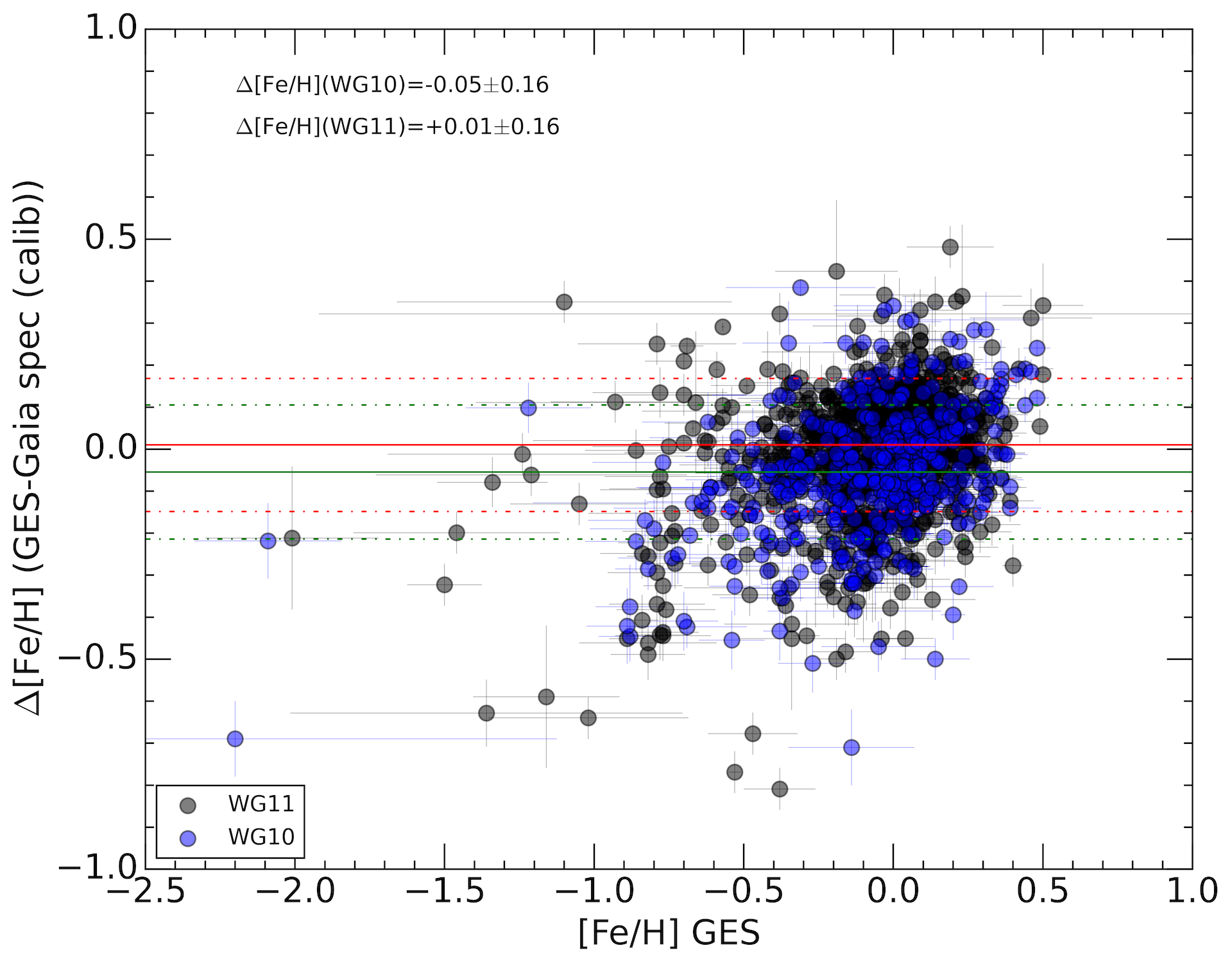}
 \caption{ [Fe/H] difference between GES and \textit{Gaia} DR3 for the stars in common. The  GES samples GIRAFFE (all setups) and U580  have been plotted in different colours. The horizontal red lines mark the median difference U580 - \textit{Gaia} DR3 calibrated metallicity (continuous line) and $\pm$1-$\sigma$ (dot-dashed lines). The green lines indicate the  median difference GIRAFFE - \textit{Gaia} DR3 calibrated metallicity (continuous line) and $\pm$1-$\sigma$ (dot-dashed lines).}
\label{fig:ges_gaia_mh}
 \end{figure}

Fig.~\ref{fig:ges_gaia_mh} shows the plot of the difference in metallicity (GES - \textit{Gaia} DR3) as a function of GES [Fe/H]. For \textit{Gaia}, we used the calibrated spectroscopic metallicities, as described in \citet{recioblanco22}. As for the radial velocity, we have separated the sample observed with U580 and the sample observed with the GIRAFFE setups. The latter contains observations with HR10, HR15N, HR10:HR21, HR21, HR9B. The median differences for both samples are close to zero, with a dispersion of 0.16~dex, indicating, on average, a very good agreement in terms of accuracy of \textit{Gaia} compared to GES, and, as expected, a lower precision. 
 
\subsection{Conclusion}

By concept, the  Gaia-ESO Survey is based on a heterogeneous set of data, medium resolution spectra with different wavelength ranges (GIRAFFE) and high resolution spectra (UVES).  One of its  strengths is that the spectra have been acquired with high efficiency multiplex spectrographs attached to an 8m class telescope. 
In contrast to other large surveys,  the sample has not been  limited to FGK stars, it includes cool pre-main sequence and hot stars (OBA). The originality of the GES survey is that it does not rely on a single pipeline for the analysis of the spectra. 
The multi-method, multi-pipeline design of the GES analysis, implemented through the analysis node and WG structure, means that multiple results are delivered for many GES stars. This includes both parallel analyses of the same stellar samples within a WG, and the analysis
of common calibration samples across WGs. 
The homogenisation process is strongly based on a set of benchmark stars and calibration open and 
globular clusters able to cover the vast range of stellar parameters of the whole sample of stars of this survey.
To provide a final consistent set of results, a dedicated working group (WG15) was set up within GES to provide the recommended results from
Working  Groups 10 to 13 and to put these on to a common scale. With this set of homogenised stellar parameters, detailed abundances have been computed by the Nodes and merged for each WG.  The resulting abundances have been analysed by WG15 to set them, element by element,  on a common scale.
This paper  describes the numerous steps followed by the WG15 that led to the final homogenised set of  stellar parameters, abundances and velocities for more than 110000 stars. The numerous figures of this article give a good overview of the quality  of the  homogenisation process and of the final abundance results compared to literature data and  large spectroscopic survey results. 

Many of the surveys currently underway and in preparation have been inspired by the Gaia-ESO structure and approach, e.g. the FITS format for data exchange is currently adopted by most Galactic spectroscopic surveys, which was not previously common in this field, the Gaia-ESO line list is widely used and the use of the various categories of calibrators has been widely adopted, particularly that of star clusters.  The experiment with multiple pipelines was helpful in understanding the inherent limitations of spectral analysis, and the origin of systematic errors in the various approaches. It has been very useful and the community has learnt a lot from it,  but at the same time, it has been necessarily very time-consuming, and has not been repeated. It has been a very unique test which has informed many choices for subsequent surveys. 

\begin{acknowledgements}

Based on data products from observations made with ESO
Telescopes at the La Silla Paranal Observatory under programmes 188.B-3002,
193.B-0936, and 197.B-1074.

These data products have been processed by the Cambridge Astronomy Survey Unit (CASU) at the Institute of Astronomy, University of Cambridge, and by the FLAMES/UVES reduction team at INAF/Osservatorio Astrofisico di Arcetri. These data have been obtained from the Gaia-ESO Survey Data Archive, prepared and hosted by the Wide Field Astronomy Unit, Institute for Astronomy, University of Edinburgh, which is funded by the UK Science and Technology Facilities Council. 

This work was partly supported by the European Union FP7 programme through ERC grant number 320360 and by the Leverhulme Trust through grant RPG-2012-541. We acknowledge the support from INAF and Ministero dell’Istruzione, dell’Universit\`a e della Ricerca (MIUR) in the form of the grant "Premiale VLT 2012". The results presented here benefit from discussions held during the Gaia-ESO workshops and conferences supported by the ESF (European Science Foundation) through the GREAT Research Network Programme.

L. Magrini and M. Van der Swaelmen acknowledges support by the WEAVE Italian consortium, and by the INAF Grant "Checs'. 

A.J.Korn acknowledges support by the Swedish National Space Agency (SNSA).

A.Lobel acknowledges support in part by the Belgian Federal Science Policy Office under contract No. BR/143/A2/BRASS and by the European Union Framework Programme for Research and Innovation Horizon 2020 (2014-2020) under the Marie Sklodowska-Curie grant Agreement No. 823734.

D.K.Feuillet was partly supported by grant No. 2016-03412 from the Swedish Research Council.

D.Montes acknowledges financial support from the Agencia Estatal de Investigaci\'on of the Ministerio de Ciencia, Innovaci\'on through project PID2019-109522GB-C54 /AEI/10.13039/501100011033

E.Marfil acknowledges financial support from the European Regional Development Fund (ERDF) and the Gobierno de Canarias through project ProID2021010128.

J.I.Gonz\'alez Hern\'andez acknowledges financial support from the Spanish Ministry of Science and Innovation (MICINN) project PID2020-117493GB-I00.

M.Bergemann is supported through the Lise Meitner grant from the Max Planck Society and acknowledges support by the Collaborative Research centre SFB 881 (projects A5, A10), Heidelberg University, of the Deutsche Forschungsgemeinschaft (DFG, German Research Foundation).  This project has received funding from the European Research Council (ERC) under the European Union‚ Horizon 2020 research and innovation programme (Grant agreement No. 949173).

P.Jofr\'e acknowledges financial support of FONDECYT Regular 1200703 as well as N\'ucleo Milenio ERIS NCN2021\_017.

R.Smiljanic acknowledges support from the National Science Centre, Poland (2014/15/B/ST /03981).

S.R.Berlanas acknowledges support by MCIN/AEI/10.13039/501100011033 (contract FJC 2020-045785-I) and NextGeneration EU/PRTR and MIU (UNI/551/2021) through grant Margarita Salas-ULL.

T.Bensby acknowledges financial support by grant No. 2018-04857 from the Swedish Research Council.

T.Merle is supported by a grant from the Foundation ULB.

T.Morel are grateful to Belgian F.R.S.-FNRS for support, and are also indebted for an ESA/PRODEX Belspo contract related to the Gaia Data Processing and Analysis Consortium and for support through an ARC grant for Concerted Research Actions financed by the Federation Wallonie-Brussels.

W.Santos acknowledges  FAPERJ for a Ph.D. fellowship.

H.M. Tabernero acknowledges financial support from the Agencia Estatal de Investigación of the Ministerio de Ciencia, Innovación through project PID2019-109522GB-C51 /AEI/10.13039/501100011033.

\end{acknowledgements}


\bibliographystyle{aa} 
\bibliography{wg15} 

\begin{appendix} 
\section{Simplified flags }
\begin{table*}
\begin{center}
\caption{ simplified flags} 
\noindent\begin{tabularx}{\textwidth}{rlX}
\hline\hline
 Acronym &   Meaning   & Comments: conditions for raising the flag \\
\hline

SNR  &
No or inaccurate results due to low SNR &
This flag is raised if the SNR is lower than 50 and if the object has an incomplete set of parameters. \\
\hline
SRP &

Spectral Reduction Problem & 
This flag is raised if there are no parameters nor abundances\\
\hline
SDS&

Some Discarded Spectra &
This flag is raised if there are some parameters and abundances despite a reduced amount of usable data. For example, it is raised in case spectral reduction problems affected some settings, preventing from getting all the results, but allowing some parameters and abundances to be nevertheless determined. \\
\hline

IPA &

Incomplete Parameters &
This flag is raised, typically, when a key set-up for a given parameter is missing, or when the node experienced an issue for converging to a consistent set of parameters, or, alternatively, when the parameters were out of the parameter grid of model atmospheres used by a specific node. \\
\hline
SSP &

Some Suspicious Parameters &
This flag is raised when some parameters, but not all, could be determined. This can occur when re-normalisation failed, when the code did not converge to a consistent set of parameters, or, again, because the parameters fell out of the node's grid. It also occurs when a parameter was derived outside the group of validated nodes for this parameter. It is also raised in case of spectroscopic multiplicity with at least two visible components (SBn, n$\ge$2). \\

\hline
NIA&

No Individual Abundance (except Fe) &
This flag is usually raised when there are too few available lines for abundance determinations (except Fe).\\

\hline
SSA &

Some Suspicious Abundances &
This flag can be raised for metallicity, e.g. when the Fe I and Fe II lines are discrepant, or for other elements. It is raised in case of high v sini values, or in case of SBn, n$\geq$2, or when the node was uncertain about this abundance.\\

\hline
PSC &

Parameter space coverage &
This flag is typically raised when the parameters are not within the model atmosphere grid parameters of the node, or are on the node's grid edge. Some abundances might then be missing.\\

\hline
SRV &

Suspicious or unreliable Radial Velocity &
This flag is raised in case the CCF was corrupted, or if the RV was discrepant between set-ups, or in case the object was identified as an SBn.\\
\hline
SRO &
Suspicious ROtational velocity &
This flag is raised in case of no rotational velocity determination, or in case of a too high, or revised, rotational velocity. It is also raised  in case of SBn, n$\ge$2. \\

\hline

BIN & 

Detected BINary : SB1 or SBn$\ge$ 2. & \\
\hline

EML &
EMission Line: any line, not only Halpha& \\

\hline \hline
\end{tabularx}
\label{simplified} \\
\end{center}
\end{table*}
\end{appendix}

\

\end{document}